\documentclass[a4paper,fleqn,usenatbib]{mnras}

\usepackage[T1]{fontenc}
\usepackage{ae,aecompl}

\DeclareRobustCommand{\VAN}[3]{#2}
\let\VANthebibliography\thebibliography
\def\thebibliography{\DeclareRobustCommand{\VAN}[3]{##3}\VANthebibliography}

\usepackage{graphicx}	
\usepackage{amsmath}	
\usepackage{amssymb}	
\usepackage{newtxtext,newtxmath}
\usepackage{orcidlink}
\usepackage{multirow}
\usepackage{lastpage}

\newcommand{\Te}{{T_{\rm e}}}
\newcommand{\Tcmb}{{T_{\rm CMB}}}
\newcommand{\Tff}{{T_{\rm ff}}}
\newcommand{\tauff}{{\tau_{\rm ff}}}
\newcommand{\gff}{{g_{\rm ff}}}
\newcommand{\alphasyn}{{\alpha_{\rm syn}}}
\newcommand{\SNRAME}{{\rm SNR_{\rm AME}}}

\newcommand{\nuame}{{\nu_{\rm AME}}}
\newcommand{\Wame}{{W_{\rm AME}}}
\newcommand{\taud}{{\tau_{\rm353}}}
\newcommand{\betad}{{\beta_{\rm d}}}
\newcommand{\Td}{{T_{\rm d}}}
\newcommand{\kb}{{k_{\rm B}}}
\newcommand{\h}{{\rm h}}
\newcommand{\GHz}{{\,GHz}}
\newcommand{\deltaSCMB}{\Delta S_{\nu}^{\rm CMB}}
\newcommand{\Iame}{{I_{\rm AME}}}
\newcommand{\Aame}{{A_{\rm AME}}}
\newcommand{\emmAME}{{\epsilon_{\rm AME}^{\rm 28.4\,GHz}}}
\newcommand{\EM}{{{\rm EM}}}

\newcommand{\nside}{{$N_{\rm side}$}}
\newcommand{\SPDUST}{{\tt SPDUST}}

\title[Galactic AME spatial variations with QUIJOTE]{QUIJOTE 
scientific results --- X. Spatial variations of Anomalous 
Microwave Emission along the Galactic plane} 

\author[Fern\'{a}ndez-Torreiro et al.]
{M. Fern\'{a}ndez-Torreiro$^{\orcidlink{0000-0002-6805-9100}}$,$^{1,2}$\thanks{E-mail: mateo.fernandez@iac.es (MFT)}
{J.~A. Rubi\~{n}o-Mart\'{\i}n$^{\orcidlink{0000-0001-5289-3021}}$,$^{1,2}$}\thanks{E-mail: jalberto@iac.es}
{C.~H. L\'{o}pez-Caraballo$^{\orcidlink{0000-0002-6439-5385}}$,$^{1,2}$}
\newauthor
{R.~T. G\'{e}nova-Santos$^{\orcidlink{0000-0001-5479-0034}}$,$^{1,2}$}
{M.~W. Peel$^{\orcidlink{0000-0003-3412-2586}}$,$^{1,2}$}
{F. Guidi$^{\orcidlink{0000-0001-7593-3962}}$,$^{1,2,3}$}
{S.~E. Harper$^{\orcidlink{0000-0001-7911-5553}}$,$^{4}$}
{E. Artal$^{\orcidlink{0000-0002-2569-1894}}$,$^{5}$}
M. Ashdown,$^{6,7}$
\newauthor
{R.~B. Barreiro$^{\orcidlink{0000-0002-6139-4272}}$,$^{8}$}
{F.~J. Casas$^{\orcidlink{0000-0002-2217-5843}}$,$^{8}$}
{E. de la Hoz$^{\orcidlink{0000-0002-5066-816X}}$,$^{8,9}$}
{D. Herranz$^{\orcidlink{0000-0003-4540-1417}}$,$^{8}$}
{R. Hoyland$^{\orcidlink{0000-0001-5346-0519}}$,$^{1,2}$}
{A. Lasenby$^{\orcidlink{0000-0002-8208-6332}}$,$^{6,7}$}
\newauthor
{E. Mart\'{i}nez-Gonzalez$^{\orcidlink{0000-0002-0179-8590}}$,$^{8}$}
L. Piccirillo,$^{4}$
{F. Poidevin$^{\orcidlink{0000-0002-5391-5568}}$,$^{1,2}$}
{R. Rebolo$^{\orcidlink{0000-0003-3767-7085}}$,$^{1,2,10}$}
{B. Ruiz-Granados$^{\orcidlink{0000-0003-3229-2725}}$,$^{1,2,11}$}
\newauthor
D. Tramonte,$^{12,1,2}$
F. Vansyngel,$^{1,2}$
{P. Vielva$^{\orcidlink{0000-0003-0051-272X}}$,$^{8}$}
{R.~A. Watson$^{\orcidlink{0000-0002-5873-0124}}$.$^{4}$}
\\
$^{1}$Instituto de Astrof\'{\i}sica de Canarias, E-38200 La Laguna, Tenerife, Spain\\
$^{2}$Departamento de Astrof\'{\i}sica, Universidad de La Laguna,
E-38206 La Laguna, Tenerife, Spain\\
$^{3}$Institut d'Astrophysique de Paris, UMR 7095, CNRS \& Sorbonne Universit\'e, 98 bis boulevard Arago, 75014 Paris, France\\
$^{4}$Jodrell Bank Centre for Astrophysics, Alan Turing Building, Department of Physics \& Astronomy, School of Natural Sciences,  The University of Manchester, \\ Oxford Road, Manchester, M13 9PL, U.K. \\
$^{5}$Universidad de Cantabria, Departamento de Ingeniería de Comunicaciones, Edificio Ingenieria de Telecomunicación, Plaza de la Ciencia nº 1, 39005 Santander, Spain\\
$^{6}$Astrophysics Group, Cavendish Laboratory, University of Cambridge, 
J J Thomson Avenue, Cambridge CB3 0HE, UK\\
$^{7}$Kavli Institute for Cosmology, University of Cambridge, Madingley Road, Cambridge CB3 0HA, UK\\
$^{8}$Instituto de F\'{\i}sica de Cantabria (IFCA), CSIC-Univ. de Cantabria, Avda. los
Castros, s/n, E-39005 Santander, Spain\\
$^{9}$Departamento de F\'{\i}sica Moderna, Universidad de Cantabria,
Avda. de los Castros s/n, 39005 Santander, Spain\\
$^{10}$Consejo Superior de Investigaciones Cient\'{\i}ficas, E-28006
Madrid, Spain\\
$^{11}$Departamento de F\'{\i}sica. Facultad de Ciencias. Universidad de C\'ordoba. Campus de Rabanales, Edif. C2. Planta Baja.  E-14071 C\'ordoba, Spain\\
$^{12}$Department of Physics, Xi'an Jiaotong-Liverpool University, 111 Ren'ai Road, \\  \quad Suzhou Dushu Lake Science and Education Innovation District, Suzhou Industrial Park, Suzhou 215123, P.R. China \\
}

\date{Accepted 2023 August 04. Received 2023 June 14; in original form 2023 February 07}

\pubyear{2023}

\begin{document}
\label{firstpage}
\pagerange{\pageref{firstpage}--\pageref{LastPage}}
\maketitle
\defcitealias{LambdaOrionis}{CA21}
\defcitealias{AMEwidesurvey}{P23}
\defcitealias{planck2015galacticcloudsAME}{PL14b}
\defcitealias{planck2016Xforegroundmaps}{PL16}

\begin{abstract}
Anomalous Microwave Emission (AME) is an important emission 
component between 10 and 60\,GHz that is not yet
fully understood. It seems to be ubiquituous in our Galaxy and 
is observed at a broad range of angular scales.
Here we use the new QUIJOTE-MFI wide survey 
data at 11, 13, 17 and 19\,GHz to constrain the AME 
in the Galactic plane ($|b|<10\degr$) on degree scales. 
We built the spectral energy distribution between 
0.408 and 3000\,GHz for each of the 5309 0.9\degr 
\,pixels 
in the Galactic plane, and fitted a parametric 
model by considering five 
emission components: synchrotron, free-free, AME, 
thermal dust and CMB anisotropies. We show that 
not including QUIJOTE-MFI data points leads to the 
underestimation (up to 50\%) of the
AME signal in favour of free-free emission.
The parameters 
describing these components are then intercompared, 
looking for relations that help to understand 
AME physical processes.
We find median values for the AME width, $\Wame$, 
and for its peak frequency, $\nuame$,
respectively of $0.560^{+0.059}_{-0.050}$ and 
$20.7^{+2.0}_{-1.9}$\,GHz,
slightly in tension with current theoretical models.
We find spatial variations throughout the Galactic plane 
for $\nuame$, but only with reduced
statistical significance.
We report correlations of AME parameters with 
certain ISM properties, such as that between the AME 
emissivity (which shows variations with the Galactic longitude)
and the interstellar radiation field, and
that between the AME peak frequency and dust 
temperature. Finally, we discuss 
the implications of our results on the possible molecules 
responsible for AME.
\end{abstract}

\begin{keywords}
radiation mechanisms:general -- ISM: general -- radio 
continuum: ISM -- \textit{(cosmology)}: diffuse 
radiation -- Galaxy: general
\end{keywords}


\section{Introduction}
The first detections of Galactic Anomalous Microwave 
Emission (AME) were carried out less than 30 years 
ago~\citep{kogut1996,leitch1997}: the Differential Microwave Radiometers 
\citep[DMR,][]{dmr} onboard of the Cosmic Bakground Explorer 
\citep[COBE,][]{cobe}, recorded an unexpected excess emission at
31\,GHz. This excess was first thought to be 
due to free-free or synchrotron components. However, 
this emission did not correlate with H$\alpha$ emission, 
which is expected for free-free\footnote{In absence of extreme 
conditions, such as in compact HII regions.}, and was not
polarized, as synchrotron is. This 
supported a scenario with a fresh new emission component that was important 
important through the 10--60\,GHz frequency 
range~\citep{deOliveira1999, cosmosomas2005, cosmosomas2007}. 
This new component was named 
``anomalous microwave emission'', or AME. Further Galactic~\citep{RCW175, 
VSAGalacticplanesurvey, rgspleiades, planck2015galacticcloudsAME, 
m31AME, comapVI} and extragalactic~\citep{NGC6946AME, NGC6946AME2,
NGC6946AME3, NGC4125AME, radiostarformationAMEsurvey} AME sources 
have since been identified, with studies also providing
upper levels when a detection was not 
achievable~\citep{peelAMEgalaxies, planckM31, planckM33, SRTsearches}. 

The emission mechanism for this new component is not 
 yet clear, but the most popular hypothesis states that 
electric dipole emission from spinning dust grains 
is responsible for this excess 
signal~\citep{historicalAME,draine1998a,draine1998b}. 
This would explain the 
correlation between 
AME and mid-infrared dust emission~\citep{firstcorrelationwdust, 
davies2006}. Nevertheless, recent results have proposed 
that polycyclic aromatic hydrocarbons (or PAHs) may not be as 
important as first thought, with 
generic very small grain (VSG) emission being more 
prominent \citep{hensley2016, hensley2017}. This would 
be supported by a higher correlation between AME and 
the dust emission at 24--60\,$\mu$m, instead of that
measured at 8--12\,$\mu$m (dominated by PAHs) \citep{pah1, pah2}. 
Large amorphous carbon or silicates \citep[$a_0\in(1,100)$\,nm][]{
compiegne2011}, generically referred to as dust big grains (BGs)
have been proposed also as AME carriers \citep{chuss2022}: in that case,
AME would correlate more strongly with 100--350\,$\mu$m emission
bands. Finally, the other main hypothesis for AME states that it
could also be due to dust grains inside a magnetic 
field, which aligns the grains that emit radiation when 
their minimum energy state is reached~\citep{draine1999magnetic}. 
This implies that, unlike in the spinning dust theory, the emission would be thermal. 

However, both hypotheses have their disadvantages. In the magnetic 
field scenario, these are the current upper limits on 
polarisation for AME emission. According to recent
 data~\citep{caraballo2011, JARM2012, Perseus, Taurus, 
W51} the AME polarized emission fraction is $\leq$\,5\,\%, with the
strongest constrains being $\leq$0.5\,\% \citep{W44}.
A higher value is predicted in most magnetic models
\citep{draine1999magnetic, drainehensley2013magneticpolarization, 
hoanglazarian2016magneticpolarization}. The problem with the 
spinning dust hypothesis, on the other hand, is the difficulty involved in the 
study of grain theory, where several parameters have 
a direct influence on the spectral shape emission 
\citep{spdust1, ysard2011, ali-hamoud_review}. For a more 
detailed and comprenhesive review on AME, see~\cite{reviewclive2017}.

In this paper, using the new data between 10 and 20\,GHz from the
Multi-Frequency Instrument (hereafter, MFI) mounted on the Q-U-I 
JOint Tenerife Experiment (from now on, QUIJOTE), we aim to analyse 
how the parameters describing the AME vary along the Galactic plane. 
Spatial variations of AME properties have been hinted at in the
past for sub-degree scales
\citep{dickinsonorionspvariations, tibbsspvariations, 
battistellispvariations, arcetord2020, rhoOphiuchi_arcmin}, although mainly 
using interferometers. However, variations at degree 
scales have been measured only recently~\citep{LambdaOrionis}, 
with the addition of medium-size telescopes focused on 
frequencies just below those studied by the WMAP
 and \textit{Planck} satellites. 
The addition of ancillary data will further allow
 us to build a map of diffuse Galactic emission  while
assuming more relaxed priors than previous studies~
\citep{planckbetamm, planck2016Xforegroundmaps, beyondplanckintensity}. 
These data also allows us better to understand the AME 
from the phenomenological point of view, along with the 
other foregrounds. Precise knowledge concerning 
these foregrounds (especially synchrotron, which greatly 
benefits from the addition of more frequency points below
 20\,GHz and is also present in polarisation) will be 
essential for future missions focusing on CMB B-mode 
studies, such as the LiteBIRD satellite~\citep{litebird2020}.

This paper is organized as follows. Section~\ref{section:region_definition} 
gives an overview of the data set used to build the 
intensity spectral energy distributions (SEDs). The 
components describing these SEDs are then explained 
in Section~\ref{section:Methodology}, together with the 
Monte Carlo Markov Chain (MCMC) analysis applied.
 Section~\ref{section:results} presents the main results 
of the paper, and Section~\ref{section:discussion} 
compares these results with previous studies. Finally, 
 in Section~\ref{section:conclusions} we summarize 
the work done.

\section{Input data}
\label{section:region_definition}
\begin{figure*}
    \centering
    \includegraphics[width=1\linewidth]{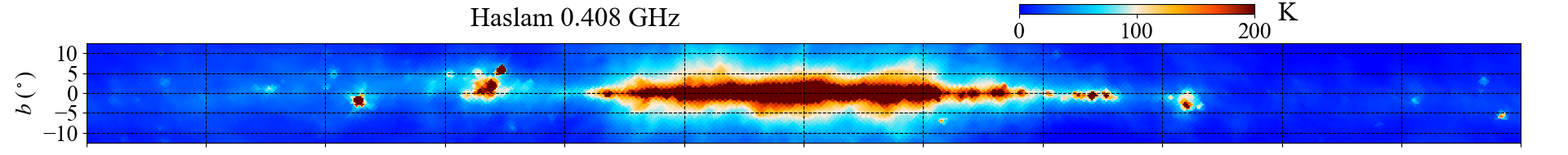}
    \includegraphics[width=1\linewidth]{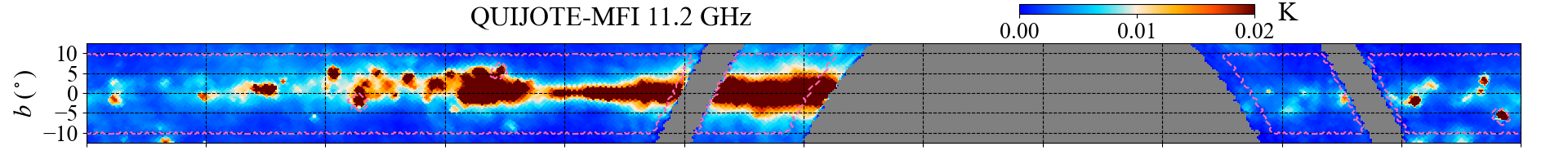}
    \includegraphics[width=1\linewidth]{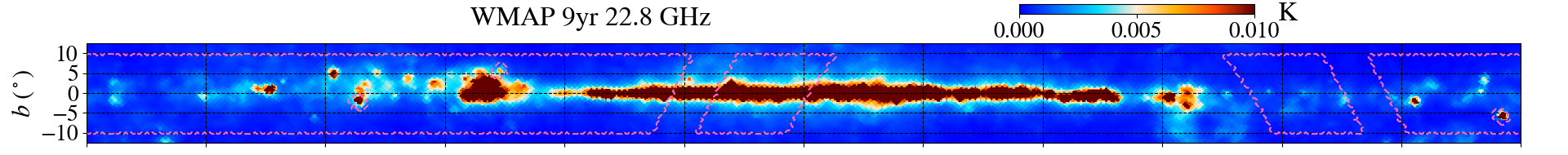}
    \includegraphics[width=1\linewidth]{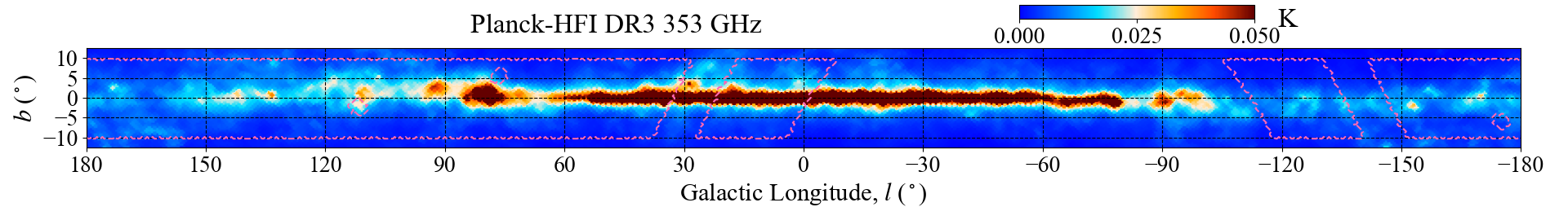}
    \caption{Example of some of the frequency maps used 
in this study. From top to bottom: \protect\cite{haslam1982} 
at 0.408\,GHz, QUIJOTE-MFI at 11.2\,GHz, WMAP 9yr at 
22.8\,GHz and \textit{Planck-HFI} DR3 at 353\,GHz. The 
first one and last two are good tracers for synchrotron, 
AME and thermal dust emission respectively. We also 
indicate the areas (shown with dashed-lines)  studied 
in this study: those with 
$|b|<10\degr$ and $\delta\in(-32\degr, -10\degr)$ and 
$\delta>0\degr$. Some pixels with lower declinations are
also observable by QUIJOTE-MFI (down to $\delta\approx-35\degr$),
but they are greatly affected by $1/f$ noise.}
    \label{fig:frequency_maps}
\end{figure*}

This paper is part of the QUIJOTE-MFI Wide Survey
\citep{mfiwidesurvey} Release and exploits the survey's new 
10--20\,GHz data. We built intensity spectral energy 
distributions (SED) in the radio domain for the 
Galactic plane $b\in(-10\degr, +10\degr)$ region. The 
parts of the Galactic plane that are not visible in 
all QUIJOTE-MFI bands (i.e. the equatorial band 
$\delta\in(-10\degr, 0\degr)$\footnote{Due to Radio 
Frequency Interference (RFI) contamination from 
geostationary satellites, which emit close to 11-13\,GHz \citep{mfiwidesurvey}.} and the southern sky 
$\delta<-32\degr$) were not studied. Together with 
the QUIJOTE-MFI data, another 19 maps, 
listed in Table~\ref{table:maps}, were used and are described 
in the following subsections. Figure~\ref{fig:frequency_maps} 
shows four examples of these maps. 

All maps were smoothed to $1\degr$ beam resolution at their 
native pixelization, and then downgraded to 
{\tt HEALPix}\footnote{https://healpix.sourceforge.io/} 
\citep{Healpix} \nside = 64 pixelization, where each pixel 
has an angular size of 
$\simeq$0.9$\degr$. 
We can assume then that each pixel on the maps is 
almost uncorrelated with its neighbours, as the resolution of
the maps matches the pixel size. Therefore, the area of 
study corresponds to 5309 regions, one per pixel 
($f_{\rm sky}\approx11\%$), after discarding the southern 
sky and the equatorial band. 

\begin{table*}
\centering
\hspace*{-0.75cm}
\caption{Surveys used in this study. Effelsberg and Parkes 
(SPASS) surveys have been used only for the Stockert/Villa-Elisa 
and HartRAO surveys recalibrations, respectively. Each 
pixel SED uses only the maps covering that pixel. Under the
column `Calibration' we quote the values used in this study 
to estimate the calibration uncertainty.}
\begin{tabular}{|c|c|c|c|c|c|}
\hline
Telescope & Frequency & Calibration & FWHM & Sky Coverage & Reference \\ 
 & (GHz) & (\%) & (arcmin) & & \\ \hline
Various & 0.408 & 10 & 51 & All-sky &~\cite{haslam1982},~\cite{remazeilleshaslam} \\ 
Dwingeloo & 0.82 & 10 & 72 & $\delta>-7\degr$ &~\cite{dwingeloo} \\ 
\multirow{2}{*}{Effelsberg} & \multirow{2}{*}{1.408} & \multirow{2}{*}{10} & \multirow{2}{*}{9.4} & $l\in[240,\ 357]\degr$ &\multirow{2}{*}{~\cite{effelsberg1},~\cite{effelsberg2}} \\
 & & & & $b\in[-4,\ 4]\degr$ & \\
\multirow{2}{*}{Stockert/Villa-Elisa} & \multirow{2}{*}{1.42} & \multirow{2}{*}{20} & \multirow{2}{*}{34.2} & \multirow{2}{*}{All-sky} &~\cite{reich1982},~\cite{reich1986} \\ 
 & & & & &~\cite*{villaelisa},~\cite{CADE} \\
Parkes (SPASS) & 2.303 & 5 & 8.9 & $\delta<-1\degr$ &~\cite{carretti2019} \\
HartRAO & 2.326 & 20 & 20 & $\delta<13\degr$ &~\cite*{hartrao},~\cite{plataniamaps} \\ 
QUIJOTE-MFI & 11.2 & 5 & 53.2 & $\delta>-32\degr$ &~\cite{mfiwidesurvey} \\ 
QUIJOTE-MFI & 12.9 & 5 & 53.5 & $\delta>-32\degr$ &~\cite{mfiwidesurvey} \\ 
QUIJOTE-MFI & 16.8 & 5 & 39.1 & $\delta>-32\degr$ &~\cite{mfiwidesurvey} \\ 
QUIJOTE-MFI & 18.7 & 5 & 39.1 & $\delta>-32\degr$ &~\cite{mfiwidesurvey} \\ 
WMAP K 9yr  & 22.8 & 3 & 51.3 & All-sky &~\cite{wmap} \\ 
\textit{Planck-LFI} DR3 & 28.4 & 3 & 33.1 & All-sky &~\cite{planck} \\ 
WMAP Ka 9yr & 33 & 3 & 39.1 & All-sky &~\cite{wmap} \\ 
WMAP Q 9yr & 40.7 & 3 & 30.8 & All-sky &~\cite{wmap} \\ 
\textit{Planck-LFI} DR3 & 44.1 & 3 & 27.9 & All-sky &~\cite{planck} \\ 
WMAP V 9yr & 60.7 & 3 & 21.0 & All-sky &~\cite{wmap} \\ 
\textit{Planck-LFI} DR3 & 70.4 & 3 & 13.1 & All-sky &~\cite{planck} \\ 
WMAP W 9yr & 93.5 & 3 & 14.8 & All-sky &~\cite{wmap} \\ 
\textit{Planck-HFI} DR3 & 143 & 3 & 7.3 & All-sky &~\cite{planck} \\ 
\textit{Planck-HFI} DR3 & 353 & 3 & 4.9 & All-sky &~\cite{planck} \\ 
\textit{Planck-HFI} DR3 & 545 & 6.1 & 4.8 & All-sky &~\cite{planck} \\ 
\textit{Planck-HFI} DR3 & 857 & 6.4 & 4.6 & All-sky &~\cite{planck} \\ 
COBE-DIRBE 240 ZSMA & 1249 & 11.6 & 37.1 & All-sky &~\cite{cobe-dirbe} \\ 
COBE-DIRBE 140 ZSMA & 2141 & 10.6 & 38.0 & All-sky &~\cite{cobe-dirbe} \\ 
COBE-DIRBE 100 ZSMA & 2998 & 13.5 & 38.6 & All-sky &~\cite{cobe-dirbe} \\ 
\hline
\end{tabular}
\label{table:maps}
\end{table*}

\subsection{QUIJOTE-MFI data}
\label{section:QUIJOTE}
The QUIJOTE CMB experiment~\citep{quijote2010} operates from  
Teide Observatory (OT) of the Instituto de 
Astrofísica de Canarias (IAC), located at 
latitude $28\degr18^\prime04^{\prime\prime}$ North
and longitude $16\degr30^\prime38^{\prime\prime}$ West.
This latitude allows the telescopes to reach declinations 
as low as $-$32$\degr$, hence permitting partial coverage of
 the South Hemisphere 
sky. A collaboration between the IAC, the Instituto 
de Física de Cantabria (IFCA), Cambridge University, 
Manchester University, the Departamento de Ingenieria de 
Comunicaciones (DICOM) from the Universidad de Cantabria 
and IDOM, QUIJOTE consists of two identical telescopes on
Cross-Dragone optics and 2.25 m primary apertures. 
An altazimuth mount was chosen to allow the telescope to spin fast at 
a constant elevation while observing (the so-called 
`nominal' mode) with a scanning speed 
of 12$\degr$\,s$^{-1}$.

The MFI was the first science instrumentation mounted on the
QUIJOTE experiment (specifically, on its first telescope). 
It observed simultaneously in four distinct bands with 2\,GHz 
bandwidths and central frequencies 11, 13, 17 and 19\,GHz. 
The MFI consisted of four horns or antennas, each observing
at two frequencies: horns 1 and 3 observed at
the lower frequencies (11 and 13\,GHz), and horns 2 and
4 observed at 17 and 19\,GHz. It came into operation in 2012, 
while the wide survey observations were run between 2013
and 2018. These observations were done in the `nominal' 
configuration introduced before for a total 
of $\sim$10000 hours \citep{mfiwidesurvey}, during which all 
the sky accessible from the OT (more than 29000 squared 
degrees) with scans of elevation higher or equal to $30\degr$ 
was observed. 

The QUIJOTE-MFI first data release is presented
in detail in \cite{mfiwidesurvey}. It consists of a wide
survey of the northern sky in four bands at 11.2, 12.9, 
16.8 and 18.7\,GHz, for both intensity and polarisation. 
Sensitivities are better for polarisation than for 
intensity: $\rm 35-40\,\mu K\ deg^{-1}$ vs. 
$\rm 60-200\,\mu K\ deg^{-1}$ respectively, owing to 
the lower 1/$f$ noise in polarisation. 
We use the combined maps between horn 2 
and horn 4 (which are the publicly available ones) for 
16.8 and 18.7\,GHz, as 
the signal-to-noise ratio (SNR) is improved significantly 
than in the non-combined case. Only the maps from horn 3 are 
used for the 11.2--12.9\,GHz pair, as it has much better noise 
properties than horn 1. 
The calibration uncertainty is 5\,\% for all bands. 
Finally, we take into account that every pair 
of frequencies (11.2--12.9\,GHz or 16.8--18.7\,GHz) 
observed with the same horn have correlations close 
to 80\,\% in intensity~\citep{mfiwidesurvey}. This will 
be further explained when building the reconstructed
 spectral energy distribution for the pixels, in 
Section~\ref{section:correlations_explained}.

\subsection{Ancillary low frequency surveys}
\label{section:low_frequency_surveys}
We used several low frequency surveys in this work, 
which are summarized at the beginning of Table~\ref{table:maps}. 
In order to use them consistently with the full data-set, 
some of them require a series of corrections, as explained in detail
in this section.

The \cite{dwingeloo} survey at 0.820\,GHz accounts 
for an uncertainty equal to $0.3$\,K for 
systematic effects (unrelated to the determination
of the zero level) between different areas of 
the sky. This transforms to 1.58\,Jy at 0.820\,GHz 
and {\tt HEALPix} \nside = 64 pixel size. We have also 
increased its calibration uncertainty from 
6\,\% to 10\,\%, owing to the clear presence of 
stripes in the map and its 72\arcmin\,angular 
resolution, which is slightly larger than the $1\degr$ 
used in the following analyses. 

Moreover, both \cite{reich1982, reich1986, villaelisa} survey at 1.42\,GHz 
and \cite{hartrao} survey at 2.326\,GHz are calibrated     
to the full, 4$\pi$ beam. As we are dealing with structures at the 
main-beam scale, multiplicative recalibration factors of 1.55 and 
1.2 should respectively be applied to them. This is done in order to 
account for the flux density that is lost outside the main beam. 
Factors similar to these have been applied in past studies 
\citep{reich1988, planck2015galacticcloudsAME, W44, LambdaOrionis}. 
\cite{irfanthesis} 
showed that the first value is consistent for a series
of free-free dominated emission regions. However, these factors 
are defined from observations of point sources. In the case of 
diffuse emission this correction becomes milder, and these factors 
should be smaller. Because of this, we used a factor of 
1.4 instead of 1.55 for the \cite{reich1982, 
reich1986, villaelisa} survey.
This was also done in the past, e.g. 
\cite{planckM31} when studying emission from M31. An even
lower value (1.3) was used in that case, in fact.
We estimated this value by comparing the map with that of
Effelsberg suvey at 1.408\,GHz (through TT-plot analyses)
 only for those pixels where diffuse emission is 
dominant over compact sources. For the \cite{hartrao} survey 
at 2.326\,GHz, the recalibration factor remains equal to 1.2, 
as this was already consistent with TT-plot analyses 
comparing this map and the S-band Polarization All Sky 
Survey \citep[SPASS, ][]{carretti2019} one at 2.303\,GHz. 
It is worth noting that before reducing the
\cite{reich1982, reich1986, villaelisa} survey factor to 1.4, 
we found 
systematically positive residuals at that frequency. 
This did not happen when dealing with the \cite{hartrao} survey.

Calibration uncertainties for both \cite{reich1982, 
reich1986, villaelisa} and 
\cite{hartrao} surveys were increased to $\simeq20\,\%$. 
Even though the~\cite{hartrao} 
recalibration factor was lower, this survey also had 
measured polarisation ($Q$) with intensity, which further 
increased the uncertainty. Further discussion on 
this issue is presented in 
Appendix~\ref{section:appendix_recalibration}. 
All the previously mentioned data are available in 
the Legacy Archive for Microwave Background Data 
Analysis (LAMBDA\footnote{https://lambda.gsfc.nasa.gov}).

\subsection{WMAP, \textit{Planck}, COBE-DIRBE}
The Wilkinson Microwave Anisotropy Probe 
\citep[WMAP,][]{wmap} and \textit{Planck}~\citep{planck} 
satellite produced full sky maps from 22.8 to 93.5 and 
from 28.4 to 857\,GHz during 2001--2010 and 2009--2013 
respectively. We used the WMAP 9-year and \textit{Planck} 
2018 (DR3) data releases. 
WMAP data are publicly available 
on LAMBDA, and \textit{Planck} data can be 
found through the \textit{Planck} Legacy Archive 
(PLA,\footnote{http://pla.esac.esa.int/pla/\#home}) 
hosted by the European Space Agency (ESA).
The nominal calibration uncertainties for the WMAP 
and \textit{Planck} bands calibrated against the 
CMB dipole (frequencies lower than 500\,GHz) are 
extremely low \citep[below $1\%$,][]{planck2016Xforegroundmaps}. 
However, we increased this value to 3\% to account 
for further inconsistencies, like beam uncertainties and colour 
correction uncertainties, which arise when dealing with
foregrounds. This is, in fact, common through the literature ~\citep{planck2011XXnewlight, 
planck2015galacticcloudsAME, planckancillarydata, 
LambdaOrionis}. \textit{Planck} bands calibrated using
planetary data (545, 857\,GHz) have higher calibration
uncertainties, propagated from the theoretical models.
We discarded \textit{Planck} 100 and 217\,GHz bands due
to the contamination from CO emission.

The Diffuse Infrared Background Experiment 
\citep[DIRBE,][]{cobe-dirbe}, mounted on the Cosmic 
Background Explorer (COBE) satellite, observed the sky
 between 1250\,GHz and 240\,THz during 1989--1990: we 
used the average mission maps with subtracted zodiacal 
light. This latter experiment allowed us better to 
recover the high-frequency side of the thermal 
dust distribution. The data are available on LAMBDA.

\subsection{Further map pre-processing}
\label{section:introduce_dipole}
All the maps used in this study were 
filtered with the same filter used to remove the QUIJOTE-MFI
residual RFI signal. This filter removes the zero mode
in lines of constant declination, effectively reducing the
large scale power of the map ($\ell < 30$).  This is done
to ensure that all maps have the same effective window 
function. Further information on this correction is available 
in~\cite{mfiwidesurvey} [Section 2.4.2 and appendix B].

\section{Methodology}
\label{section:Methodology}
\subsection{Foreground modelling}
\label{section:foreground_model}
We considered five different emission components in our 
frequency range, between 0.4 and 3000\,GHz: synchrotron, 
free-free, anomalous microwave emission (AME), thermal 
dust and CMB anisotropies. In this study, we adopted a 
parametric description for the flux density of these
five components that is based on a set of 10 independent
parameters, $\theta$, as:
\begin{equation}
\begin{split}
  S_\nu^{\rm total}(\theta) = {}&S_{\nu}^{\rm syn}(I_{\rm1\,GHz}, 
\alphasyn) + S_\nu^{\rm ff}(\EM) \\
 &+S_{\nu}^{\rm AME}(\Iame, \nuame, \Wame) \\ 
&+S_{\nu}^{\mathrm{dust}}(\tau_{353}, \betad, \Td) 
+ \deltaSCMB (\Delta \Tcmb)
\end{split}
\end{equation}
which are described in the following subsections. 

\subsubsection{Synchrotron emission}
Synchrotron emission is radiated by ultrarelativistic
electrons accelerated by a magnetic field. Its 
spectral shape can be fitted as a power law 
\citep[e.g.,][]{rybickilightman, nraobook}:
\begin{equation}
  S_{\nu}^{\rm syn}(I_{\rm1\,GHz}, \alphasyn) = 
I_{\rm1\,GHz} \left(\frac{\nu}{1\ \mathrm{GHz}}\right)^{\alphasyn} \Omega,
\end{equation}
where $I_{\rm1\,GHz}$ is the synchrotron flux intensity 
evaluated at 1\,GHz, $\alphasyn$ the synchrotron spectral 
index (for flux units) and $\Omega$ the solid 
 angle covered. 

\subsubsection{Free-free emission}
Unlike synchrotron, free-free emission is radiated by 
electrons accelerated by electric fields 
\citep[e.g.][]{rybickilightman, nraobook}. Its almost flat 
spectrum implies that it can be important at both low 
(below $10$\,GHz) and medium (between 10 and 100\,GHz) 
frequencies. Because 
of this, important degeneracies between the free-free and 
other components (mostly AME and synchrotron) can arise. 
A function of the absorption along the line of sight, or
 opacity, $\tauff$, is needed to describe free-free. We
 use the parametrization from~\cite{draine2011}:
\begin{equation}
  S_\nu^{\rm ff}(\EM) = \frac{2 \kb\nu^2}{c^2} \Omega \Tff 
\end{equation}
with 
\[ \Tff = \Te \left(1 - \mathrm{e}^{-\tauff}\right)\] 
\[\tauff = 5.468\cdot10^{-2}\cdot \left(\frac{\EM}{\rm pc\ cm^{-6}}\right)\cdot \left(\frac{\Te}{\rm K}\right)^{-3/2}\cdot\left(\frac{\nu}{\rm\,GHz}\right)^{-2}\cdot g_{\rm ff}(\nu)\]
\[\gff(\nu) = \ln\left[\exp\left(5.960-\frac{\sqrt{3}}{\pi}\cdot\ln\left[\frac{\nu}{\rm\,GHz}\left(\frac{\Te}{10^4\,{\rm K}}\right)^{-3/2}\right] \right)+\mathrm{e}\right],\]
where we take only the emission measure, $\EM$, as a 
free parameter. The electron temperature, on the other 
hand, is fixed at $\Te=8000$\,K\footnote{The selection
of $\Te$ is not relevant, as it is importantly degenerated
with $EM$ and does not have a significant effect on the 
shape of the spectrum. 
Thus, only one of the two are needed to fix the amplitude.} 
\citep[as in e.g.][]{planck2011XXnewlight, Perseus}. 
This component dominates over synchrotron as the frequency 
increases. There are few data studying diffuse 
emission at frequencies between 3 and 10\,GHz (because of 
the need for large telescopes), where we expect 
synchrotron and free-free to overlap, so both 
are usually strongly degenerate. However, current
 and future experiments will help solve this 
issue~\citep{cbass, LambdaOrionis}.

\subsubsection{AME}
The physical processes responsible for AME are not 
clear yet. Even though limits on polarisation~
\citep{caraballo2011, Perseus, W51} partly discard 
the possibility of a magnetic origin, it is still 
not clear whether they are linked to PAHs or not~
\citep{hensley2016, reviewclive2017, ysard2022}. 
Theoretical models for the electric dipole emission
from spinning dust depend on a large number of 
parameters \citep{spdust1, spdust2}, so we chose
instead to use a simpler, phenomenological model.
This model consists of a log-normal distribution,
which mimics well enough the spinning dust models
\citep{stevenson2014, LambdaOrionis, AMEwidesurvey}:
\begin{align}
\begin{aligned}
   S_{\nu}^{\rm AME}(\Iame, \nuame, \Wame) & \\ 
 = \Iame \mathrm{exp}&\left[-\frac{1}{2\Wame^2} \ln^2\left(\frac{\nu}{\nuame}\right) \right] \Omega,
\end{aligned}
\end{align}
where $\Iame$ is the maximum flux intensity due 
to AME, $\nuame$ the correspondent frequency for 
that maximum and $\Wame$ the width of the 
distribution on the log-log plane.

\subsubsection{Dust emission}
Thermal dust dominates the spectrum at higher 
frequencies, and we fitted its emission to a 
single modified blackbody (MBB):
\begin{equation}
  S_{\nu}^{\mathrm{dust}}(\tau_{353}, \betad, \Td) 
= \frac{2\h\nu^3}{c^2}  \left(\frac{\nu}{353\,\mathrm{GHz}}\right)^{\betad}\tau_{353} \frac{1}{\mathrm{e}^ x-1}\Omega,
\end{equation}
where $\tau_{353}$ is the optical depth normalized 
at 353\,GHz, $\betad$ is the dust emissivity and 
$x=h\nu / \kb \Td$, with $\Td$ being the dust temperature.

\subsubsection{CMB}
Finally, we accounted for a contribution from CMB
anisotropies in our photometry method. 
We estimated the flux density from CMB anisotropies as:
\begin{equation}
\begin{split}
  \deltaSCMB (\Delta \Tcmb) = \frac{2k_{\rm B}\nu^2}{c^2}\frac{x^2\mathrm{e^x}}{(\mathrm{e}^x-1)^2} \Delta\Tcmb\Omega,
\label{eq:CMB}
\end{split}
\end{equation}
where $x=h\nu/\kb T_{\rm CMB}$ and $\Delta\Tcmb$ models the CMB 
anisotropies. $T_{\rm CMB}$ is fixed to 2.72548\,K \citep{fixsen2009}.
In this study, the amplitude of this CMB component 
was consistent with the expected value at these angular scales
($80\,\mu$K). This value makes the CMB less bright than the 
rest of components.

\subsection{Estimation of flux densities for individual pixels}
\label{section:AP}
We built a SED between 0.408\,GHz and 3000\,GHz for 
each {\tt HEALPix} \nside$ = $64 pixel in the region described 
in Section~\ref{section:region_definition}. The flux 
densities for each pixel were computed only for the maps listed
in Table~\ref{table:maps} that covered that pixel 
(e.g. pixels with $\delta>0\degr$ did not have data from 
the \citealt{hartrao} survey, for example).
These flux densities were calculated by subtracting the 
signal from an aperture outside the Galactic plane (the so-called
zero-level reference aperture, see below) from the signal from each pixel 
aperture, which accounts only for the pixel itself:
\begin{equation}
  S_\nu = a(\nu)\Omega T = 
\frac{2\kb\nu^2}{c^2}\frac{x^2\mathrm{e}^x}{(\mathrm{e}^x-1)^2} \Omega T,
\label{eq:AP}
\end{equation}
where $S_\nu$ is the flux density at $\nu$ frequency, 
$a(\nu)$ is the conversion factor between thermodynamic
temperature and intensity, 
$x=\h\nu/\kb T_{\rm CMB}$ and $\Omega$ is the 
solid angle covered by a single pixel. $T$ 
is the difference between the temperature in the 
pixel we want to study, $T_{\rm aper}$, and the 
median temperature in the zero-level reference region, 
$\mathrm{med}(T_{\rm BG})$: 
$T = T_{\rm aper} - \mathrm{med}(T_{\rm BG})$. 
This zero-level reference region is defined here as a 1 degree 
radius aperture outside the Galactic plane and 
centred on $\rm(RA,\delta)=(157.5\degr, +8\degr)$.
 Its uncertainty ($\sigma$) is estimated as the 
quadratic sum of the statistical and calibration 
uncertainties (the latter can be checked in 
Table~\ref{table:maps}), as in the following 
equations~\citep{JARM2012, Perseus}:
\begin{equation}
\sigma_{\rm AP} = a(\nu)\Omega\,\sigma(T_{\rm BG})\sqrt{\frac{1}{n_{\rm aper}} + \frac{\pi}{2}\frac{1}{n_{\rm BG}}}
\label{eq:sigma_AP}
\end{equation}
\begin{equation}
\sigma_{S_\nu} = \sqrt{\sigma_{\rm AP}^2 + \mathrm{cal}^2\cdot S_\nu^2},
\label{eq:sigma_AP_calibration}
\end{equation}
where $n_{\rm aper}$ and $n_{\rm BG}$ are the number 
of pixels embedded in each aperture, constant and 
equal to 1 and 11 pixels respectively, and 
$\sigma(T_{\rm BG})$ is the temperature standard 
deviation within the background aperture. This way of estimating
uncertainties is conservative, as we are assuming calibration errors 
to be uncorrelated across frequencies, which is not true.

\begin{figure*}
\includegraphics[width=1\linewidth]{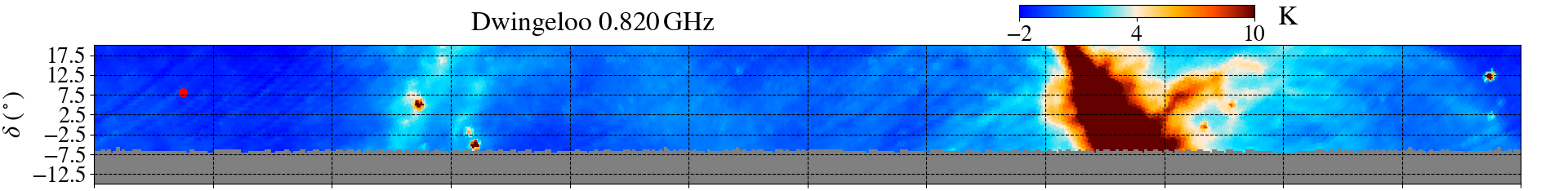}
\includegraphics[width=1\linewidth]{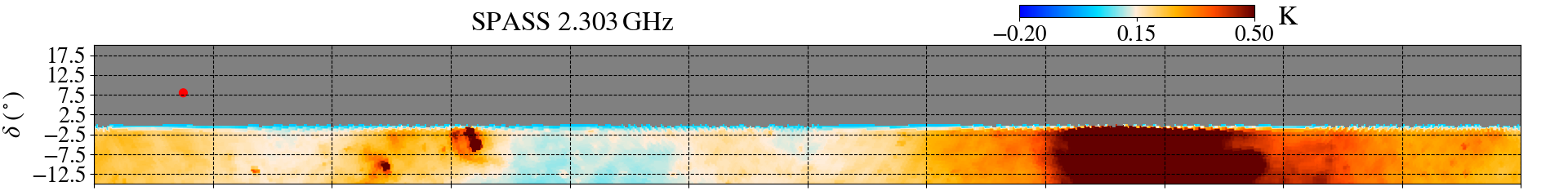}
\includegraphics[width=1\linewidth]{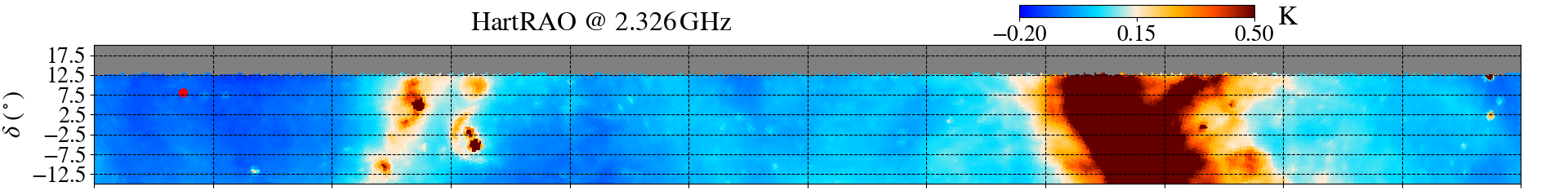}
\includegraphics[width=1\linewidth]{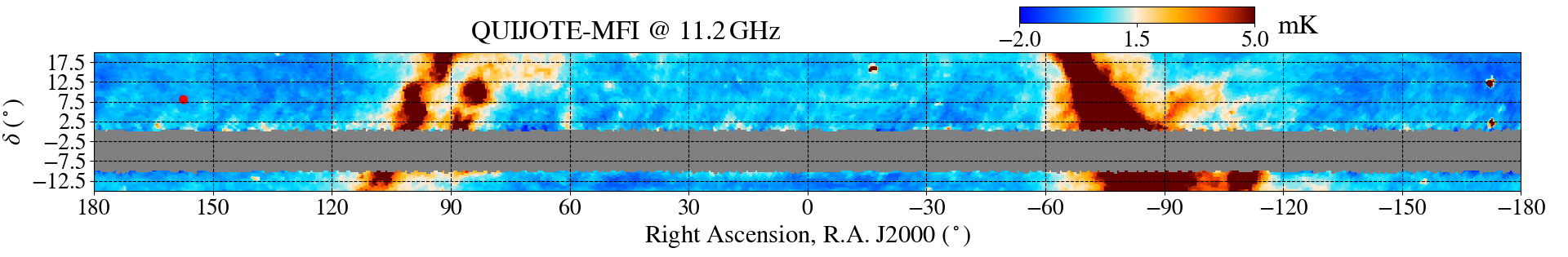}
\caption{Equatorial view of the overlapping region 
between \protect\cite{dwingeloo}, \protect\cite{hartrao} and 
11.2--12.9\,GHz bands from QUIJOTE surveys 
($0\degr<\delta<13\degr$). The background region 
pixels, which defined the map zero-levels for the 
analyses, are those within the red dot at 
$\rm (RA, \delta) = (157.5\degr, +8\degr)$. This 
region was selected in an effort to avoid regions with
 high emission in the lower frequency surveys. 
It should also be as far as possible from both 
the satellite band ($-10\degr<\delta<0\degr$) and 
the upper declination \protect\cite{hartrao} survey
 limit. SPASS \protect\citep{carretti2019} map is 
also shown, for comparison.}
\label{fig:BG_region_selection}
\end{figure*}

It is worth noting that although we keep the notation from 
aperture photometry works, we did not perform a standard
aperture photometry analysis where a background region 
is selected to correct for the local background emission. 
Instead, we used a common region for all pixels to define a
reference zero-level for the whole Galactic plane (as in, for 
example, \citealt{planck2014LFIcalibration}).
Determining zero-levels in radio surveys is a critical step for 
component separation analyses \citep{wehus+2017}, and these are
especially difficult to define for the low-frequency ($<10$\,GHz) surveys 
used in this work. Besides, after applying the FDEC
filtering to the 
QUIJOTE-MFI data (as explained in Section 2.4) some power at large angular
scales is removed from the maps, and in particular, the mean levels of 
the maps are removed. For our analysis, all ancillary maps are filtered 
using the same FDEC approach as for QUIJOTE-MFI. The selection of a 
common aperture to set the new (common) zero level of all maps is 
thus needed for a consistent analysis.
This reference background region must be present 
in all maps, so it should be in the overlapping 
region $0\degr<\delta<13\degr$ between~
\cite{dwingeloo},~\cite{hartrao} and the 11.2--12.9\,GHz 
QUIJOTE surveys (Fig.~\ref{fig:BG_region_selection}). 
We selected the aperture with radius $r=1\degr$ 
located at $\delta=8\degr$ that 
minimized WMAP $K$-band flux, as this band is one of the 
best AME detection proxies. This aperture is located 
at RA = 157.5$\degr$. We tested several 
apertures with fixed $\delta=8\degr$ and variable RA, 
and none of them introduced biases greater than 1\,Jy
on the WMAP K band map. We did the same for \textit{Planck-HFI}
143 and 353\,GHz bands, in order to show that the region
selection was not biasing the thermal dust calculation.
We found that there were no regions without evident emission from
Galactic structures that had a median value further than 1$\sigma$ 
away from the value obtained from our chosen background region. 
This $\sigma$ is calculated as the quadratic sum of the uncertainties
from both our chosen aperture and the tested ones. Therefore, our method
is robust against changes when choosing the background region, maintaining 
the following results.

Finally, we ran several tests, applying variations to 
the zero levels of the flux density values from the low frequency 
surveys, compatible with their photometric uncertainties.
These are mostly dominated by calibration uncertainties, which are 
conservatively defined. We found that the distributions for the AME 
parameters are not affected by these changes on the zero-levels. 
$\Iame$ median values from the marginalized posteriors were
less than $\pm0.1\sigma$ away from their real values, where 
$\sigma$ is computed as the quadratic sum of the dispersions from
the real and simulated cases. The effect
was slightly larger for $\nuame$, but still under $\pm0.2\sigma$, 
while negligible for $\Wame$. On the other hand, the changes for
$I_{\rm 1\,GHz}$ could be as important as $\pm0.6\sigma$. 
This proved that our results on AME are robust against 
variations in the zero levels of the low frequency surveys.

\begin{table}
\centering
\caption{Top-hat priors for the parameters during 
the MCMC. Please notice that we are referring to the 
synchrotron spectral index, $\alphasyn$, in flux 
density, not in temperature ($\beta_{\rm syn}=\alphasyn-2$).}
\begin{tabular}{|l|c|c|}
\hline
Parameter & Lower prior & Upper prior \\ \hline
$I_{\rm1\,GHz}$ (Jy/sr) & $0$ & --- \\ 
$\alphasyn$ & $-2$ & 1 \\ 
$EM$ (pc\,cm$^{-6}$) & 1 & --- \\ 
$\Iame$ (Jy/sr) & 0 & --- \\ 
$\nuame$ (GHz) & 10 & 60 \\ 
$\Wame$ & 0.2 & 1 \\ 
$\taud$ & 0 & --- \\ 
$\betad$ & 0 & 3 \\ 
$\Td$ (K) & 10 & 40 \\ 
$\Delta \Tcmb$ ($\mu$K) & $-600$ & 600 \\ \hline
\end{tabular}
\label{table:priors}
\end{table}

\subsubsection{QUIJOTE data uncertainties assessment}
Classical aperture photometry studies (from 
the QUIJOTE collaboration~\citealt{Perseus, 
W44, Taurus} and previous 
experiments --- e.g.~\citealt{planck2015galacticcloudsAME}) 
estimated the aperture flux density uncertainty as described 
above by using the scatter between 
pixels in the background aperture. In 
that case, we rely on the assumption that the 
background fluctuation level is similar between 
the aperture and the background regions. 
However, we found large-scale residuals due 
to $1/f$ in the QUIJOTE-MFI intensity maps 
(particularly at 17 and 19\GHz), which implied 
that this assumption might not be correct for
 that data set. This is especially severe for 
this kind of analysis, where the aperture and
 background regions are far apart. 
We therefore generated a set of $N=1000$ simulations 
for the QUIJOTE-MFI bands to quantify the 
correlated noise plus instrumental and systematic 
effects. Further description of these noise 
simulations is provided in Section 6 of 
\cite{mfiwidesurvey}. We used the simulations 
to compute an additional contribution to the aperture 
photometry uncertainty to account for 
large-scale variations between parts of the maps: 
\begin{equation}
  \sigma_{\rm sims} = \sqrt{\sum_i^{N}\frac{(S_{0+i} - S_0)^2}{N}},
\label{eq:sigma_simulations}
\end{equation}
where $S_0$ is the aperture flux density computed on 
the QUIJOTE-MFI map alone, and $S_{0+i}$ is the 
same result when the $i$-sm simulation is added 
to that QUIJOTE-MFI map. Statistical and 
calibration uncertainties still needed to be
added quadratically when using this estimator. 
Thus, the final uncertainty estimates for QUIJOTE-MFI 
flux densities\footnote{The rest remain as in 
equation~\ref{eq:sigma_AP_calibration}.} increased from that described in 
equation~\ref{eq:sigma_AP_calibration} to:
\begin{equation}
 \sigma_{S_\nu} = \sqrt{\sigma_{\rm AP}^2 + 
\sigma_{\rm sims}^2 + \mathrm{cal}^2\cdot S_\nu^2}.
\label{eq:sigma_AP_calibration_simulations}
\end{equation}

\subsection{SED fitting through MCMC}
For each pixel, we used a Maximum Likelihood Estimator 
(MLE) to obtain the posterior distribution for the (ten) 
free parameters described in Section~\ref{section:foreground_model}:
\begin{equation*}
\begin{split}
\theta = (I_{\rm1\,GHz},\, \alphasyn,\, \EM, \,\Iame,\, &\nuame,\, \Wame,\, \\ &\tau_{353},\, \betad,\, \Td,\, \Delta T_{\rm CMB}).
\end{split}
\end{equation*}
We apply flat priors on these parameters, which are listed
in Table~\ref{table:priors}. These are defined 
to be as little restrictive as possible. $\chi^2$ 
depends on the sum of the differences between measured,
 $\textbf{\textit{S}}$, and expected,  
$\textbf{\textit{S}}^{\rm total}(\theta)$, flux densities 
across the frequency domain, and their covariance, 
$\textbf{\textit{C}}$:
\begin{equation}
\chi^2 = (\textbf{\textit{S}} - 
\textbf{\textit{S}}^{\rm total} (\theta))^{\rm T}\textbf{\textit{C}}^{-1}(\textbf{\textit{S}} 
- \textbf{\textit{S}}^{\rm total} (\theta)).
\label{eq:chi2}
\end{equation}
When every combination of surveys has negligible 
covariance, the previous equation turns into:
\[
  \chi^2 = \sum_{\nu}\left[\frac{S_\nu - 
S_\nu^{\rm total}(\theta)}{\sigma_{S_\nu}}\right]^2,
\]
where the measured flux density ($S_\nu$) and its uncertainty
($\sigma_{S_\nu}$) are obtained through aperture photometry, 
as described in Equations~\ref{eq:AP} and 
\ref{eq:sigma_AP_calibration_simulations}. However, QUIJOTE-MFI 
frequencies observed by the same horn (11.2--12.9\,GHz 
and 16.8--18.7\,GHz) are highly correlated, so this 
assumption is no longer valid. The required correction 
will be explained in Section~\ref{section:correlations_explained}.

We therefore built a log-likelihood MLE using 
$\chi^2$ from equation~\ref{eq:chi2}:
\begin{equation}
\begin{split}
    \log\, \mathcal{L} = -0.5\cdot\chi^2 =& \\ 
-0.5 &\cdot \left(\textbf{\textit{S}} - 
\frac{\textbf{\textit{S}}^{\rm total}(\theta)}{\textbf{\textit{cc}}}\right)^{\rm T}\textbf{\textit{C}}^{-1}\left(\textbf{\textit{S}} - \frac{\textbf{\textit{S}}^{\rm total}(\theta)}{\textbf{\textit{cc}}}\right),
\label{eq:chi2_w_CC}
\end{split}
\end{equation}
where $\textbf{\textit{cc}}$ accounts for the 
required colour corrections to be applied to the 
surveys. Colour corrections must be applied to account 
for the fact that the measured flux densities are 
integrated on the bandpass of each experiment detector, 
and are explained in Section~\ref{section:CC}. This 
log-likelihood is used within the MCMC sampler ensemble 
from the \textbf{\tt emcee} package~\citep{emcee}. 

The initial values for the fit parameters were drawn 
from the respective COMMANDER~
\citep{planck2016Xforegroundmaps} (downgraded to 
\nside = 64) pixel posteriors. We then ran the chains 
to build the posteriors using the priors in 
Table~\ref{table:priors} until they converged. We 
produced an additional set of results by adding a 
Jeffrey's ignorance prior (e.g.~\citealt{eriksen2008}) 
to prevent $\alphasyn$ parameter being biased towards 
steeper values: however, the differences with our results
were compatible within our uncertainties. We relied 
on the autocorrelation time from the \textbf{\tt emcee} 
sampler to assess whether convergence had been achieved, so
 the number of required chain steps changes from pixel 
to pixel. This issue is further discussed in 
Appendix~\ref{section:appendix_convergence}. Once 
convergence was achieved, we recovered the median 
value from the parameter posteriors as their final 
value. Their uncertainties were estimated as half 
the difference between their 84th and 16th percentiles. 

\begin{figure}
    \centering
    \includegraphics[width=1\linewidth]{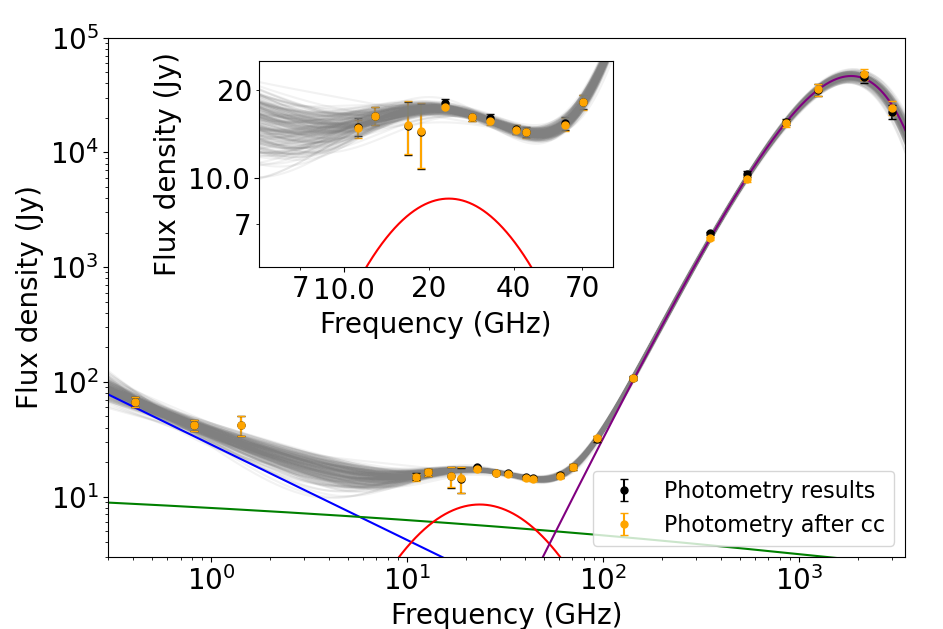}
    \caption{Example of a spectral energy 
distribution (SED) for one pixel --- the one centred 
on $(l, b)=(111.1\degr,3\degr)$ specifically. 
Photometry values are plotted twice: before and 
after applying their colour corrections. The embedded
panel shows in detail the region where AME dominates
over the rest of foregrounds. Random realizations 
from the MCMC are also shown in grey. It is clear 
that there is a larger dispersion for the fitted
models between 1 and 10\,GHz,
where the degeneracy between synchrotron and 
free-free appears. The foreground component
SEDs defined by their median parameter values
(from Fig.~\ref{fig:example_corner_plot}) are
also displayed.}
    \label{fig:example_SED}
\end{figure}

An example SED computed with this method can be 
seen in Fig.~\ref{fig:example_SED}, where the 
fit was obtained as the sum of all the components 
described in Section~\ref{section:foreground_model}.
 This is a pixel dominated by AME, with more than 
50\,\% of its flux density between 20 and 30\,GHz coming 
from that component. It is clearly visible how the 
spectrum rises at QUIJOTE-MFI's lower frequencies due 
to AME. There are no large residuals across the full
frequency domain, the largest one being the one from 
the WMAP $K$ band, because of its low uncertainty. 

\begin{figure*}
    \centering
    \includegraphics[width=1\linewidth]{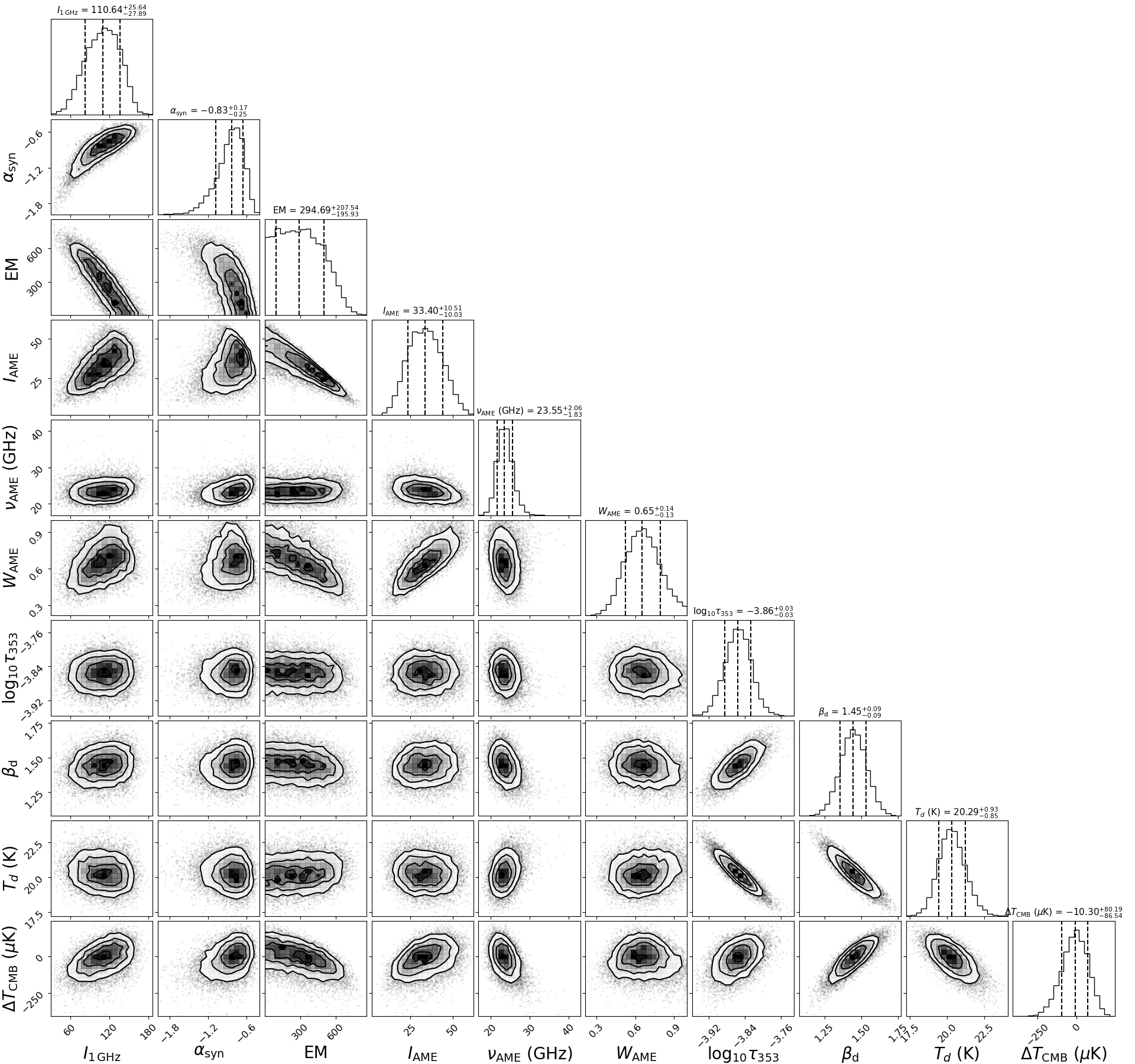}
    \caption{Corner plot containing the marginalized posteriors 
and correlation plots for the parameters describing the SED from 
Fig.~\ref{fig:example_SED}. Their median and 
16th and 84th percentile values are plotted with 
vertical dashed lines in the 1D marginalized posteriors: 
final parameter values and their uncertainties are 
obtained from those values. In the case of asymmetric
distributions, such as the one for $\alphasyn$,
there is a displacement between the median and the 
peak of the posterior.
We see degeneracies that are common for most of the 
pixels in the map, mainly
 those between synchrotron, free-free and AME 
amplitudes ($I_{\rm1\,GHz}$, $\EM$, $\Iame$) or the 
dust parameters (especially $\betad$ and $\Td$). 
$I_{\rm1\,GHz}$ and $\Iame$ units are 10$^3$ Jy/sr, 
while $EM$ units are $\rm pc\ cm^{-6}$.}
    \label{fig:example_corner_plot}
\end{figure*}

The corner plot showing the parameter posteriors 
obtained for this pixel is shown in 
Fig.~\ref{fig:example_corner_plot}, where some 
degeneracies are clearly visible. The most important 
ones are those involving the synchrotron ($I_{\rm1\,GHz}$), 
free-free ($\EM$) and AME ($\Iame$) amplitudes. 
The synchrotron index, $\alphasyn$, 
is also correlated with $\EM$. The degeneracies 
between the dust parameters 
are well-known, especially that between $\betad$ 
and $\Td$ (e.g.\ \citealt{planckbetamm}). These 
behaviours are common for most of the pixels in 
this study: it is also usual that $\EM$ is the worst 
defined parameter. This was expected, owing to the 
flat free-free behavior, which makes it strongly degenerate 
with both the synchrotron emission and AME.

In order to validate our fitting procedure, and 
to show that we produced unbiased estimates for 
the various parameters, we generated synthetic 
SEDs for the pixel shown in Figures~\ref{fig:example_SED} 
and \ref{fig:example_corner_plot}. We built a multivariate
Gaussian distribution taking into account the flux
density values obtained using aperture photometry,
\textbf{\textit{$S_\nu$}}, and the full covariance,
\textbf{\textit{C}}. We took random guesses from this
distribution, building a simulated set of flux densities,
\textbf{\textit{$S_\nu^{\rm sim}$}}. We then ran the MCMC
and fitted the SED formed by these simulated flux densities
in the same
way as described before in this section for the real data,
and checked whether the results had changed. 
We found no such variations: when combining the parameter 
posteriors obtained from all the simulated SEDs, 
we recovered the same posteriors as when directly 
studying the real pixel SED.

The complete final maps took $\simeq10700$ (more than 1.2 years)
hours of CPU time to compute 
the parameter posteriors for the 5309 independent pixels. 
We used the HTCondor distributed system at the Instituto de 
Astrofísica de Canarias (IAC): the median and 
standard deviation computation times were 
$1.66_{-0.46}^{+1.04}$ hours for each pixel.

\subsubsection{Correlations between pairs of frequencies}
\label{section:correlations_explained}
As previously stated in Section~\ref{section:QUIJOTE} 
and~\cite{mfiwidesurvey}, frequencies obtained with 
the same horns from QUIJOTE, i.e. 11.2--12.9\,GHz and
 16.8--18.7\,GHz, are highly correlated in intensity 
(up to $80\,\%$). Thus, those pairs of points could
 not be taken as independent when building the SED, 
so the covariance matrix was no longer diagonal when 
calculating $\chi^2$ for the likelihood estimate.
The covariance matrix components were then defined 
by the following equation:
\begin{equation}
c_{ij} = (\rho_{ij} + \delta_{ij})\sigma_i\sigma_j
\end{equation}
where $\delta_{ij}$ is the Kronecker delta and
\begin{equation}
 \rho_{ij}=\begin{cases} 
      0.8 &  (i,j) \ | \ (j,i)={\rm11.2,12.9\,GHz} \\
      0.8 &  (i,j) \ | \ (j,i)={\rm16.8,18.7\,GHz} \\
      0 & {\rm otherwise}
   \end{cases}
\end{equation}

\subsubsection{Colour corrections}
\label{section:CC}
Colour corrections (cc) were performed iteratively 
through the MCMC for every point above 10\,GHz: 
QUIJOTE, WMAP, \textit{Planck-LFI} and \textit{HFI} 
and COBE-DIRBE experiments. It was assumed to be unnecessary 
for lower frequency surveys owing to their narrower 
bandpasses. Depending on the frequency point studied, 
one of two approximation methods was used:
\begin{itemize}
    \item Frequencies below 100\,GHz: we used a 
power law approximation where, for each
frequency ($\nu$) a spectral index ($\alpha_\nu$)
was obtained while assuming $S_{\nu}^{\rm total}(\theta)$ 
was linear in its log-log space vicinity.
    As the colour correction was embedded inside 
the MCMC and computed every time we performed a 
step, reducing the computation time was critical. 
That is why we produced a second-order polynomial 
fit, tabulating the colour correction, $cc$, as a 
function of the spectral index, $\alpha_\nu$. This 
was done for every experiment before 
running the MCMC. For every step of the MCMC, we
computed $\alpha_\nu$, and re-scaled the
flux densities obtained from aperture photometry
with the appropiate $cc(\alpha_\nu)$. This was done 
using {\tt fastcc}~\citep{fastcc}, as explained 
in~\cite{mfiwidesurvey} and Section 8.2 from~\cite{mfipipeline}.
    \item For frequencies above 100\,GHz, 
where the thermal dust emission dominates the 
SED, the former approximation was no longer 
valid, as spectra had a more pronounced curvature
and bandwidths were larger. We switched to a greybody model 
described by the thermal dust index and 
temperature, $cc(\betad, \Td)$. This implied that we
had to built 2D grids to tabulate $cc$
values against $\betad$ and $\Td$, instead of having a 
polynomial fit that could be evaluated at a certain value, 
as happened for $cc(\alpha_\nu)$ for frequencies below 
$100$\,GHz. Within the MCMC, we then took the appropiate 
$cc(\betad, \Td)$ factor from the grid taking into account 
the value of $\betad$ and $\Td$ on each step of the MCMC, and 
re-scaled the flux density estimate accordingly.
\end{itemize}

Normally, these $cc(\alpha)$ and $cc(\betad, \Td)$ 
factors should multiply the flux densities obtained 
from aperture photometry. However, that would imply
applying the factors also to re-scale the respective
uncertainties. To avoid doing so, we introduced the
colour-corrections to the likelihood calculation
dividing the estimated flux density instead. This is already
applied in equation~\ref{eq:chi2_w_CC}.

\section{Results}
\label{section:results}
The five emission components considered 
(synchrotron, free-free, AME, thermal dust 
and CMB anisotropies) are defined by ten 
parameters, 
$\theta = (I_{\rm1\,GHz},\, \alphasyn,\, \EM,\,
 \Iame,\, \nuame,\, \Wame,\, \taud,\, \betad,\, 
\Td,\, \Delta T_{\rm CMB})$, as explained in
 Section~\ref{section:foreground_model}. 
The maps for these parameters are shown in 
Fig.~\ref{fig:parameter_maps_1}. For some of the 
analyses, we discarded pixels with 
$\SNRAME=\Iame/\sigma(\Iame)<2$, as AME is 
the emission component we are mainly interested 
in. We have also applied a $\SNRAME<3$ threshold 
and seen that the results do not change between 
the two cases. Thus, keeping $\SNRAME>2$ allows us
to increase the sample (by almost a factor 3)
 while maintaining the results. Moreover, in this 
way we prevent the presence of a positive $\Iame$
 bias from pixels with no AME emission. The 
previous maps with those pixels masked are shown 
in Appendix~\ref{section:appendix_snr}. We also
 masked the pixels located less than 
$1.5\degr$ away from the SIMBAD\footnote{http://simbad.u-strasbg.fr/simbad/~\citep{simbad}} positions of Tau A, Cas A and 
Cyg A. 

\begin{figure*}
    \centering
    \includegraphics[width=1\linewidth]{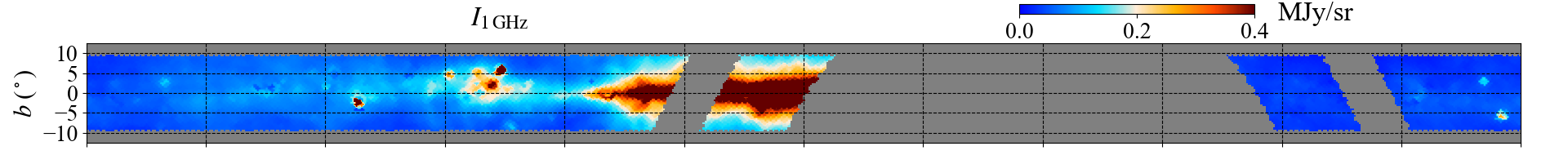}
    \includegraphics[width=1\linewidth]{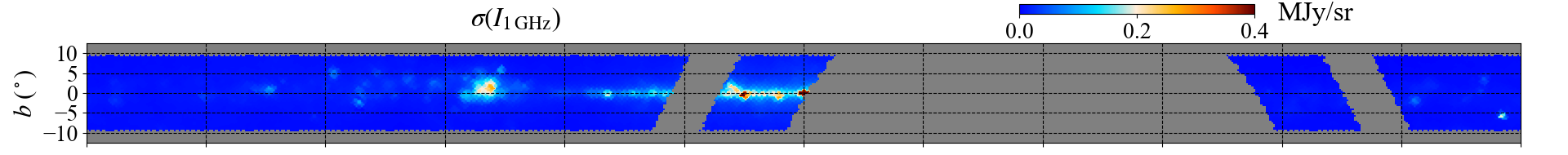}
    \includegraphics[width=1\linewidth]{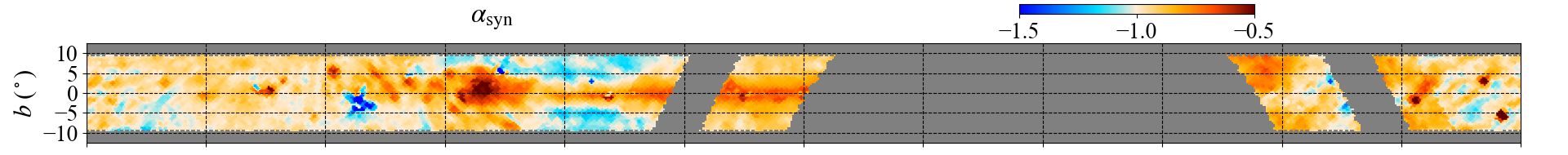}
    \includegraphics[width=1\linewidth]{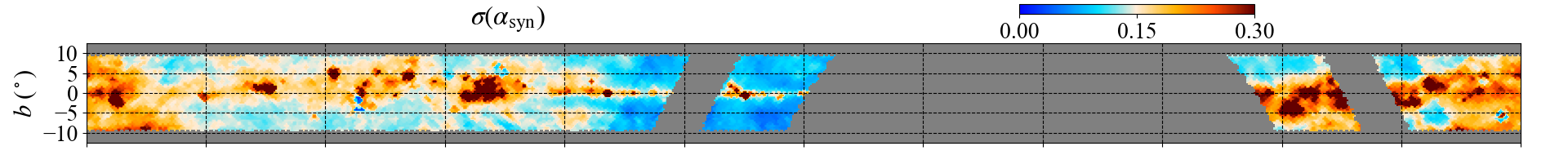}
    \includegraphics[width=1\linewidth]{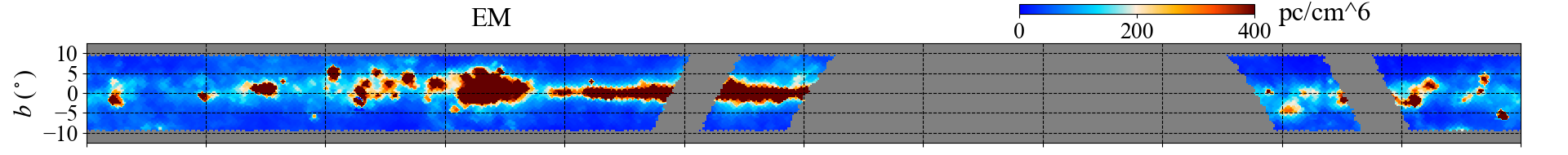}
    \includegraphics[width=1\linewidth]{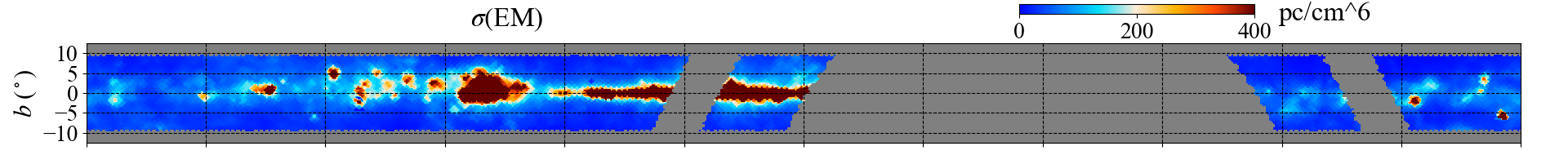}
    \includegraphics[width=1\linewidth]{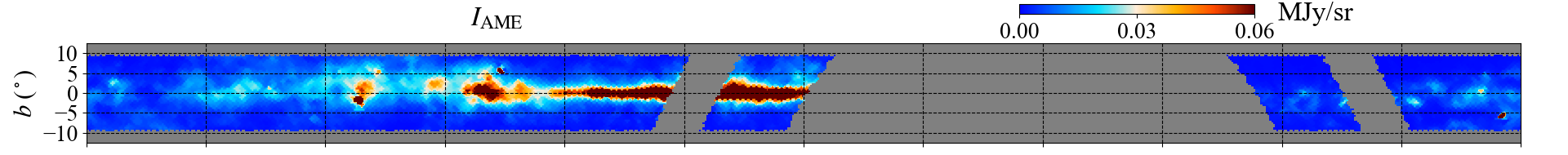}
    \includegraphics[width=1\linewidth]{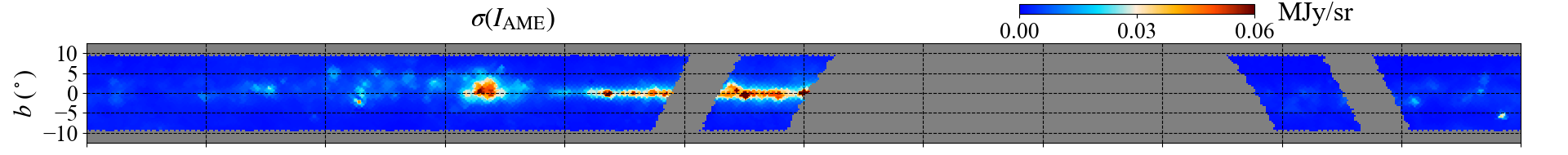}
    \includegraphics[width=1\linewidth]{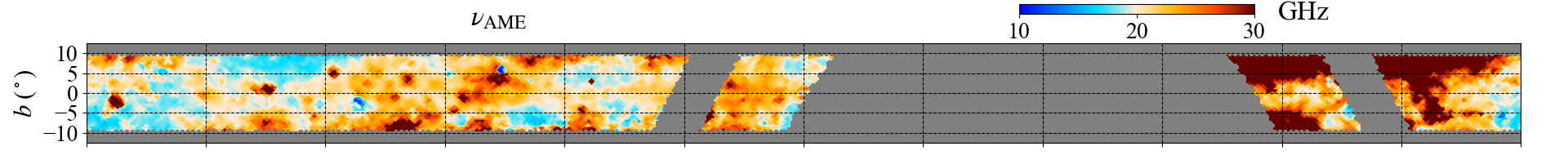}
    \includegraphics[width=1\linewidth]{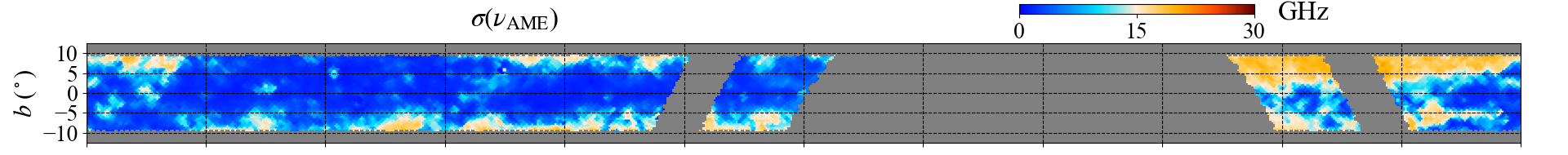}
    \includegraphics[width=1\linewidth]{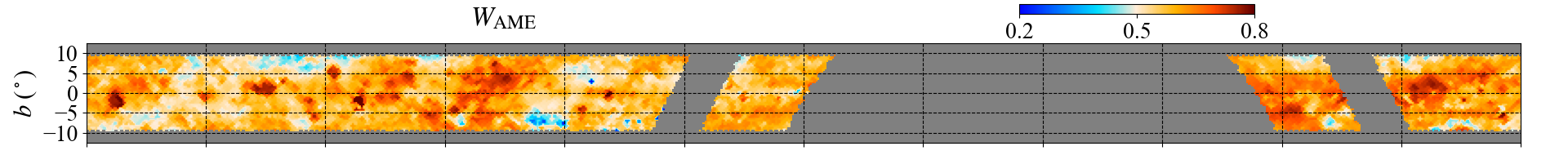}
    \includegraphics[width=1\linewidth]{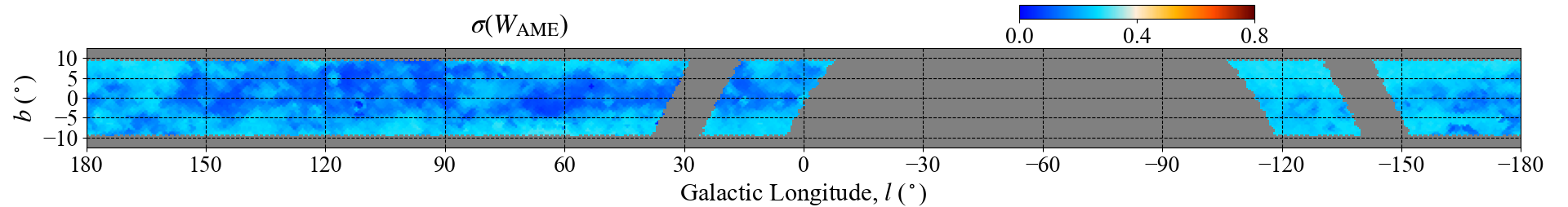}
    \caption{Reconstructed maps for the 
parameters describing
synchrotron, free-free and AME. 
We can see how the $\alphasyn$ 
uncertainty decreases for the band 
$\delta\in(0\degr, 13\degr)$, owing to the 
addition of a fourth point (from the
\protect\citealt{hartrao} survey) to the
low-frequency (0.4--3\,GHz) regime. It is also
clear how the $\EM$ uncertainties remain
high for most of the pixels in the plane. 
Regarding the AME parameters, $\nuame$ and $\Wame$ 
have high signal-to-noise ratios along the plane, 
and both decrease as we get farther from the 
plane.}
    \label{fig:parameter_maps_1}
\end{figure*}
\begin{figure*}
    \includegraphics[width=1\linewidth]{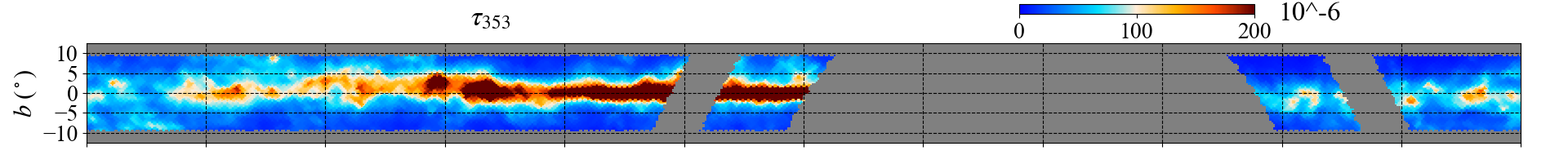}
    \includegraphics[width=1\linewidth]{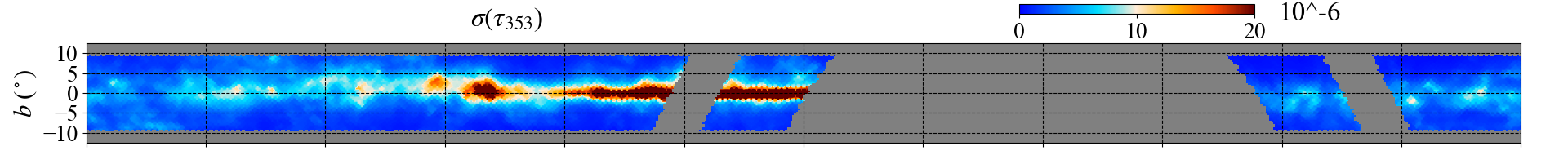}
    \includegraphics[width=1\linewidth]{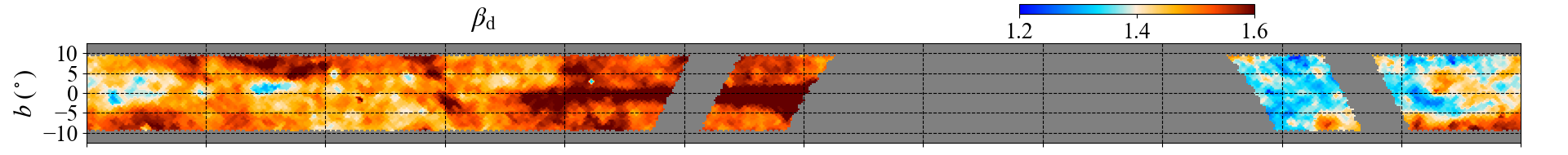}
    \includegraphics[width=1\linewidth]{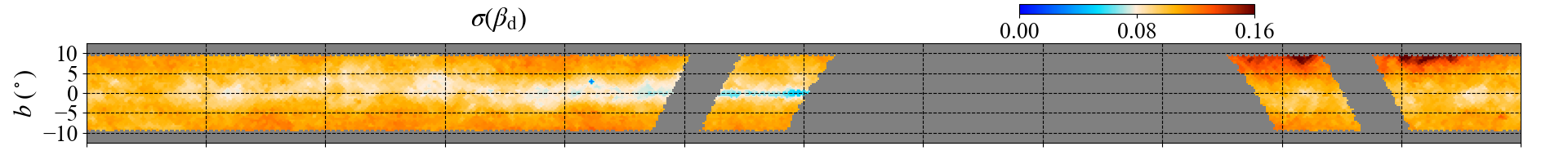}
    \includegraphics[width=1\linewidth]{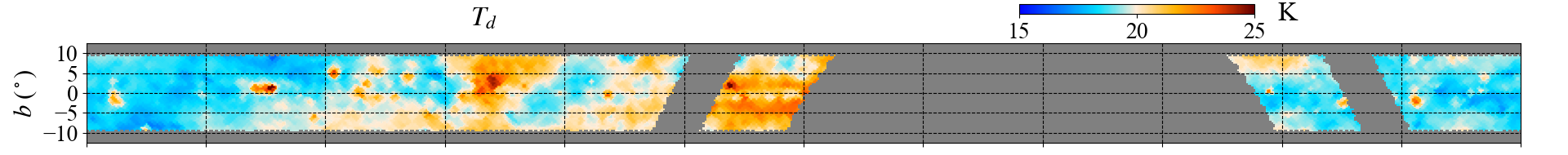}
    \includegraphics[width=1\linewidth]{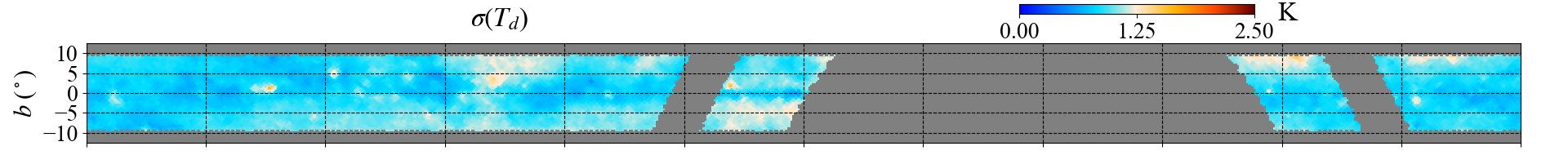}
    \includegraphics[width=1\linewidth]{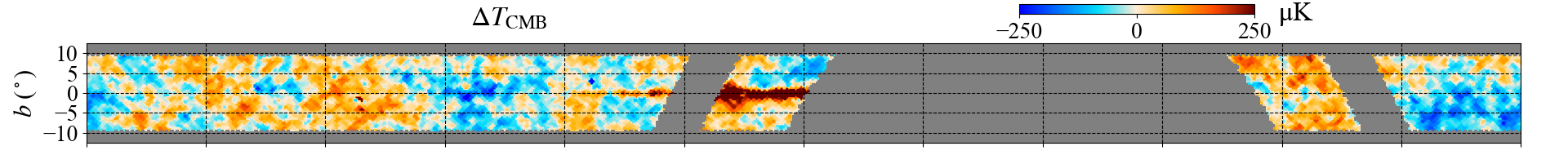}
    \includegraphics[width=1\linewidth]{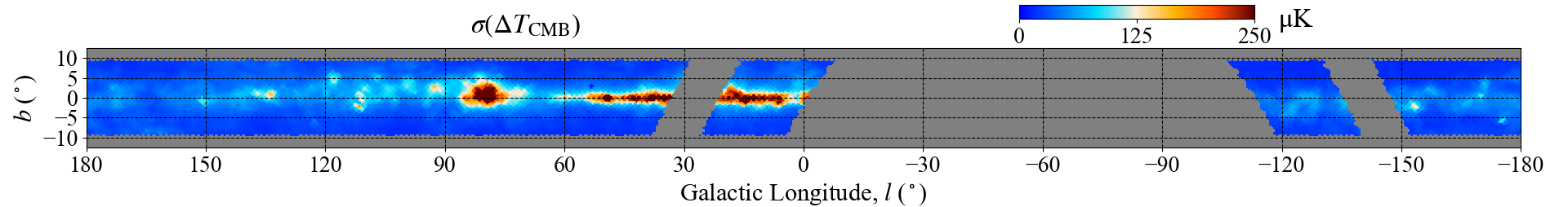}
    \contcaption{Reconstructed maps for the dust 
and CMB parameters. The colorbar limits for 
the dust parameters uncertainties have been fixed 
to 10\% of those for the parameters themselves. 
$\Delta T_{\rm CMB}$ shows a clear residual near
the Galactic Centre, due to the model failing to
reproduce all the measured flux density between
100 and 200\,GHz with just one thermal
dust component. The residual thermal dust emission
is accounted for by the CMB component instead.}
\end{figure*}

None of the priors introduced in Section~\ref{section:Methodology} 
was too restrictive, according to the posterior 
distributions for the parameters. This is demonstrated in 
Fig.~\ref{fig:example_corner_plot}, where it is seen that 
in most cases the 95\,\% confidence intervals of the 2D 
posteriors lie well within the flat priors. 
In the opposite scenario, we would expect to have the 
peak of the posterior close to any of the flat prior 
edges\footnote{This happens for the $\rm EM$ in those pixels where
the free-free emission is not large, but $\rm EM>0$ is a
physical prior, as emission cannot be negative.}. 
For example, $\Wame$ is one of the most complex parameter to constrain:
small inconsistencies between adjacent flux 
densities in the SED can be compensated by a really
narrow AME component, so $\Wame$ would be biased to low values.
Also, large $\Wame$ values turn the spinning dust contribution
almost into a power-law in the 10--60\,GHz domain, so it
would be replacing free-free. However, $\Wame$ has median values 
between 0.4 and 0.8 for almost every pixel, while the lower and upper 
priors were 0.2 and 1, avoiding any of the previous possible issues.

Apart from studying the pixel set presented in 
Section~\ref{section:region_definition} in its entirety, 
we defined a series of regions to be studied 
independently, which we named ``sectors''. 
The first two correspond to regions 
with different data information, while Sectors 
3 to 6 study various galactic longitude cuts:
\begin{itemize}
    \item Sector 1: $\delta<-10\degr$. Covers 
the area below the QUIJOTE satellite band at low 
longitudes, i.e.\ the Galactic centre and the pixels 
located at $l\simeq-120\degr$. For this region we 
have no data from the Dwingeloo survey at 0.820\,GHz, but 
we do have data from HartRAO survey at 2.326\,GHz.
    \item Sector 2: $\delta>13\degr$. Covers 
the area where we have the complementary configuration 
to Sector 1: we have data at 0.820\,GHz, but not 
at 2.326\,GHz.
    \item Sector 3: $|l|<50\degr$. Covers 
the Galactic centre and some pixels above the 
satellite band. $l\simeq50\degr$ is, approximately, 
the point where synchrotron emission begins to be 
less important (as shown in 
Figure~\ref{fig:parameter_maps_1}).
    \item Sector 4: $50\degr\leq l < 90\degr$. 
This region hosts the feature with the highest 
$\SNRAME$ from all the plane, at $l\simeq60\degr$, 
$b\in(-5\degr,0\degr$), as can be seen in 
Figure~\ref{fig:snr_map_Aame_gt_3}. The Cygnus 
region, located at $l\simeq80\degr$ and dominated 
by free-free, is also embedded in this region.
    \item Sector 5: $90\degr\leq l < 160\degr$. 
This region has the longest QUIJOTE-MFI integration time 
(as shown in the figures in  Appendix
 A from~\citealt{mfiwidesurvey}).
    \item Sector 6: $160\degr\leq l < 200\degr$. 
This area covers the Galactic anticentre, where 
the total emission is lower. This region has 
limited interest in this work (the number of 
pixels with $\SNRAME>2$ is low).
\end{itemize}

\subsection{Spatial variations for AME parameters}
\label{section:results_spatial_variations}
\begin{figure}
    \centering
    \includegraphics[width=1\linewidth]{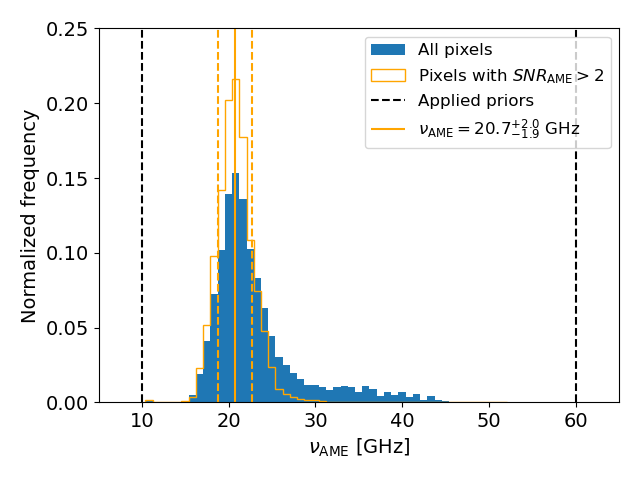}
    \includegraphics[width=1\linewidth]{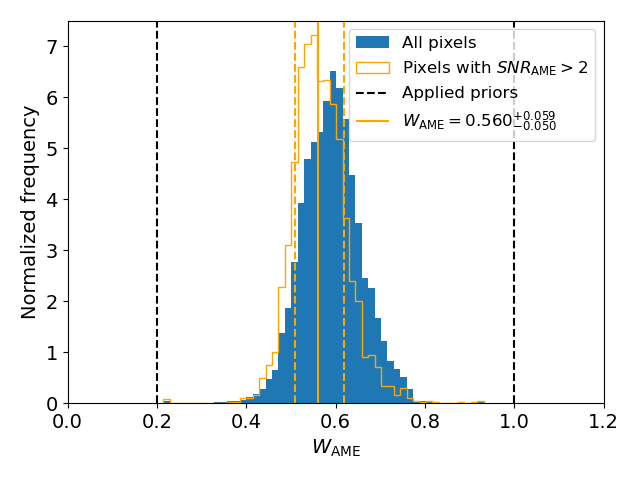}
    \caption{Top: Distribution of $\nuame$ values along 
the Galactic plane. We see that, when focusing on pixels 
with high $\SNRAME$, the tail of pixels towards high 
$\nuame$ values is suppressed. This greatly decreases 
the variability in the histogram (from 
$\nuame=21.6^{+5.8}_{-2.6}$ to $\nuame=20.7^{+2.0}_{-1.9}$).
 Bottom: Distribution of $\Wame$ values along the Galactic 
plane. The width of the distribution remains the same after
removing those pixels with $\SNRAME<2$, but its median value
decreases a little 
($0.591^{+0.070}_{-0.069}$ vs. $0.560^{+0.059}_{-0.050}$ 
from Tables~\ref{table:parameter_variation_w_regions} 
and~\ref{table:parameter_variation_w_regions_SNRgt2}). 
In both cases, the applied flat priors are shown as 
vertical dashed lines.}
    \label{fig:Wame_nuame_distribution}
\end{figure}

The distribution of the reconstructed maps 
for $\nuame$ and $\Wame$ is shown in 
Fig.~\ref{fig:Wame_nuame_distribution}. The 
median values are $\nuame=21.6^{+5.8}_{-2.6}$\,GHz, 
$\Wame=0.591^{+0.070}_{-0.069}$, where upper and 
lower confidence intervals were obtained as half 
the difference between the distribution percentiles
 84th and 16th. These boundaries account for the 
variation in the parameter values along the Galactic 
plane, and are not related to the uncertainties 
for those same parameters from the individual pixels. 

After discarding pixels with $\SNRAME<2$, the 
previous values turn to $\nuame=20.7^{+2.0}_{-1.9}$\,GHz 
and $\Wame=0.560^{+0.059}_{-0.050}$. When this threshold is 
applied, the long tail towards high values of $\nuame$
visible in Fig.~\ref{fig:Wame_nuame_distribution} is 
suppresed. This 
last $\nuame$ value is almost 2.5$\sigma$ away from 
that obtained by the {\tt BeyondPlanck} collaboration
\cite{beyondplanckintensity}, 
$25.3\pm0.5$\,GHz, but that value was obtained as 
a joint fit for all sky pixels, and not just those 
with highest AME significance from the Galactic plane. 
Our median value is also lower than, but consistent with, 
that of \cite{AMEwidesurvey} (hereafter, 
\citetalias{AMEwidesurvey}): $23.6\pm3.6$\GHz. 
\cite{planck2015galacticcloudsAME} (hereafter, 
\citetalias{planck2015galacticcloudsAME}) reported 
a weighted mean value of 
$\nuame=27.9$\GHz, slightly discrepant from the 
results mentioned above, but they lacked data between 
2.3 and 22.8\,GHz\footnote{However, the methodology was similar to
the one in this study, as the AME was fitted with a single 
distribution. This distribution was obtained using \SPDUST, and then
variations in amplitude and peak frequency were allowed,
although the shape (thus, the width of the distribution)
remained the same.}. Therefore, in 
\citetalias{planck2015galacticcloudsAME} they were unable to constrain
both sides of the AME distribution, as we can do after 
the addition of QUIJOTE-MFI data. However, both works
study compact Galactic sources, while our study is focused
on the diffuse emission from the Galaxy. \cite{cbassSH2022}
reported a consistent value  of $\nuame=19.4\pm6.4$\,GHz. 
A log-normal AME distribution 
similar to the one used in this study was applied, 
although in that case the high-latitude sky was 
the focus of the study. This suggests that, on 
1$\degr$ scales, both studies could be most sensitive
to AME at similar phases of the ISM. \cite{comapVI} 
studied a series of HII regions on arcminute scales 
and found $\nuame$ values above 40\GHz, although 
with large (around 15\GHz) uncertainties.

\begin{figure}
    \centering
    \includegraphics[width=1\linewidth]{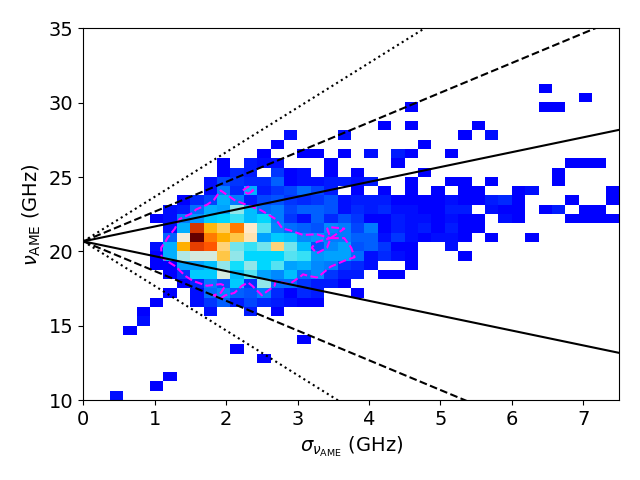}
    \includegraphics[width=1\linewidth]{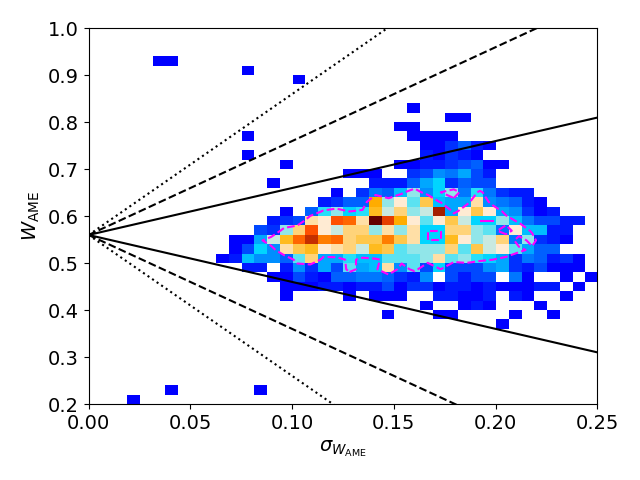}
    \caption{Spatial variations of $\nuame$ and $\Wame$ 
along the Galactic plane. The black lines show the median 
value of the distribution plus/minus 1 (solid), 2 (dashed),
 3 (pointed) times the dispersion from the parameter distribution
 (as shown in Fig.~\ref{fig:Wame_nuame_distribution}). 
 Spatial variability is more important as more data 
are located outside the regions enclosed by those lines.
 We therefore see how the spatial variations along the Galactic 
plane are more important for $\nuame$ than for $\Wame$: the latter 
is, in fact, almost entirely embedded within 1$\sigma$ (see 
text). Magenta lines enclose 68\,\% of the data. Both figures
 use just $\SNRAME>2$ pixels.}
    \label{fig:new_ame_variations_residuals}
\end{figure}

We compared the dispersion of the parameters to their median 
value along the Galactic plane. Spatial variations are 
well detected for the intensity, $\Iame$, with 66.4\% and 
64.5\% (when the $\SNRAME>2$ thershold is applied) of the 
points showing residuals greater than 1$\sigma$. However, 
variations are not statistically significant for either
$\nuame$ or $\Wame$ at 1 degree scales. This can be seen in 
Fig.~\ref{fig:new_ame_variations_residuals}, where we 
plot the results for $\SNRAME>2$ pixels versus their 
uncertainty, and compare with the $1\sigma$, $2\sigma$ and $3\sigma$ 
variation levels. Only 17.9\% and 22.9\% of the pixels 
show differences greater than $1\sigma$ for $\nuame$ when 
all pixels and just those with $\SNRAME>2$ are considered
 respectively. These two values decrease even more, down 
to 1.5\% and 1.6\%, for $\Wame$. However, this does not 
mean that these parameters do not vary pixel-to-pixel; 
those variations could be smaller than the statistical 
uncertainty from our analysis. For $\nuame$, because of the 
small span of frequency values (97\% of the pixels with
 $\SNRAME>2$ have $\nuame$ values between 15 and 25\,GHz) 
and high uncertainties (92\% of the pixels with 
$\SNRAME>2$ have $\sigma(\nuame)$ larger than $1.5$\,GHz),
it is difficult to have low residual values. 
Reducing the uncertainties would improve the detection of 
variability; this could 
also be achieved by studying smaller, correlated regions, 
as did \cite{LambdaOrionis} (hereafter, 
\citetalias{LambdaOrionis}), where less smooth variations 
can be present.

We also qualitatively compare these results with 
theoretically proposed models. The \SPDUST\,software 
\citep{spdust1, spdust2} is commonly used to build 
spinning dust SEDs: we
compared its models with a log-normal distribution.
We found that those models obtained using typical parameter
values tend to have width ($\Wame$) values below 0.6.
As mentioned before, the median 
$\Wame$ value in our results is $0.560^{+0.059}_{-0.050}$ 
when only those pixels with $\SNRAME>2$ are considered. Thus, 
almost $50\%$ of the pixels show values higher than this 
limit. We repeated our analysis by reducing our flat prior to 
$\Wame$ within $[0.2, 0.6]$ to see if the wider prior
artificially increased the values for $\Wame$. $\Wame$ individual
pixel posteriors continued to have median values close to 0.6, 
the new prior, thus invalidating it. 
These higher-than-expected $\Wame$ values were also measured 
by~\citetalias{AMEwidesurvey} and~\citetalias{LambdaOrionis}, 
for unresolved sources and the $\lambda$Orionis ring 
respectively. The latter is particularly significant, as 
the use of C-BASS data (plus the absence of synchrotron) 
provides a better description of free-free. This directly 
improves its disentangling from AME: $\Wame$ is still greater, however,
 than 0.5 for most of the regions with high AME significance. 
These results suggest that it may be convenient to revisit 
theoretical AME models to investigate how they could predict wider
spectra.

\subsubsection{Other emission components}
For the other components, we expect a spatial 
distribution of amplitudes resembling the maps normally 
used as tracers of each type of emission. The 0.408\,GHz~\cite{haslam1982} 
map describes synchrotron, while free-free can be described 
both by H$\alpha$~\citep{WHAM, finkbeinerHalpha} or radio-recombination 
lines (RRLs) maps~\citep{RRLs}, the latter being less affected
by extinction than the former\footnote{We have tested how 
well the full-sky H$\alpha$ map from the Wisconsin H-alpha Mapper 
\citep[WHAM,][]{WHAM} compares to both $\EM{}$ and $\Td$ maps, as the
three should be linked to star formation. We found low $SRCC$ values of
0.45--0.55; however, the maps show similar morphologies on high free-free
emission regions, such as the Cygnus region or clouds in $l\in(180\degr,
240\degr)$. This low SRCC value is probably an effect of uncorrected extinction in the H$\alpha$ map. Nevertheless, we must bear in mind that $EM$ is the worst defined 
parameter throughout the plane, with a significance lower than 2 for most 
of the pixels.}. Finally, thermal dust is commonly 
described by any of the \textit{Planck} highest-frequency bands 
or~by the \cite{100micronmap} map. Thermal dust templates are 
sometimes used as templates for the AME, as both are expected to be 
highly spatially correlated~\citep{draine1998a, draine1998b}. In fact, we
can see that the \cite{haslam1982} map at 0.408\,GHz from 
Fig.~\ref{fig:frequency_maps} present a similar morphology to 
that of the $I_{\rm1\,GHz}$ map in Fig.~\ref{fig:parameter_maps_1}. The 
same happens for the \textit{Planck-HFI} map at 353\,GHz from 
Fig.~\ref{fig:frequency_maps} and the $\tau_{353}$ map from 
Fig.~\ref{fig:parameter_maps_1}. The $\tau_{353}$ map is also 
similar to the  $\Iame$ map of Fig.~\ref{fig:parameter_maps_1}
supporting the presence of a correlation between and the AME 
and thermal dust components. CMB anisotropies are hard 
to constrain on top of the Galactic plane emission, as the 
other components are much brighter. Thus, the recovered uncertainties
 for this parameter ($\Delta\Tcmb$) are large, as can be seen 
in the bottom panel of Figure~\ref{fig:parameter_maps_1}.

\subsection{Parameter variations with longitude}
\label{sect:longitude_variations}
\begin{figure}
    \centering
    \includegraphics[width=1\linewidth]{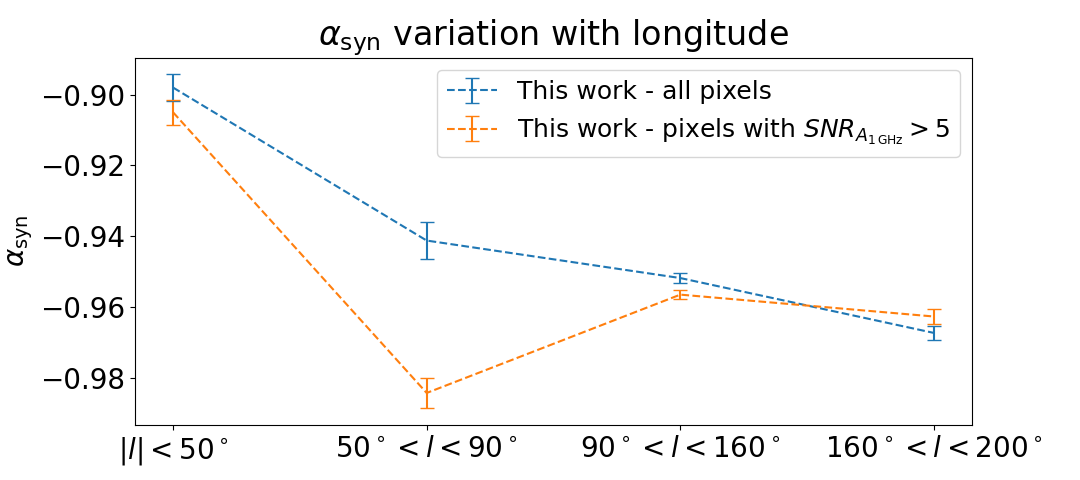}
    \includegraphics[width=1\linewidth]{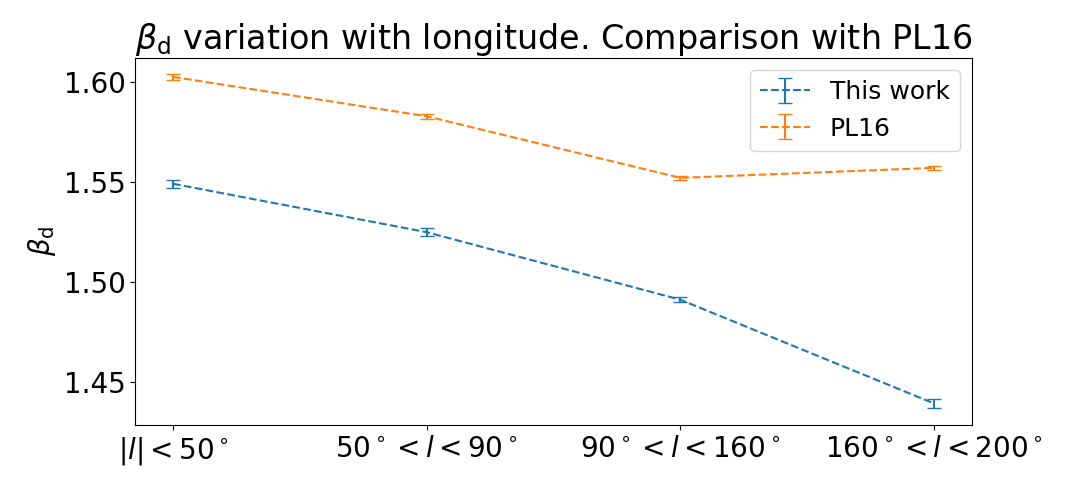}
    \includegraphics[width=1\linewidth]{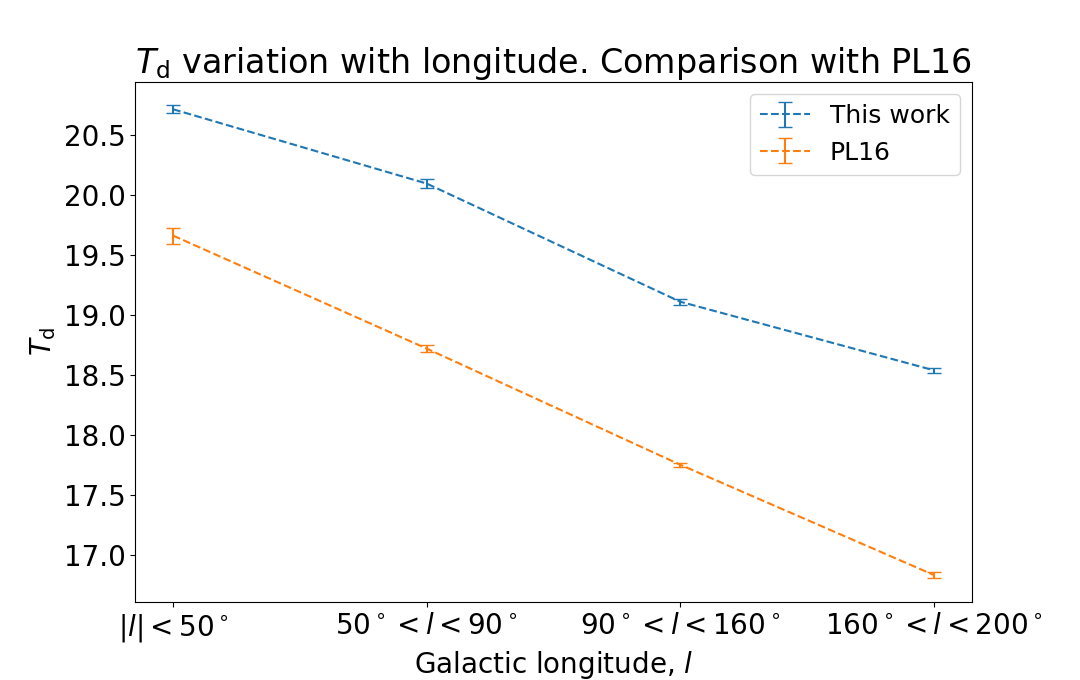}
    \caption{Top: Variation of $\alphasyn$ along 
the regions described in Section~\ref{section:results}.
 Middle: Variation of $\betad$ along the regions 
described in Section~\ref{section:results}. Bottom: 
same as previous plot, but for $\Td$. In all cases
 we are plotting the error of the mean instead of the 
histogram variability (which is reported in 
Tables~\ref{table:parameter_variation_w_regions}, 
\ref{table:parameter_variation_w_regions_SNRgt2} 
and \ref{table:summary_SRCC_new}): the difference
between the two is a factor of $N_{\rm pix}^{-1/2}$.
The values from COMMANDER code results described
in \citetalias{planck2016Xforegroundmaps} are
shown for comparison purposes.}
    \label{fig:latitude_variations}
\end{figure}

Following \cite{planckancillarydata}, 
we studied the average value of spectral indices 
(and other parameters) in the sectors defined at 
the beginning of Section~\ref{section:results}. 
The results are presented in 
Table~\ref{table:parameter_variation_w_regions} 
for the full pixel set. Higher (flatter) values 
for $\alphasyn$ are obtained for the regions 
closer to the Galactic centre. There is also a
 decreasing trend for the dust index and 
temperature, $\betad$ and $\Td$, as we get farther
 from the Galactic centre.  These types of
behavior can be seen in 
Fig.~\ref{fig:latitude_variations}.

Our $\betad$ estimates are close to the 
$\beta_{\rm mm}=1.60\pm0.06$ value obtained 
by~\cite{planckbetamm} for the $l\in(20\degr,44\degr)$, 
$b\in(-4\degr,4\degr)$ region. This area coincides with 
that of the satellite band, so it is not observed by
QUIJOTE-MFI, but the result is more similar to the value for 
Sector 3: $1.56\pm0.06$. $\Td$ values 
are also comparable between the two studies; however, 
 \textit{Planck} was fitting for a bimodal 
distribution, with a break at 353\,GHz. When 
performing further analyses, constant $\Td=19$\,K 
and $\beta_{\rm mm}=1.52$ values were assumed by 
\cite{planckbetamm}, which 
are in fact very close to our median results for 
the full plane ($19.35\pm1.23$\,K and $1.50\pm0.09$ 
respectively).The decrease in $\betad$ values for $200\degr
<l<240\degr$ shown in Fig.~\ref{fig:parameter_maps_1} 
is also recovered in $\betad$ maps from both 
\cite{planck2013XI} and \cite{planck2016Xforegroundmaps}, 
although in the latter case the use of a strong prior on 
$\betad$ reduces the amplitude of the decrement (see 
Section~\ref{section:COMMANDER_comparison}).

The ratio between AME and total flux densities at 
28.4\,GHz increases as we get farther from 
the Galactic centre. This is also consistent 
with previous results. \cite{planckancillarydata}
focused on the regions with $|b|<4\degr$ and Galactic longitude in the ranges $l\in(20\degr, 40\degr)$ and $l\in(320\degr, 340\degr)$, and obtained $S_{\rm AME}^{\rm28.4\,GHz}/S_{\rm
total}^{\rm28.4\,GHz}=0.44\pm0.03$ on average, 
while in this work we get $0.27\pm0.12$ for Sector 3 
and $0.40\pm0.14$ for Sector 4. This 
last region is more adequate to be used as a reference 
for comparison because the first one contains the Galactic 
Centre. The AME to free-free emission ratio is 
also consistent, but has a much higher uncertainty 
in our case: $\simeq1.0\pm0.5$ compared
to the $\simeq0.85\pm0.10$ value from \cite{planckancillarydata}. 
These larger uncertainties are due to the wider latitude 
range in our study ($|b|<10\degr$ compared to $|b|<4\degr$), 
which translates into higher dispersion.

\begin{figure*}
    \centering
    \includegraphics[width=1\linewidth]{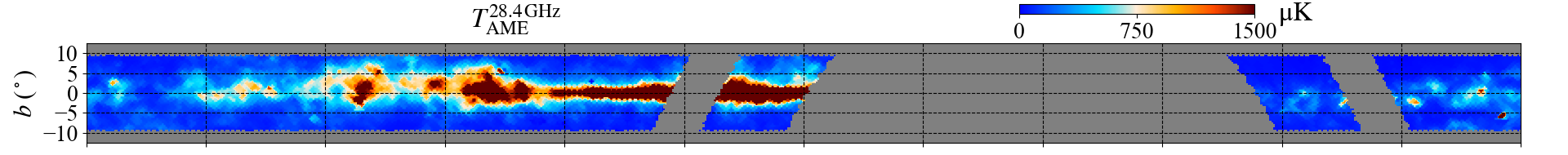}
    \includegraphics[width=1\linewidth]{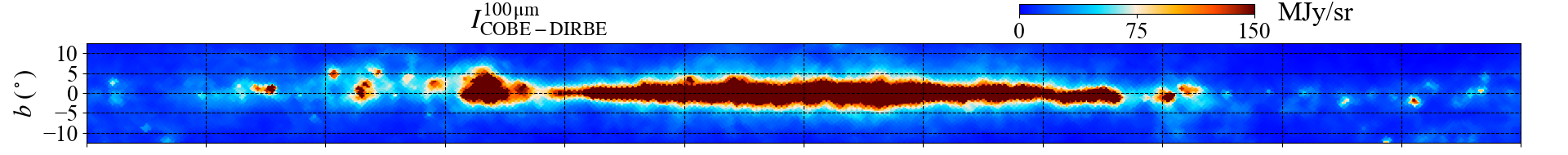}
    \includegraphics[width=1\linewidth]{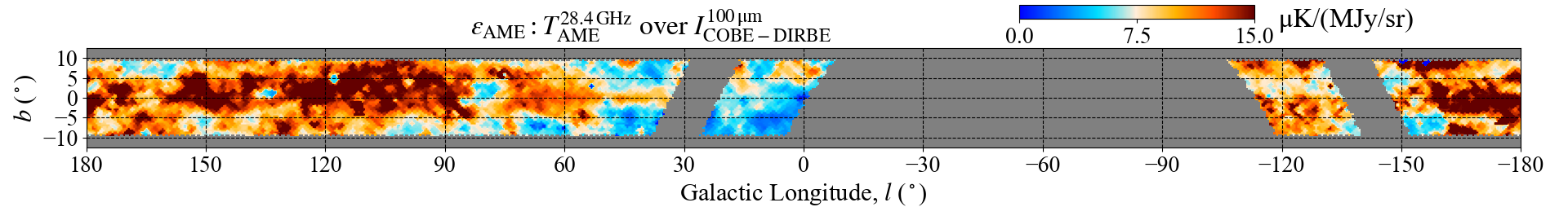}
    \caption{Top: Map of the AME emission at 
28.4\,GHz. We have converted the map from intensity 
to temperature units to facilitate comparison 
with the literature. Middle: COBE-DIRBE map at 
100\,$\mu$m, showing thermal dust emission. Bottom: 
ratio between the former two, or AME emissivity. 
The values for the Galactic centre are lower than 
in other areas and suggest that AME emission is 
much less efficient in that region.}
    \label{fig:emissivity_at_28.4Ghz}
\end{figure*}

The AME emissivity is generally defined as the ratio between an AME 
tracer and a thermal dust one. Previous works used 
different parameters as tracers. In this study, we keep the notation 
from each of the papers we compare to. We show the variations along
the Galactic plane for the ratio between the AME amplitude 
at 28.4\,GHz and the COBE-DIRBE emission map at 100\,$\mu$m, which
we define as $\emmAME$.
As both observables in the numerator and denominator
depend on the column density, the emissivity 
cancels the density dependence, showing only sensitivity 
to the physics of the emission mechanism instead. 
The $\emmAME$ map is shown in the 
bottom panel of Fig.~\ref{fig:emissivity_at_28.4Ghz}. 
We find $\emmAME=11.62\pm3.45$ $\mu$K\,MJy$^{-1}$\,sr 
when taking into account all pixels with $\SNRAME>2$. 
This estimate is consistent with the values of 
$10.9\pm1.1$ $\mu$K\,MJy$^{-1}$\,sr of
\cite{davies2006} (although in that case $\epsilon_{\rm AME}$ 
referred to the 31\,GHz WMAP $K$ band) ; 
$9.8\pm0.5$ $\mu$K\,MJy$^{-1}$\,sr of \cite{planckancillarydata}: 
and $14.0\pm3.5$ $\mu$K\,MJy$^{-1}$\,sr of \cite{cbassSH2022}. 
When focusing on 
variations with the longitude, it is clear that the
 Galactic centre has a much lower $\emmAME$ 
than the rest of the plane. 

The \cite{planckemissivityestimates} also showed how the 
emissivity changes in different regions and environments 
in the sky. The ratio between the AME 
amplitude at 22.8\,GHz and $\taud$ was therefore used: we find 
$T_{\rm AME}^{\rm 22.8\,GHz}/\taud=9.84\pm3.57$\,K 
for those pixels with $\SNRAME>2$. This value is really 
close to the high latitude ($|b|>10\degr$) cut from the 
previous \cite{planckemissivityestimates}, $T_{\rm AME}^{\rm 22.8\,GHz}/\taud=9.7\pm1.0$\,K, 
and lower (but consistent, because of the large dispersion) 
to the value of $11.5_{-1.5}^{+4.2}$\,
found by \cite{cbassSH2022}. For the full sky, the
\cite{planckemissivityestimates} found 
$T_{\rm AME}^{\rm 22.8\,GHz}/\taud=8.3\pm0.8$\,K, 
which is also consistent with our result. At 30\,GHz, we find 
$T_{\rm AME}^{\rm 30\,GHz}/\taud=4.66\pm2.18$\,K,
 much lower than the value of $7.9\pm2.6$\,K
obtained at high latitudes by~\cite{hensley2016}. At 
28.4\,GHz, the 
$T_{\rm AME}^{\rm 28.4\,GHz}/\taud=5.51\pm2.43$\,K ratio 
is really similar to that of $5.8_{-0.5}^{+2.9}$\,K
obtained by \cite{cbassSH2022}. The AME emissivity 
increment with Galactic longitude is more noticeable 
than for the AME fraction. But the difference between 
the anticentre (Sector 6) and the centre 
(Sector 3) is still not statistically significant 
(just 1.75$\sigma$).

In Table~\ref{table:parameter_variation_w_regions_SNRgt2} 
we show the same values as in
 Table~\ref{table:parameter_variation_w_regions}, but 
only taking into account those pixels with high 
synchrotron, AME or dust significances, depending on 
the parameter studied. The results are similar to those 
for the whole pixel set, apart from a lower dispersion 
on AME parameter distributions and higher AME fraction 
when the $\SNRAME$ threshold is introduced (as expected).
 There is also an inconsistency ($2.9\sigma$) for $\alphasyn$ 
in Sector 4 when 
introducing the $\rm SNR_{\rm syn}>5$ threshold (see 
Fig.~\ref{fig:latitude_variations}). This is because
 for this region the threshold masks most of the 
pixels in the plane. These pixels come mainly from the Cygnus region, 
which is dominated by free-free emission (therefore 
their ${\rm SNR}_{\rm syn}$ values are low). The 
weight of the pixels outside the Galactic plane (which in 
general show steeper values) for the determination of 
$\alphasyn$ is therefore higher. 

\begin{table*}
\caption{Median values, plus their dispersion, for a selected 
group of parameters along the sectors described at the beginning
of Section~\ref{section:results}. Every pixel in our $|b|<10\degr$ 
map is taken into account. We see variations mostly 
between the regions closest to the Galactic centre 
($\delta<-10\degr$; $|l|<50\degr$)  and the rest of 
the plane. Both the synchrotron and dust indices 
($\alphasyn$, $\betad$), and the dust temperature 
($\Td$) show a decreasing trend as we get farther from 
the Centre. Uncertainties in this table and in
Table~\ref{table:parameter_variation_w_regions_SNRgt2} 
account for the histogram dispersions, not errors 
of the mean.}
\begin{tabular}{lccccccc}
\hline
 \multirow{2}*{Parameter} & \multirow{2}*{All pixels}     & Sector 1   & Sector 2   & Sector 3      & Sector 4   & Sector 5   & Sector 6   \\
 & & ($\delta<-10\degr$) & ($\delta>13\degr$) & ($|l|<50\degr$) & ($50\degr\leq l<90\degr$) & ($90\degr\leq l < 160\degr$) & ($160\degr\leq l < 200\degr$) \\
\hline
 $\alphasyn$                            & $-0.94\pm0.10$ & $-0.88\pm0.09$  & $-0.96\pm0.08$ & $-0.90\pm0.12$ & $-0.94\pm0.16$    & $-0.95\pm0.06$       & $-0.97\pm0.06$        \\
 $\betad$                              & $1.49\pm0.08$  & $1.49\pm0.12$   & $1.49\pm0.06$  & $1.55\pm0.06$  & $1.52\pm0.06$     & $1.49\pm0.05$        & $1.44\pm0.07$         \\
 $\Td$ (K)                            & $19.37\pm1.19$ & $20.17\pm1.52$  & $19.20\pm1.07$ & $20.72\pm1.03$ & $20.10\pm1.07$    & $19.11\pm0.95$       & $18.54\pm0.64$        \\
 $\nuame$ (GHz)                        & $21.63\pm3.68$ & $23.54\pm6.70$  & $20.95\pm2.76$ & $21.61\pm2.30$ & $21.80\pm2.94$    & $20.72\pm2.24$       & $20.89\pm3.56$        \\
 $\Wame$                               & $0.59\pm0.06$  & $0.61\pm0.06$   & $0.58\pm0.06$  & $0.58\pm0.05$  & $0.59\pm0.08$     & $0.57\pm0.06$        & $0.60\pm0.06$         \\
 $\Iame/S^{\nuame}_{\rm total}$        & $0.35\pm0.15$  & $0.21\pm0.11$   & $0.41\pm0.12$  & $0.27\pm0.12$  & $0.42\pm0.13$     & $0.43\pm0.11$        & $0.38\pm0.12$         \\
 $\Iame/S^{\rm 28.4\,GHz}_{\rm total}$ & $0.34\pm0.14$  & $0.21\pm0.10$   & $0.39\pm0.12$  & $0.28\pm0.12$  & $0.41\pm0.14$     & $0.40\pm0.11$        & $0.36\pm0.12$         \\
 $\Iame/S^{\nuame}_{\rm ff}$           & $1.07\pm0.58$  & $0.70\pm0.50$   & $1.20\pm0.58$  & $1.18\pm0.46$  & $1.26\pm0.69$     & $1.40\pm0.53$        & $0.85\pm0.34$         \\
 $\Iame/S^{\rm 28.4\,GHz}_{\rm ff}$    & $0.90\pm0.50$  & $0.65\pm0.44$   & $1.01\pm0.49$  & $1.05\pm0.40$  & $1.14\pm0.60$     & $1.19\pm0.48$        & $0.72\pm0.25$         \\
 $\emmAME$ ($\mu$K MJy$^{-1}$ sr) & $8.84\pm3.77$ & $6.79\pm2.48$ & $10.12\pm3.65$ & $5.70\pm2.28$ & $8.84\pm3.01$ & $11.43\pm3.58$ & $9.95\pm3.88$ \\
\hline
\end{tabular}
\label{table:parameter_variation_w_regions}
\end{table*}

\begin{table*}
\caption{Similar to Table~\ref{table:parameter_variation_w_regions}, 
but taking into account only  pixels with 
${\rm SNR}_{\rm syn}=I_{\rm1\,GHz}/\sigma(I_{\rm1\,GHz})>5$ 
for $\alphasyn$, ${\rm SNR}_{\rm dust}=\tau_{353}/\sigma(\tau_{353})>5$ 
for $\betad$ and $\Td$, and $\SNRAME>2$ for AME parameters, 
fractions and $\emmAME$. These selections account for 
80\,\%, 100\,\% and 47\,\% of the pixels respectively
 and are applied only to the corresponding 
component studied. The latter selection returns similar results 
as using $\SNRAME>3$, while increasing the sample.} 
\begin{tabular}{lccccccc}
\hline
 \multirow{2}*{Parameter} & \multirow{2}*{All pixels}     & Sector 1   & Sector 2   & Sector 3      & Sector 4   & Sector 5   & Sector 6   \\
 & & ($\delta<-10\degr$) & ($\delta>13\degr$) & ($|l|<50\degr$) & ($50\degr\leq l<90\degr$) & ($90\degr\leq l < 160\degr$) & ($160\degr\leq l < 200\degr$) \\
\hline
 $\alphasyn$                            & $-0.94\pm0.09$ & $-0.88\pm0.09$  & $-0.96\pm0.07$ & $-0.90\pm0.11$ & $-0.98\pm0.12$    & $-0.96\pm0.05$       & $-0.96\pm0.05$        \\
 $\betad$                              & $1.49\pm0.08$  & $1.49\pm0.12$   & $1.49\pm0.06$  & $1.55\pm0.06$  & $1.52\pm0.06$     & $1.49\pm0.05$        & $1.44\pm0.07$         \\
 $\Td$ (K)                            & $19.37\pm1.19$ & $20.17\pm1.52$  & $19.20\pm1.07$ & $20.72\pm1.03$ & $20.10\pm1.07$    & $19.11\pm0.95$       & $18.54\pm0.64$        \\
 $\nuame$ (GHz)                        & $20.69\pm1.91$ & $22.84\pm2.03$  & $20.43\pm1.78$ & $21.50\pm2.20$ & $21.10\pm1.73$    & $20.39\pm1.72$       & $19.93\pm1.99$        \\
 $\Wame$                               & $0.56\pm0.05$  & $0.56\pm0.04$   & $0.56\pm0.06$  & $0.56\pm0.04$  & $0.57\pm0.06$     & $0.56\pm0.06$        & $0.56\pm0.06$         \\
 $\Iame/S^{\nuame}_{\rm total}$        & $0.46\pm0.08$  & $0.34\pm0.05$   & $0.48\pm0.07$  & $0.36\pm0.06$  & $0.49\pm0.07$     & $0.47\pm0.07$        & $0.48\pm0.06$         \\
 $\Iame/S^{\rm 28.4\,GHz}_{\rm total}$ & $0.44\pm0.09$  & $0.34\pm0.05$   & $0.45\pm0.08$  & $0.36\pm0.06$  & $0.48\pm0.08$     & $0.44\pm0.09$        & $0.47\pm0.10$         \\
 $\Iame/S^{\nuame}_{\rm ff}$           & $1.54\pm0.45$  & $1.48\pm0.50$   & $1.57\pm0.47$  & $1.54\pm0.32$  & $1.75\pm0.58$     & $1.60\pm0.44$        & $1.22\pm0.26$         \\
 $\Iame/S^{\rm 28.4\,GHz}_{\rm ff}$    & $1.35\pm0.45$  & $1.35\pm0.49$   & $1.36\pm0.47$  & $1.39\pm0.30$  & $1.50\pm0.51$     & $1.38\pm0.45$        & $1.02\pm0.23$         \\
 $\emmAME$ ($\mu$K MJy$^{-1}$ sr) & $11.62\pm3.45$ & $7.32\pm2.07$ & $12.18\pm2.99$ & $7.15\pm2.27$ & $10.76\pm2.27$ & $12.74\pm2.86$ & $14.15\pm3.28$ \\
\hline
\end{tabular}
\label{table:parameter_variation_w_regions_SNRgt2}
\end{table*}

\subsection{Correlations between the model parameters}
\label{section:correlations_parameters}
We have used the Spearman Rank Correlation Coefficient 
(hereafter, SRCC) to compare the correlations between 
parameters, as in \citetalias{planck2015galacticcloudsAME}, 
\citetalias{LambdaOrionis} and \citetalias{AMEwidesurvey}. 
To maintain the notation of those articles, we have 
used the AME amplitude, $\Aame$, instead of its intensity, 
$\Iame$: $\Iame$ was better for the representations 
on the maps. The relation between the two depends on the 
solid angle of our aperture ($\Omega$), which 
is equal and constant to one \nside=64 pixel: 
$\Omega=2.56\times10^{-4}{\rm\,sr\ }$. Some of the 
main correlations between the reconstructed parameters
 (and some others derived from them) are summarized 
in Table~\ref{table:summary_SRCC_new}. If AME is
 expected to be the result of spinning dust 
\citep{draine1998a, draine1998b}, its emission will
 be correlated with the thermal emission from that 
same dust. This is explored in 
Figures~\ref{fig:correlation_bw_Aame_and_dustopacity} 
and~\ref{fig:correlation_bw_Aame_and_dustradiance},
 where the AME amplitude is compared with the dust 
opacity ($\taud$) and radiance respectively. The dust
radiance is obtained as follows:
\begin{equation}
\mathfrak{R}_{\rm dust}(\tau_{353}, \betad, \Td)=
\int_{-\infty}^{\infty} S_{\nu}^{\mathrm{dust}}(\tau_{353}, \betad, \Td) {\rm d}\nu.
\end{equation}
The SRCC is a bit higher for $\mathfrak{R}_{\rm dust}$ 
than for $\taud$, confirming the claim from \cite{hensley2016}
that $\mathfrak{R}_{\rm dust}$ is the best AME predictor
(although is larger than $0.9$ in both cases). 
These correlations were also found by 
\citetalias{planck2015galacticcloudsAME}, 
\citetalias{LambdaOrionis} and \citetalias{AMEwidesurvey}.
The uncertainties for these estimates have been 
computed as in~\cite{SRCCmcmc}, using 1000 different 
realizations. This is further explained in 
Appendix~\ref{section:appendix_SRCC}. When studying 
these correlations, we selected only the pixels with 
$\SNRAME>2$, which account 
for $\simeq47\,\%$ of the sample. We chose the $\SNRAME$
as a selection proxy to be sensitive to pixels high AME
significance, instead to those with high flux density residuals
between 20 and 60\GHz{}. The difference 
between the two estimates is important, as we are covering 
a large variety of regions. For example, the Cygnus region
 has large flux densities, but the AME fraction is low 
(below $25\,\%$), as most of the emission is 
free-free. Nevertheless, these regions showing
strong free-free emission are interesting to study the 
expected correlation between the dust temperature, $\Td$,
and star formation ratio, SFR, and therefore $\EM$. 
The two do not appear correlated ($SRCC=0.013\pm0.026$)
when studying all the pixels together due to the bad $\EM$
definition in most pixels of the plane. However, when using
just those pixels with ${\rm SNR}_{\EM}= \EM / \sigma(\EM)>2$,
the correlation between $\EM$ and $\Td$ greatly increases to 
$SRCC=0.77\pm0.03$.

There are no big differences in SRCC values between 
the regions described in Section~\ref{section:results},
 but we find important differences (larger than $10\sigma$) 
for the slopes from the corresponding fits. 
These are noticeable for the $\Aame$ vs.\  
$\mathfrak{R}_{\rm dust}$ or $\tau_{353}$ relations 
(Fig.~\ref{fig:correlation_bw_Aame_and_dustradiance}):
 the slope flattens (steepens) with increasing 
$\mathfrak{R}_{\rm dust}$ ($\tau_{353}$) values 
(as for the Galactic centre). This could have 
several causes: maybe there are different dust populations 
along the lines of sight that complicate the dust 
physics. In that case, our single MBB model would not
 completely solve that component. Or maybe the 
interstellar radiation field (ISRF) is too strong in
 this area, hence preventing the spinning grains from emitting AME 
as expected and partially destroying them instead (as
 happens in unresolved sources [\citealp{oct21PAHinQSOs}]
 and photoionized gas [\citealt{dongdraine2011fromhensley2021}]).
This could imply different correlations between the 
ISRF and the AME as we get close to the Galactic 
Centre, or generally between different regions. However, 
sources in low resolution (degree scales) analyses (such as 
those of \citetalias{planck2015galacticcloudsAME},
~\citetalias{LambdaOrionis}, ~\citetalias{AMEwidesurvey}, 
and this study) appear too faint owing to dilution in the 
beam: the $\Aame$ vs.\ $\mathfrak{R}_{\rm dust}$ relation 
seems to bend at $\mathfrak{R}_{\rm dust}=10^9$ Jy Hz. 
Observations at higher resolution of compact AME sources 
are needed to fully sample this behavior, 
as stronger interstellar radiation fields (high $G_0$) 
or $\mathfrak{R}_{\rm dust}$ values 
are required.

We report another correlation between the dust 
temperature, $\Td$, and the AME peak frequency, 
$\nuame$, as showed in Fig.~\ref{fig:correlation_bw_Td_and_nuame}.
 A similar dependence is recovered between $\nuame$ 
and the ISRF proxy, $G_0$, obtained as in \cite*{G0mathis1983}:
\begin{equation}
G_0 \equiv \left(\frac{\Td}{T_0}\right)^{4+\betad}
\end{equation}
with $T_0=17.5$\,K. We see that $\betad$ has little 
impact on the behaviour of $G_0$, so $\Td$ dominates it: in 
fact, the correlations of $G_0$ and $\Td$ with other
components are often very similar. This is the 
case for the correlations between $\Td$ and $G_0$ 
with $\nuame$, which return identical results ($0.63\pm0.11$). 
This result is consistent with those obtained 
by~\citetalias{planck2015galacticcloudsAME} and~
\citetalias{LambdaOrionis} ($0.65\pm0.15$ and 
$0.60\pm0.07$ respectively), while~\citetalias{AMEwidesurvey}
founds a correlation only for its semi-significant
AME sample ($0.60\pm0.15$). \cite{beyondplanckintensity} 
found a positive correlation between $\nuame$ and 
$\betad$ (SRCC=0.85) instead of $\Td$. 
However, in that case $\Td$ is fixed to the results obtained
with the NPIPE pipeline on \textit{Planck} DR4 \citep{npipe}. 
Both cases probably indicate the same relation between 
the location of spinning and thermal dust emission peaks in the 
frequency range, but in one case the correlation is found
when comparing to $\Td$ and in the other when comparing 
with $\betad$. Neither did we find any correlations 
between $\nuame$ and $\betad$ ($0.00\pm0.04$), nor 
between $\betad$ and $\Td$ ($0.13\pm0.04$).  
The full set of correlations between the 
parameters studied can be seen in Table~\ref{table:appendix_full_SRCC_table}, 
the most interesting feature being the 
lack of correlations between $\Wame$ and all 
the other parameters.

Other differences between this study, 
~\citetalias{LambdaOrionis} and~\citetalias{AMEwidesurvey} are:
\begin{itemize}
    \item the $\EM$ vs.\ $\nuame$ correlation,
 which is $0.42\pm0.08$ in this study, $0.80\pm0.03$ 
in~\citetalias{LambdaOrionis} ($>2\sigma$ away;
 with no synchrotron present) and negative or
 absent in~\citetalias{AMEwidesurvey} (depending
 on the exact sample);
    \item the correlation between the AME 
emissivity, defined as 
$\Aame/\tau_{353}$,\footnote{This emissivity 
definition is analogous to that in 
Sect.~\ref{sect:longitude_variations}, as 
we still compare AME to thermal dust emission. 
We use now this definition to be consistent 
with the analyses of \citetalias{LambdaOrionis} 
and~\citetalias{AMEwidesurvey}, both using
QUIJOTE-MFI data too.} and 
$\Wame$, which is less important here ($\sim0.44\pm0.20$) 
than in~\citetalias{AMEwidesurvey} ($\sim0.60\pm0.15$); 
\item the absence of correlation between $\EM$ and $\Td$ 
in this study, $0.25\pm0.03$, 
partly compatible with~\citetalias{AMEwidesurvey} 
(depending on the exact sample), 
while~\citetalias{LambdaOrionis} finds a higher 
degree of correlation ($0.65\pm0.07$).
\end{itemize} 

In any case, it is difficult to compare the 
free-free related correlations between the 
previous two studies and this one, as no synchrotron 
component was used in \citetalias{LambdaOrionis}. 
\citetalias{AMEwidesurvey} introduced it only 
for a few sources, mostly supernova remnants, 
where the presence of synchrotron was evident. 

On the other hand,  this study, \citetalias{LambdaOrionis}
 and~\citetalias{AMEwidesurvey} find similar
 results for the correlation between the AME
 emissivity and $\Td$ ($0.76\pm0.12$, 
$0.82\pm0.06$ and $\simeq0.68\pm0.08$ respectively; see 
Figures~\ref{fig:correlation_bw_AMEemissivity_and_Td} 
and~\ref{fig:correlation_bw_AMEemissivity_and_G0}).
 \citetalias{planck2015galacticcloudsAME} 
found a lower value, $0.63\pm0.07$, which is 
still consistent within $1\sigma$ with our result.
 The linear fits describing both this relation and 
the $\Td$ vs.\ $\nuame$ one are consistent across
the regions considered. 
However,
when calculating the AME emissivity using a different
dust tracer, as could be the dust radiance ($\Aame/
\mathcal{R_{\rm dust}}$) or the previously presented
intensity at 100\,$\mu$m ($\emmAME$), this correlation
is absent ($-0.43\pm0.09$ and $-0.42\pm$0.07). 
This issue is further discussed on
Section~\ref{section:new_carriers_section}.

Finally, we find significant correlations between 
the amplitudes of synchrotron ($A_{1\,\mathrm{GHz}}$), 
free-free ($\rm EM$), AME ($\Aame$) and thermal dust ($\taud$). 
Studying a region so heavily populated as the Galactic plane,
we detect very large variations for the column density 
of the pixels. A pixel with many environments along its 
line-of-sight is more likely to have large amplitudes 
for all components.
\citetalias{LambdaOrionis} and~\citetalias{AMEwidesurvey} 
did not fully recover this, as they were not so
sensitive to the different line-of-sight densities. 
For example, the EM vs.\ $\Aame$ relation was not studied in the 
former, while bringing coherent results between 
selections in the latter ($0.59\pm0.05/0.65\pm0.11$), 
depending on the sample. On the other hand, we got a higher
 correlation value of $0.90\pm0.14$.

\begin{table*}
\centering
\caption{SRCC values for a series of selected 
variable pairs, taking into account only $\SNRAME>2$ 
pixels. Results are similar to those obtained 
placing the threshold at $\SNRAME>3$, while increasing 
the sample. $N_{\rm pix}$ accounts for the number of
 pixels considered for each region: we can see that 
it is anticorrelated with the SRCC uncertainties,
 as expected.}
\begin{tabular}{ccccccccc}
\hline
 \multirow{2}*{Variable 1} & \multirow{2}*{Variable 2} & \multirow{2}*{All pixels}     & Sector 1   & Sector 2   & Sector 3      & Sector 4   & Sector 5   & Sector 6   \\
  & & & ($\delta<-10\degr$) & ($\delta>13\degr$) & ($|l|<50\degr$) & ($50\degr\leq l<90\degr$) & ($90\degr\leq l < 160\degr$) & ($160\degr\leq l < 200\degr$) \\
\hline
 $\Aame$ (Jy)               & $\tau_{353}$                      & 0.90 $\pm$ 0.03 & 0.92 $\pm$ 0.02 & 0.89 $\pm$ 0.03 & 0.96 $\pm$ 0.02 & 0.94 $\pm$ 0.02   & 0.90 $\pm$ 0.03      & 0.84 $\pm$ 0.10       \\
 $\Aame$ (Jy)               & $\mathfrak{R}_{\rm dust}$ (Jy Hz) & 0.95 $\pm$ 0.03 & 0.98 $\pm$ 0.02 & 0.96 $\pm$ 0.03 & 0.98 $\pm$ 0.02 & 0.98 $\pm$ 0.02   & 0.98 $\pm$ 0.03      & 0.90 $\pm$ 0.11       \\
 $\Aame$ (Jy)               & $S_{\rm TD,\ peak}$ (Jy)          & 0.96 $\pm$ 0.22 & 0.98 $\pm$ 0.12 & 0.97 $\pm$ 0.28 & 0.98 $\pm$ 0.12 & 0.98 $\pm$ 0.19   & 0.97 $\pm$ 0.32      & 0.90 $\pm$ 0.49       \\
 $\Aame / \tau_{353}$ (Jy)  & $\Td$ (K)                         & 0.76 $\pm$ 0.12 & 0.83 $\pm$ 0.14 & 0.76 $\pm$ 0.14 & 0.74 $\pm$ 0.14 & 0.73 $\pm$ 0.14   & 0.72 $\pm$ 0.16      & 0.52 $\pm$ 0.15       \\
 $\Wame$                          & $\Aame / \tau_{353}$ (Jy)   & 0.44 $\pm$ 0.20 & 0.17 $\pm$ 0.09 & 0.51 $\pm$ 0.22 & 0.36 $\pm$ 0.21 & 0.68 $\pm$ 0.29   & 0.58 $\pm$ 0.27      & 0.05 $\pm$ 0.07       \\
 $\Aame$ (Jy)               & $\mathfrak{R}_{\rm AME}$ (Jy Hz)  & 0.97 $\pm$ 0.08 & 0.99 $\pm$ 0.13 & 0.96 $\pm$ 0.19 & 0.99 $\pm$ 0.12 & 0.97 $\pm$ 0.15   & 0.96 $\pm$ 0.21      & 0.93 $\pm$ 0.34       \\
 $S_{\rm TD,\ peak}$ (Jy)         & EM (pc / cm$^6$)                  & 0.88 $\pm$ 0.27 & 0.93 $\pm$ 0.20 & 0.87 $\pm$ 0.36 & 0.97 $\pm$ 0.20 & 0.91 $\pm$ 0.27   & 0.91 $\pm$ 0.44      & 0.89 $\pm$ 0.60       \\
 $\Aame$ (Jy)               & EM (pc / cm$^6$)                  & 0.90 $\pm$ 0.14 & 0.94 $\pm$ 0.09 & 0.89 $\pm$ 0.16 & 0.98 $\pm$ 0.08 & 0.93 $\pm$ 0.10   & 0.90 $\pm$ 0.18      & 0.87 $\pm$ 0.27       \\
 $\mathfrak{R}_{\rm AME}$ (Jy Hz) & $\mathfrak{R}_{\rm dust}$ (Jy Hz) & 0.92 $\pm$ 0.14 & 0.98 $\pm$ 0.05 & 0.93 $\pm$ 0.06 & 0.98 $\pm$ 0.04 & 0.94 $\pm$ 0.04   & 0.95 $\pm$ 0.07      & 0.84 $\pm$ 0.15       \\
 $\Wame$                          & $\Td$ (K)                         & 0.41 $\pm$ 0.17 & 0.25 $\pm$ 0.13 & 0.53 $\pm$ 0.20 & 0.27 $\pm$ 0.16 & 0.62 $\pm$ 0.24   & 0.65 $\pm$ 0.26      & 0.24 $\pm$ 0.15       \\
 $\nuame$ (GHz)                   & $\Td$ (K)                         & 0.63 $\pm$ 0.11 & 0.72 $\pm$ 0.15 & 0.64 $\pm$ 0.13 & 0.69 $\pm$ 0.13 & 0.66 $\pm$ 0.16   & 0.66 $\pm$ 0.15      & 0.52 $\pm$ 0.13       \\
 $\Td$ (K)                        & EM (pc / cm$^6$)                  & 0.25 $\pm$ 0.03 & 0.49 $\pm$ 0.10 & 0.07 $\pm$ 0.03 & 0.43 $\pm$ 0.08 & 0.00 $\pm$ 0.05   & 0.07 $\pm$ 0.03      & 0.09 $\pm$ 0.06       \\
 $\Aame$ (Jy)               & $\Td$ (K)                         & 0.29 $\pm$ 0.02 & 0.54 $\pm$ 0.06 & 0.13 $\pm$ 0.02 & 0.40 $\pm$ 0.06 & -0.15 $\pm$ 0.04  & 0.08 $\pm$ 0.03      & 0.04 $\pm$ 0.06       \\
 $\Aame / \tau_{353}$ (Jy)  & EM (pc / cm$^6$)                  & 0.37 $\pm$ 0.07 & 0.58 $\pm$ 0.15 & 0.29 $\pm$ 0.06 & 0.64 $\pm$ 0.12 & 0.29 $\pm$ 0.08   & 0.31 $\pm$ 0.08      & 0.18 $\pm$ 0.10       \\
 $\nuame$ (GHz)                   & EM (pc / cm$^6$)                  & 0.42 $\pm$ 0.08 & 0.69 $\pm$ 0.69 & 0.32 $\pm$ 0.32 & 0.61 $\pm$ 0.61 & 0.35 $\pm$ 0.34   & 0.30 $\pm$ 0.29      & 0.28 $\pm$ 0.27       \\ \hline
  & $N_{\rm pix}$                         & 2437            & 257             & 1985            & 417             & 513               & 1142                 & 317                   \\
\hline
\end{tabular}
\label{table:summary_SRCC_new}
\end{table*}

\section{Discussion}
\label{section:discussion}
\subsection{Impact of QUIJOTE-MFI on the characterization of 
the Galactic plane}
\label{section:no_MFI_comparison}
\begin{figure*}
    \centering
    \includegraphics[width=1\linewidth]{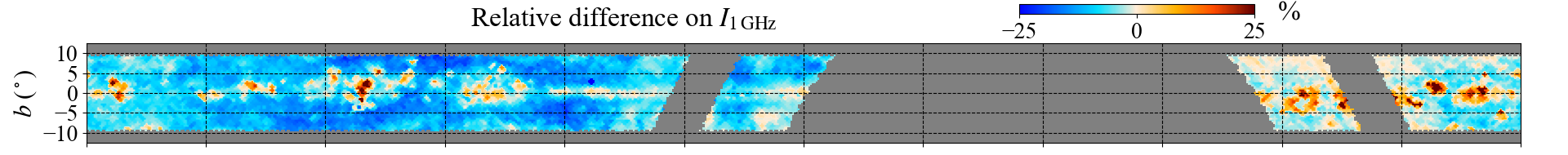}
    \includegraphics[width=1\linewidth]{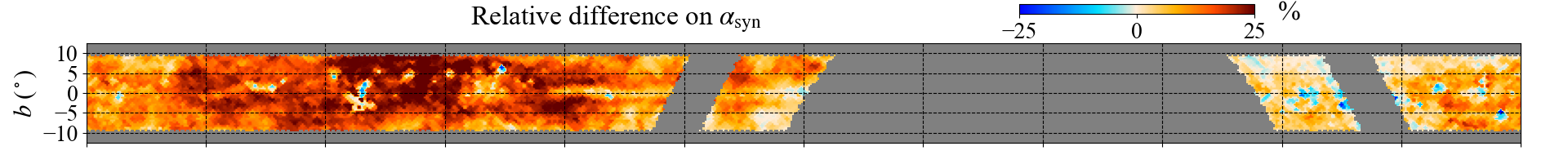}
    \includegraphics[width=1\linewidth]{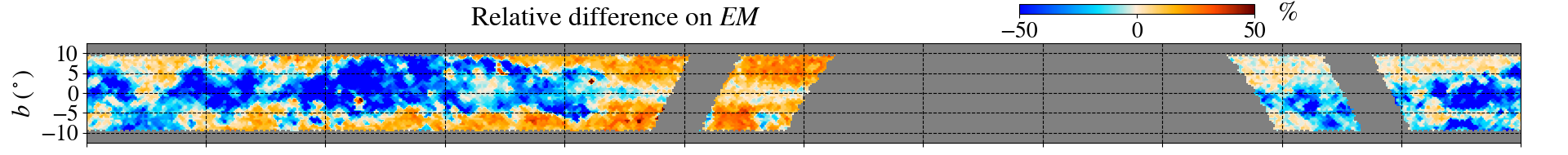}
    \includegraphics[width=1\linewidth]{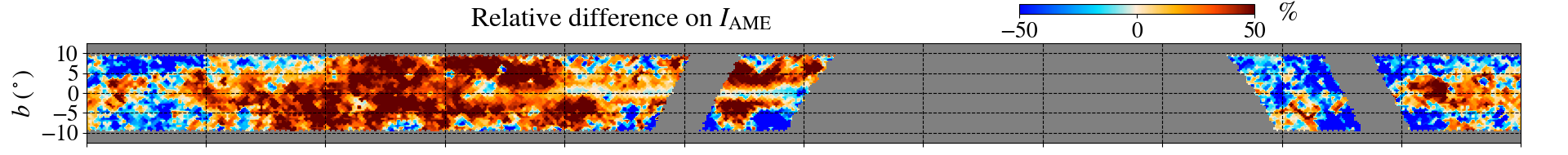}
    \includegraphics[width=1\linewidth]{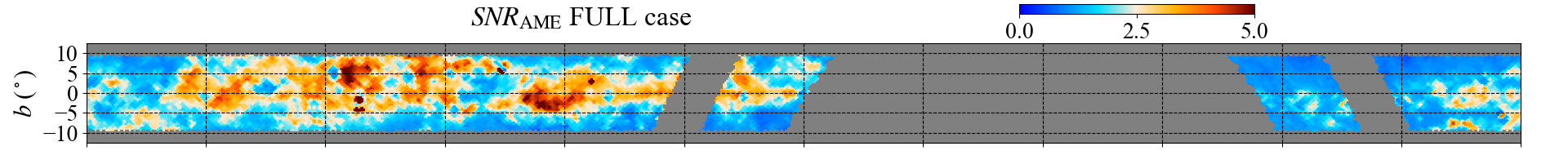}
    \includegraphics[width=1\linewidth]{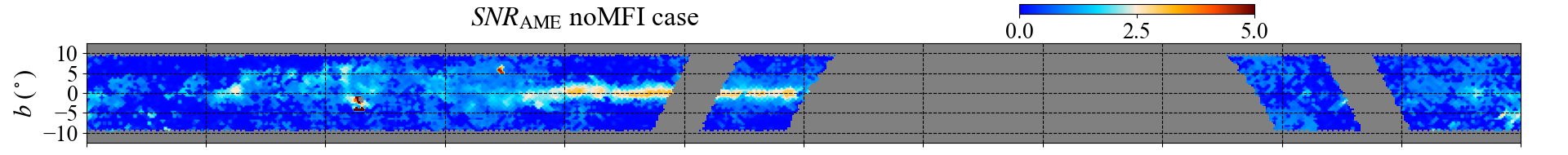}
    \includegraphics[width=1\linewidth]{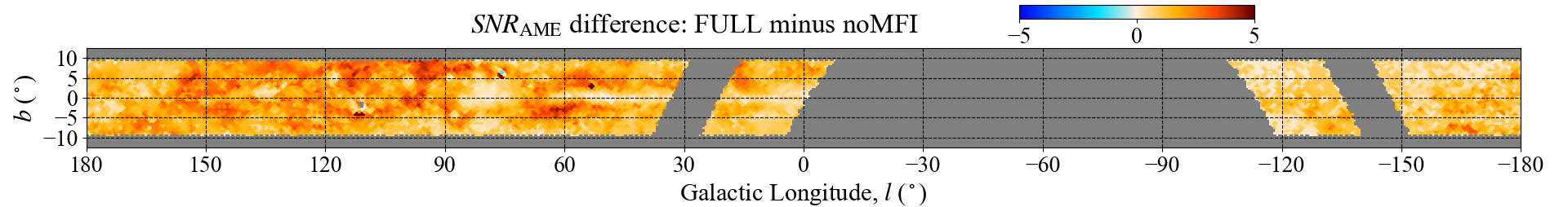}
    \caption{Top two: Relative differences for 
the two synchrotron parameters when using, or not using,
 the MFI points to build the spectral energy 
distributions (SEDs). Differences are calculated as the
estimate for the FULL case minus the one from noMFI case,
divided by the first one. Next two: Relative differences
for the parameters driving free-free and AME intensities. 
We see that AME is more important when QUIJOTE-MFI data is 
taken into account (FULL scenario), while free-free gains 
importance when it is not (noMFI scenario). Bottom three: differences in $\SNRAME$ 
when the QUIJOTE points are added to the survey set.}
    \label{fig:maps_comparison_with_without_QUIJOTE}
\end{figure*}

In order to properly assess the increased capability of recovering 
the AME properties provided by the addition of QUIJOTE-MFI 
data, we applied the same methodology, but now 
without QUIJOTE-MFI (the rest of the survey set from 
Table~\ref{table:maps} remains the same). 
For clarity, we named these two cases as `FULL' 
when the QUIJOTE-MFI points are taken into account, and 
`noMFI' otherwise. The results can be seen in 
Fig.~\ref{fig:maps_comparison_with_without_QUIJOTE}, where we show 
how $\SNRAME$ changes. $\SNRAME$ is higher for the FULL case,
 with great improvement towards the areas where AME is not 
detected at all in the noMFI case (farther away from the plane). 
In the FULL case, these regions have marginal detections, 
predominantly with $\SNRAME$ between 2 and 3.

\begin{figure}
    \centering
    \includegraphics[width=1\linewidth]{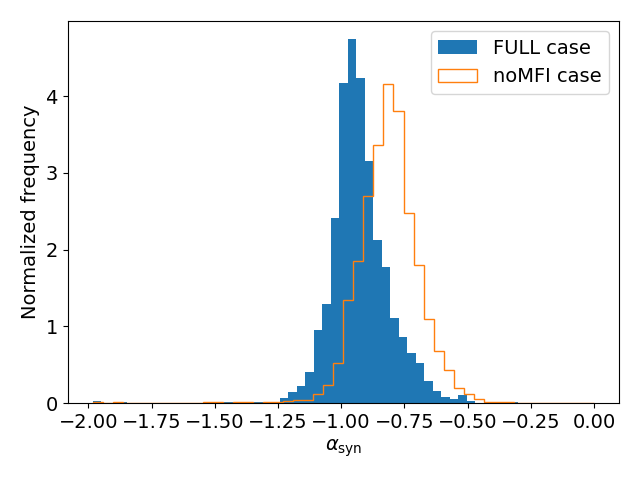}
    \caption{Distribution for $\alphasyn$, both 
with and without MFI points. We see how, in  the 
latter case, the results are displaced towards higher 
values: from $\alphasyn=-0.95_{-0.09}^{+0.15}$ to
 $\alphasyn=-0.84_{-0.09}^{+0.13}$ (both are consistent 
within $1\sigma$, though).}
\label{fig:beta_syn_distribution_comparison_wo_MFI_points_part2}
\end{figure}

Focusing on the parameters, we can see in 
Figures~\ref{fig:maps_comparison_with_without_QUIJOTE} 
and~\ref{fig:beta_syn_distribution_comparison_wo_MFI_points_part2} 
how $\alphasyn$ changes towards flatter values in the
 noMFI case, especially close to the Fan region ($l\sim110\degr$). In 
Fig.~\ref{fig:maps_comparison_with_without_QUIJOTE} we also 
show that $\Iame$ values are lower in the noMFI case,
free-free signal thus increasing. For example, the same SED 
showed in Fig.~\ref{fig:example_SED}, when computed in the
noMFI case, returns an AME flux density fraction at 28.4\,GHz
lower than half its previous value (from $53\,\%$ to $21\,\%$).
That difference is mostly accounted for by free-free. 
The clear anticorrelation pattern between $EM$ and $\Aame$ 
present on Fig.~\ref{fig:example_corner_plot} also worsens
in the noMFI case. The dust parameters do not change 
significantly, as expected: data between 10 and 20\,GHz 
should not constrain a component that rises beyond 100\,GHz.


\begin{figure}
    \centering
    \includegraphics[width=1\linewidth]{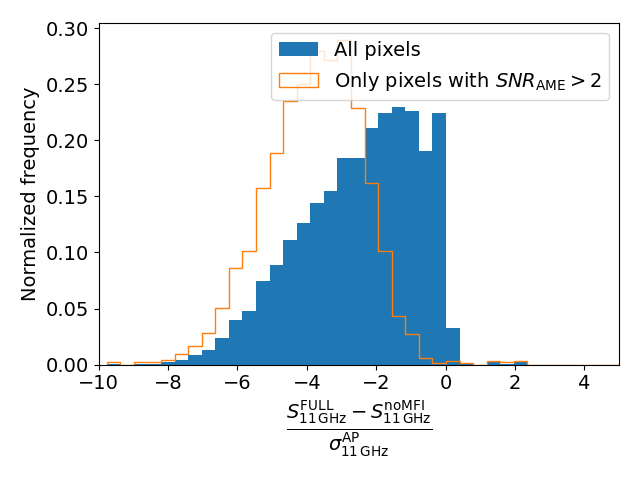}
    \includegraphics[width=1\linewidth]{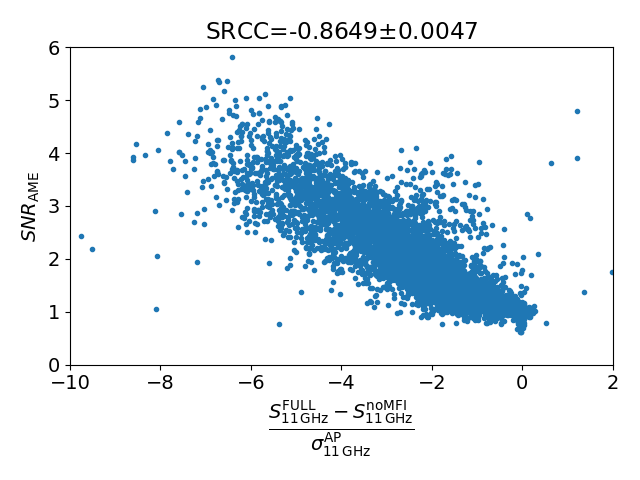}
    \caption{Top: Histograms of the difference in the flux density
predicted by the fitted model at 11.2\,GHz between the FULL
and noMFI cases, over 
the uncertainty obtained for the SED in the FULL 
scenario. We see how most of the pixels (especially 
when focusing on those with $\SNRAME>2$) have 
non-negligible differences between the two cases. Bottom: 
The same difference, but now plotted versus $\SNRAME$ 
from FULL case. There is a clear trend, the flux density
deficit being more important for those pixels with higher
 AME significance.}
    \label{fig:difference_at_11GHz_w_wo_MFI}
\end{figure}

\begin{figure}
    \centering
    \includegraphics[width=1\linewidth]{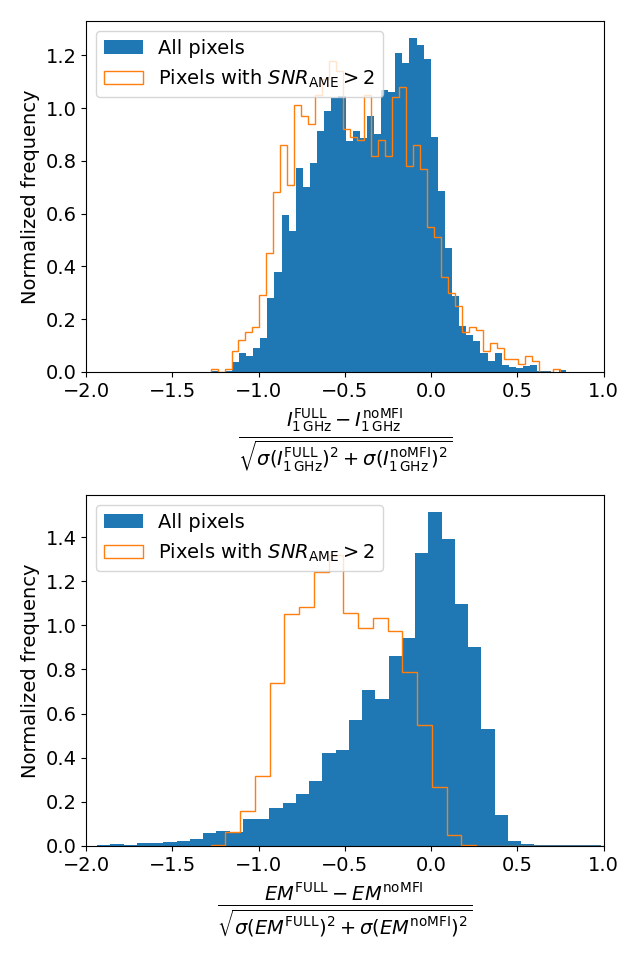}
    \caption{Top: Comparison between FULL and noMFI 
cases for $I_{\rm1\,GHz}$. We have plotted the difference
 between the two estimates over their quadratic
 uncertainty. Bottom: Equivalent histogram for $EM$. The difference
 peaks around zero when studying all the pixels, but it 
is clearly displaced towards lower values (higher free-free 
estimates for the noMFI case) when focusing on pixels with
 high AME significance.}
    \label{fig:beta_syn_distribution_comparison_wo_MFI_points_part1}
\end{figure}

There is an excess for the total flux density from the fits at 
11.2\,GHz in the noMFI case, when compared with the obtained in the FULL one. This difference is higher than 3$\sigma$ ($\sigma$ being 
the uncertainty from 
equation~\ref{eq:sigma_AP_calibration_simulations}
 at 11.2\,GHz in the FULL case) for most of the 
pixels. This is particularly evident when focusing 
on those pixels with $\SNRAME>2$, which are the ones 
primarily studied here
(Fig.~\ref{fig:difference_at_11GHz_w_wo_MFI}). 
Plotting this difference against $\SNRAME$ in the FULL case, we see
a clear trend. The excess is more important for
those pixels with higher AME significance. This
is as expected: a pixel described by synchrotron or 
free-free, or the sum of both, will have a more similar behavior 
between the low (1--10\,GHz) and 
medium (10--100\,GHz) frequency domains, compared 
to one with a rising AME component. A similar 
result can be seen in 
Fig.~\ref{fig:beta_syn_distribution_comparison_wo_MFI_points_part1}, 
where it is clear that the pixels with the highest 
AME significances show the greatest free-free excesses 
in the noMFI case. On the other hand, the synchrotron 
amplitude ($I_{\rm1\,GHz}$) estimates are also higher 
in the noMFI case, but those excesses are similar 
between pixels with high and low AME significances.

This analysis illustrates the importance of having reliable 
data in the 10--20\,GHz region. These data are required to avoid 
overestimating the free-free and/or synchrotron component (due to 
$\alphasyn$ flattening), as will happen when the flux density is higher at 
those frequencies. The increasing flux density towards WMAP and LFI 
bands should then be accounted for by the AME, thus increasing the 
expected importance of this component within the diffuse 
emission. \cite{planck2016Xforegroundmaps} (hereafter, 
\citetalias{planck2016Xforegroundmaps}) partly solved 
this issue primarily by fixing the synchrotron spectral index, thus 
preventing the low frequency foregrounds from accounting for that 
difference.

\subsection{Correlations between the AME amplitude map
and the frequency maps between 0.408\,GHz and 8 micron}
\begin{figure}
    \centering
    \includegraphics[width=1\linewidth]{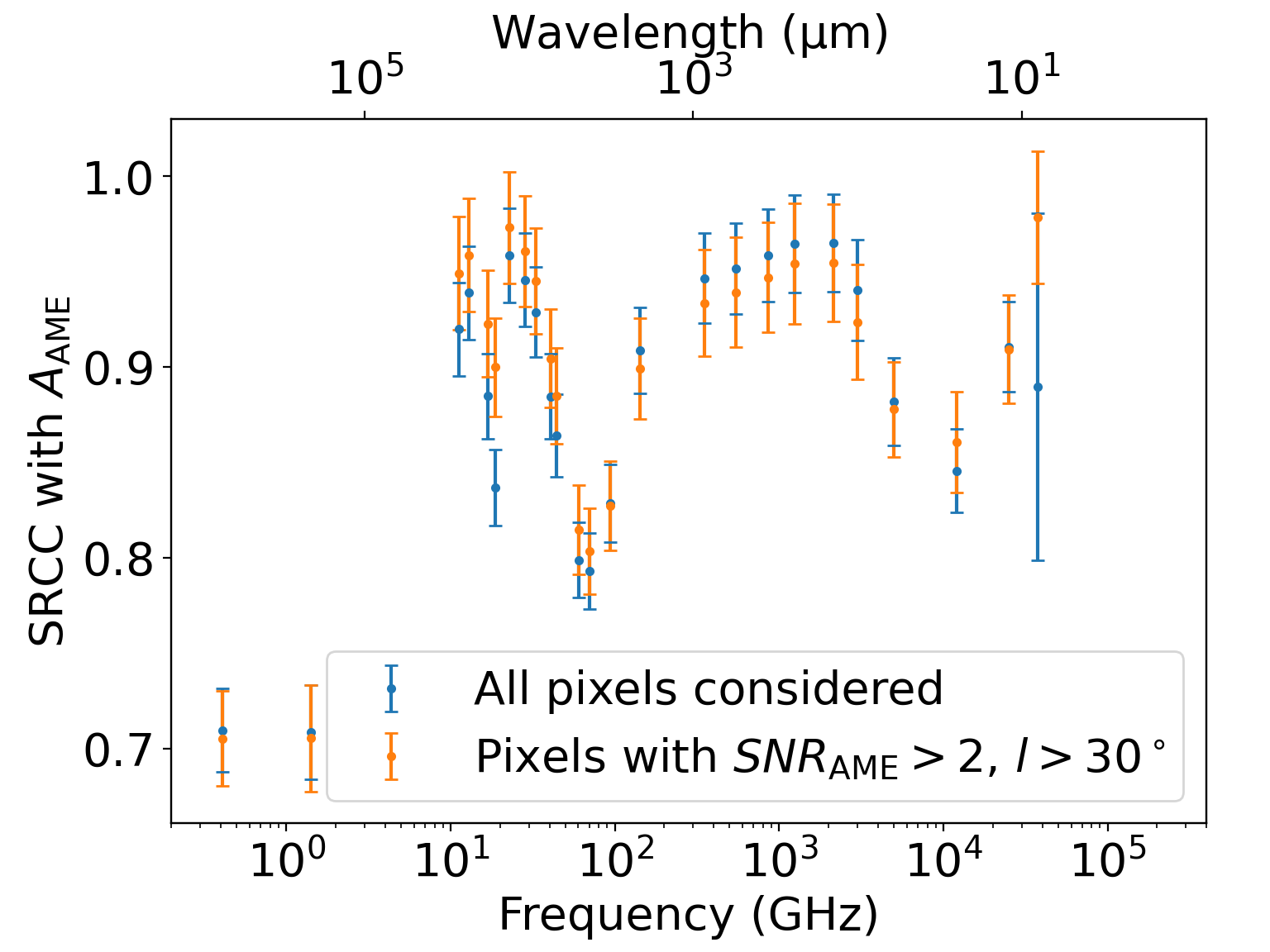}
    \caption{SRCC values between the various
frequency maps and the $\Aame$ map: we 
see that the correlation is high for most of the
 bands. We used only those pixels with high AME 
significance ($\SNRAME>2$) for this figure. We 
are highlighting the difference between studying 
all pixels and only those with $l>30\degr$. 
This difference is especially important for Spitzer 
8$\rm\,\mu m$ case, as pointed out in Figure~\ref{fig:Aame_vs_spitzer8micron}.}
    \label{fig:SRCC_with_individual_frequencies}
\end{figure}

We compared our $\Aame$ map from Fig.~\ref{fig:parameter_maps_1}
with all frequency maps present in Table~\ref{table:maps}.
Besides, we introduced the maps from the COBE-DIRBE 
\citep{cobe-dirbe} with wavelengths shorter than 100\,$\mu$m
that were not used in the SED fitting, down to 12\,$\mu$m.
Data at 8\,$\mu$m from the Spitzer \citep{spitzertelescope,
spitzeriraccamera} satellite was also introduced. For COBE-DIRBE,
we used the released version with zodiacal light subtracted. 
We also considered IRAS maps (Infrared Astronomical Satellite,
~\cite{iras1, iras2_issa}), but the zodiacal 
light emission was still present in those maps, even 
when using IRIS\footnote{Improved Reprocessing of the 
IRAS Survey.} \citep{iris} data. For this reason, we used
COBE-DIRBE data, in which a small residual from the 
zodiacal light is still visible, especially at $12$ 
and $25\rm\,\mu$m. IRAS data is also available 
through the LAMBDA, while for Spitzer we used the 
8\,$\rm\mu$m GLIMPSE \citep{GLIMPSE} data available 
in the Centre d'Analyse de Données Etendues (CADE) 
webpage and already in {\tt HEALPix} format. This last map 
does not cover the full Galactic plane: we have 
avoided pixels closer to $1\degr$ to the non-observed
part of the map, to prevent issues arising from the 
$1\degr$ smoothing and downgrading to {\tt HEALPix} \nside=64. 
The smaller number of available pixels imply greater
uncertainty estimates when computing the correlations
between this Spitzer map and the parameter maps.
This is especially important 
for $90\degr<l<160\degr$ and the anticentre regions.
In Table~\ref{table:SRCC_w_dust_maps} we show the correlations
between the maps from COBE-DIRBE and Spitzer when compared to
the $\Aame$ parameter map obtained in this study.
Figure~\ref{fig:SRCC_with_individual_frequencies} shows 
the correlation values with $\Aame$ for all the frequency maps
listed in Table~\ref{table:maps}.

\begin{figure}
    \centering
    \includegraphics[width=1\linewidth]{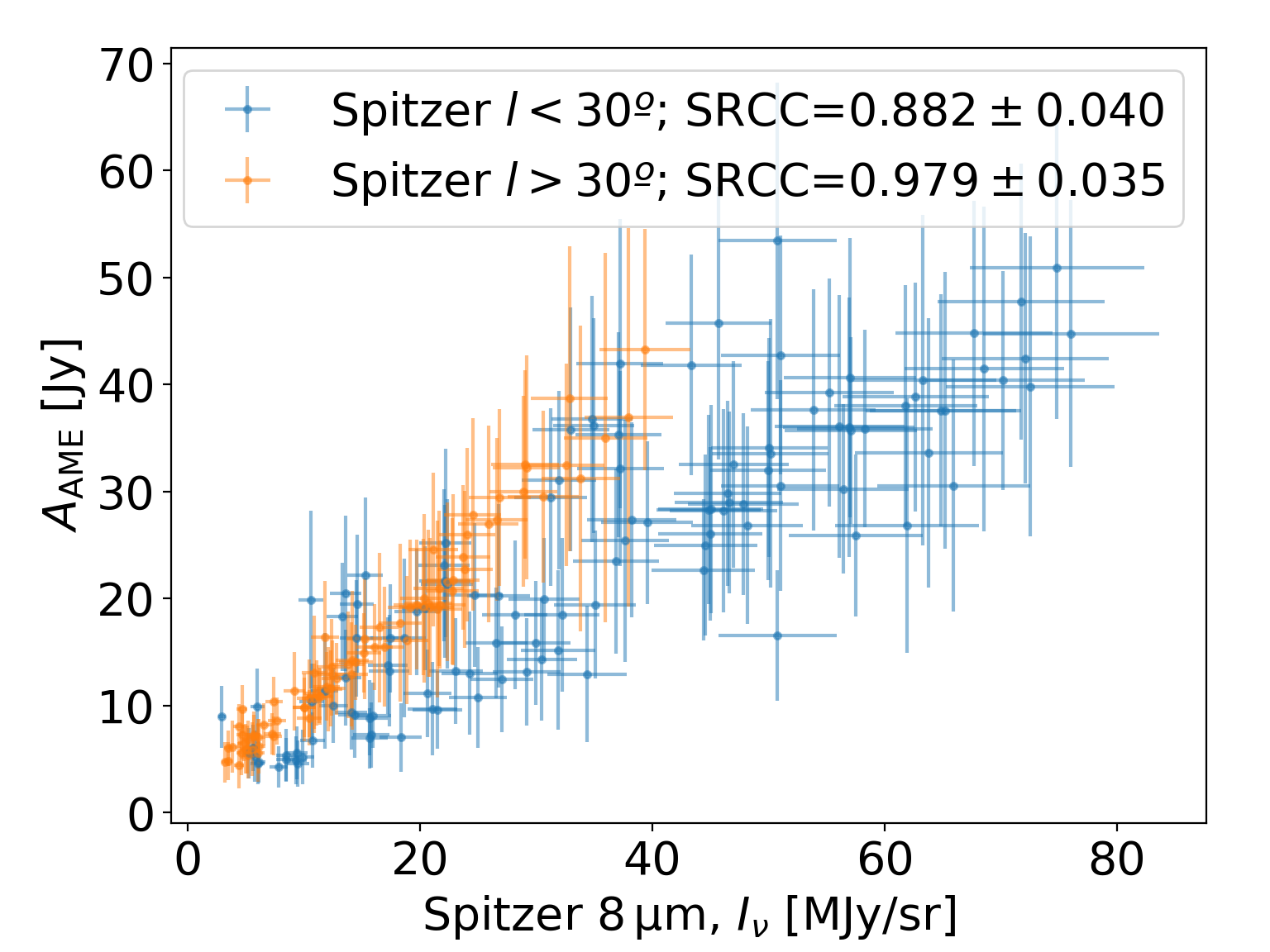}
    \caption{Correlations between AME amplitude and the 
$8\,\mu{\rm m}$ Spitzer map. Most of the data come from 
two regions: the Galactic centre and $l>30\degr$. 
It is clear that the two areas 
behave differently.}
    \label{fig:Aame_vs_spitzer8micron}
\end{figure}

The correlation between the AME amplitude, $\Aame$, 
and the maps is good from the mid- to the far-infrared 
(8--$100\rm\,\mu$m). An example for these correlations 
is shown in Fig.~\ref{fig:Aame_vs_spitzer8micron}, 
for the Spitzer $\rm 8\,\mu m$ map. This case is
particularly remarkable, as all the data comes from just
two positions of the sky: the first covers the 
Galactic centre, while the second focuses on those 
pixels with $l>30\degr$. We see that the 
behaviour is different between the two regions, 
although consistent within the uncertainties. 
As the region
covering the Galactic Centre probably hosts more
heterogeneous environments along its lines-of-sight,
the latter is better for making 
comparisons between surveys. SRCC increases from 
$0.904\pm0.028$, when all pixels are studied together,
 to $0.979\pm0.035$, when studying just the $l>30\degr$
region. On the other hand, we find that the correlation 
factor decreases for the DIRBE 60 and 25$\,\rm\mu m$ bands 
when the $l>30\degr$ limit is applied, down to 
$0.878\pm0.025$ and $0.861\pm0.027$ respectively. 
The values for 60--25\,$\mu{\rm m}$ are 2.3--2.6$\sigma$ 
lower than those recovered for the Spitzer $\rm 8\,\mu m$ band,
and lower than $\sim2\sigma$ than those from e.g. 
\textit{Planck-HFI} 857\,GHz or COBE-DIRBE 240\,$\mu$m. We note
that this high level of correlation between the dust 
and the AME in the full Galactic Plane was obtained 
introducing minimal priors.

\subsubsection{Implications on preferred AME carriers from
correlations between AME and far-infrared surveys}
\label{section:new_carriers_section}
Since the first works proposing theoretical models to
explain the AME as spinning radiation from dust grains \citep{draine1998b}, 
polyciclic aromatic hydrocarbons (PAHs) have been the preferred
proposed carriers for this emission. 
The thermal emission from dust grains with the small size of PAHs 
($a_0\sim0.64$\,nm) show up in the mid-to-near infrared region of the ISM 
spectrum, while their emission due to an hypothetical spinning mechanism
would lie in the microwave, thus possibly accounting for AME.
PAHs spectra present several emission lines and features (both
broad and narrow) between 1 and 12\,$\mu$m (see e.g. Fig.~2 from 
\citealt{compiegne2011}): we are mostly interested on the 7.7\,$\mu$m
emission line, as it is within the Spitzer 
8\,$\mu$m band. This allows us to use this band as a PAH tracer\footnote{
\cite{hensley2022} showed that a lower correlation between
the PAH 11.3\,$\mu$m line -and thus COBE-DIRBE 12\,$\mu$m band- and 
AME could be expected even if PAHs are responsible for AME.}.
Between 30 and 50\,$\mu$m wavelengths generic small amorphous
carbon grains (or very small grains, VSGs) are the main emitting molecules: 
for larger wavelengths, emission from big dust grains 
(BGs, either amorphous carbon or amorphous silicates, with 
$a_0\in(1, 100)$\,nm) dominates. Finally, it is worth noting 
that the average Galactic PAH emission per H atom and normalised by $G_0$
has very little sensitivity to $G_0$. This is because the almost linear 
dependency of PAH emissivity per H atom with $G_0$ is largely cancelled 
out after normalising by $G_0$ (see Fig. 7 of \cite{compiegne2011}, 
where $G_0$ is noted as $U$). MBB emission from large grains, 
on the other hand, shows a more complicated dependence with $G_0$.

Therefore, the higher SRCC between $\Aame$ and 8\,$\mu$m 
compared to $\Aame$ and 24-60$\,\mu$m implies a marginal preference 
for spinning dust from PAHs (or nanocarbons) over other VSGs (or
nanosilicates) as the main carrier. However, the 
correlation between the AME map and the 8\,$\mu$m one is comparable
to those between the AME map and those tracing thermal dust emission
from BGs (e.g. COBE-DIRBE 100\,$\mu$m). Besides, we
explained how PAH emission (and thus, its 7.7$\,\mu$m band) is not
correlated with $G_0$ while the emission from BGs is, as well as the
AME emissivity (Section~\ref{section:correlations_parameters}). 
This, together with
the important correlation between the AME and the BGs emission bands,
could be pointing to BGs being the main AME carrier instead of PAHs 
or VSGs \citep{chuss2022}. This apparent non-PAH origin of AME, which
contradicts many of the first works focused on AME as spinning dust
emission, has been also proposed in recent studies: both \cite{NGC6946AME3} 
and \cite{NGC4125AME} found AME estimates too large in extragalactic
regions to be solely explained by PAHs, while \cite{hensley2016} and
\cite{hensley2017} showed that other carriers -such as silicates, as 
later demonstrated by \cite{ysard2022}- could account for the entirety
of AME with no PAH contribution at all.

However, even the correlation between the AME emissivity and $G_0$ is 
unclear. Although $\Aame/\taud$ and $G_0$ show a strong correlation,
when using other dust tracers to build the AME emissivity the correlation
disappears, as introduced in Section~\ref{section:correlations_parameters}.
For example, when building the AME emissivity as in 
Section~\ref{sect:longitude_variations}, i.e. using the ratio
between the AME intensity at 28.4\,GHz (in temperature units) and the
dust intensity at 100\,$\mu$m, $\emmAME$, we find
that it is slightly anticorrelated with $G_0$, with $SRCC=-0.42\pm0.07$.
We obtain the same result when using the total dust radiance, as defined in
Section~\ref{section:correlations_parameters}, as the dust tracer: 
$\Aame/\mathcal{R}_{\rm dust}$ vs. $G_0$ return $SRCC=-0.43\pm0.09$. These
differences between the correlations with $G_0$ when using different dust 
tracers were already presented in \citetalias{LambdaOrionis} (see H and L
panels on Fig.~6). 

These apparently contradictory findings contribute to the still open 
and unclear situation on AME carriers within the community \citep[e.g.,][]{
reviewclive2017}. Considering them together with the low significant differences between the correlations when comparing $\Aame$ with the different
frequency maps (most are consistent within 1--2$\sigma$), 
it is difficult to provide a strong claim as to what is the
preferred AME carrier. The only differences  greater than 3$\sigma$ on the
$\Aame$ and frequency maps correlations are the ones comparing COBE-DIRBE 
25\,$\mu$m and 60\,$\mu$m bands, which would trace VSGs, with respect to 240 
and 140\,$\mu$m bands, tracing BGs. Even in this case, this significance 
level is only reached when applying the $l>30\degr$ selection.
\citetalias{planck2015galacticcloudsAME} and~\citetalias{AMEwidesurvey}
also found no evident preference for any far-infrared band. On the other
hand, \citetalias{LambdaOrionis} found that the SRCC was as low
as 0.4--0.7 for the 100--$25{\rm\,\mu}$m bands, and then increased
for the mid-infrared bands (12$\rm\,\mu$m and AKARI 
9$\rm\,\mu$m\footnote{Which was used instead of the 
Spitzer 8$\rm\,\mu$m band.} maps, \citealp{akariLOri}). 
This is similar to our results, but we find much less pronounced
differences, partly owing to the wider variety of regions 
studied along the Galactic plane, instead of a single and isolated 
one, such as $\lambda$Orionis. But the main difference
between~\citetalias{LambdaOrionis} and this study is the 
absence of correlation between $\Aame$ and the frequency 
bands below 100\,GHz in the former. This is due to the fact
that in that case for the frequency maps where the AME 
is brighter (between 20 and 30\,GHz) 
there is a lot of free-free
emission in the inner hydrogen shell. This is still much
brighter than the AME in the ring that surrounds the region, 
and that is smearing the correlation. Finally, we 
should mention that~\cite{vidalLDN180} found the highest 
correlation between AME and the FIR bands at 70 $\rm\mu$m when 
studying LDN 1780 on arcminute scales. This was the lowest
correlated band in the \citetalias{LambdaOrionis} analysis 
($60\,\mu$m in that case, which is also the second least correlated 
band in our analysis, after 25\,$\mu$m). Nevertheless, 
this was attributed to LDN 1780 not being in local thermal 
equilibrium, so the comparison is difficult. The difference 
between the two cases could be due to that issue instead 
of the different angular scales, which was shown in~\cite{arcetord2020}.

Because of  the strong correlations between the AME map and 
almost every emission map (SRCC is higher than 0.8 for 
all the surveys between 10\,GHz and 8\,$\mu$m), we believe 
that the full Galactic plane is  too heterogeneous a sample
 to build AME relations with different tracers. Many 
authors have proposed in the past that AME is extremely 
sensitive to the local properties within its environment 
(e.g.~\citealt{hensley2022}). Studying the Galactic plane 
as a whole implies the mixing of really different environments
not only within the lines-of-sight, but also when binning
the different longitude regions. This is extremely important
in this kind of study, but not as much when focusing on the 
high latitude sky \citep[e.g.,][]{cbassSH2022} or resolved regions 
(\citetalias{LambdaOrionis}).

\begin{table*}
    \centering
    \caption{Spearman Rank Correlation Coefficients (SRCC) 
between the AME amplitude ($\Aame$) and several surveys mapping 
the far-infrared sky emission, only for those pixels with 
$\SNRAME>2$. Every band has a high correlation degree with 
the $\Aame$ map. $N_{\rm pix}$ and $N_{\rm pix}$ Spitzer 
account for the number of pixels considered when building 
the correlations with the full-sky maps and Spitzer, 
respectively.}
    \begin{tabular}{lcccccccc}
    \hline
 \multirow{2}*{Map} & \multirow{2}*{All pixels} & Sector 1   & Sector 2   & Sector 3      & Sector 4   & Sector 5   & Sector 6   \\
  & & ($\delta<-10\degr$) & ($\delta>13\degr$) & ($|l|<50\degr$) & ($50\degr\leq l<90\degr$) & ($90\degr\leq l < 160\degr$) & ($160\degr\leq l < 200\degr$) \\
    \hline
 DIRBE 240\,$\rm\mu m$  & 0.96 $\pm$ 0.03 & 0.98 $\pm$ 0.02 & 0.96 $\pm$ 0.03 & 0.98 $\pm$ 0.02 & 0.97 $\pm$ 0.02   & 0.96 $\pm$ 0.04      & 0.88 $\pm$ 0.12       \\
 DIRBE 140\,$\rm\mu m$  & 0.95 $\pm$ 0.03 & 0.98 $\pm$ 0.02 & 0.96 $\pm$ 0.03 & 0.98 $\pm$ 0.02 & 0.98 $\pm$ 0.02   & 0.98 $\pm$ 0.03      & 0.91 $\pm$ 0.12       \\
 DIRBE 100\,$\rm\mu m$  & 0.93 $\pm$ 0.03 & 0.98 $\pm$ 0.02 & 0.93 $\pm$ 0.03 & 0.98 $\pm$ 0.02 & 0.98 $\pm$ 0.02   & 0.96 $\pm$ 0.04      & 0.90 $\pm$ 0.13       \\
 DIRBE 60\,$\rm\mu m$   & 0.89 $\pm$ 0.02 & 0.97 $\pm$ 0.02 & 0.88 $\pm$ 0.03 & 0.97 $\pm$ 0.02 & 0.97 $\pm$ 0.02   & 0.92 $\pm$ 0.03      & 0.85 $\pm$ 0.11       \\
 DIRBE 25\,$\rm\mu m$   & 0.88 $\pm$ 0.02 & 0.97 $\pm$ 0.02 & 0.87 $\pm$ 0.03 & 0.97 $\pm$ 0.02 & 0.96 $\pm$ 0.02   & 0.94 $\pm$ 0.04      & 0.32 $\pm$ 0.06       \\
 DIRBE 12\,$\rm\mu m$   & 0.92 $\pm$ 0.03 & 0.97 $\pm$ 0.02 & 0.92 $\pm$ 0.03 & 0.98 $\pm$ 0.02 & 0.97 $\pm$ 0.02   & 0.93 $\pm$ 0.03      & 0.52 $\pm$ 0.07       \\
 Spitzer 8\,$\rm\mu m$         & 0.90 $\pm$ 0.00 & 0.93 $\pm$ 0.04 & 0.96 $\pm$ 0.05 & 0.86 $\pm$ 0.04 & 0.97 $\pm$ 0.14   & 0.49 $\pm$ 0.21      & ---        \\
\hline
    \end{tabular}
    \label{table:SRCC_w_dust_maps}
\end{table*}

\subsection{Comparison with \citetalias{planck2016Xforegroundmaps}}
\label{section:COMMANDER_comparison}
We next compare our results with those 
obtained using the COMMANDER code in 
\citetalias{planck2016Xforegroundmaps}. We 
chose to compare our results with the COMMANDER ones 
instead of those from the previously 
mentioned studies \citep{planckbetamm, planckancillarydata} 
because the methodology and area to 
be studied are more similar. For example, a break
 in the dust index at 353\,GHz was introduced
 in those two works and was neither used nor 
required in our low-resolution analysis.
 There are crucial differences, though, between
 this study and that of \citetalias{planck2016Xforegroundmaps},
 the most important one being the addition of 
information from the low-frequency 
(below $20$\,GHz) surveys, especially from 
QUIJOTE-MFI, which were not available for 
\citetalias{planck2016Xforegroundmaps}. The
 methodologies also differ between the two studies: 
whereas we are using a single log-normal 
distribution to fit for the AME, \citetalias{planck2016Xforegroundmaps}
used two components, one representing cold neutral medium (CNM) 
and the other representing warm neutral medium (WNM). 
The approach behind these two distributions 
is not phenomenological, as in our case, 
but was instead driven by the physics of the dust. 
For that purpose, they used 
{\tt SPDUST} software\footnote{Which shows $\Wame$ values slightly 
incompatible with those found in this study, 
as discussed in Section~\ref{section:results_spatial_variations}.} 
to model the spinning dust emission 
in the two scenarios. 
Regarding the synchrotron fitting, 
\citetalias{planck2016Xforegroundmaps}
used a spatially constant $\alphasyn$ derived from
a model assuming a certain propagation of cosmic-rays scenario 
through the Galactic magnetic field 
\citep[{\tt GALPROP},][]{orlando2013,planckemissivityestimates}. 
$\alphasyn$ varies with frequency in this model, being 
flatter (higher than $-1$) at lower frequencies and then 
steeper (and almost constant to $-1.1$) at higher 
(above $1$\,GHz) frequencies.
On the other hand, we assume a spatially dependent
(as suggested by the analysis of the QUIJOTE-MFI maps in this
data release, \citealp[e.g.,][]{MFIcompsep_pol, mfiwidesurvey}), but 
frequency invariant, $\alphasyn$ value.
Besides, \citetalias{planck2016Xforegroundmaps} 
applied a Gaussian prior to the dust parameters 
$\betad$ and $\Td$, while we use a flat prior (from 
Table~\ref{table:priors}). The Gaussian $\betad$ prior
was particularly important, with a median and standard
deviation equal to 1.55 and 0.1 respectively; this prior 
is clearly imprinted in the reconstructed dust spectrum, 
as we show below in Fig.~\ref{fig:dust_distributions_w_COMMANDER}) 
when studying the pixel-value distribution. Finally, the 
full intensity signal is fitted in~\citetalias{planck2016Xforegroundmaps}, 
while in our study a background level is subtracted.

Our recovered $\alphasyn$ median value is $-0.94\pm0.10$,
which is flatter than the usual $-1.1$ value used to
model the synchrotron. This is due to the
fact that \cite{haslam1982} and \cite{dwingeloo} data points 
have the lowest calibration uncertainties among all our low 
frequency data, so the two drive the fit at those frequencies. 
We mentioned previously that the {\tt GALPROP} model used by 
\citetalias{planck2016Xforegroundmaps} expects a flatter 
synchrotron component at those frequencies below $1$\,GHz. 
Therefore, our synchrotron estimates at higher frequencies 
are larger than those from 
\citetalias{planck2016Xforegroundmaps}. This causes free-free 
estimates to be lower, due to the important degeneracy between 
the two components.

\begin{figure*}
    \centering
    \includegraphics[width=1\linewidth]{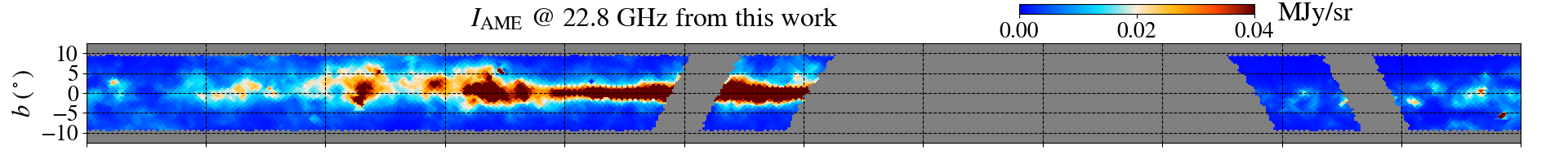}
    \includegraphics[width=1\linewidth]{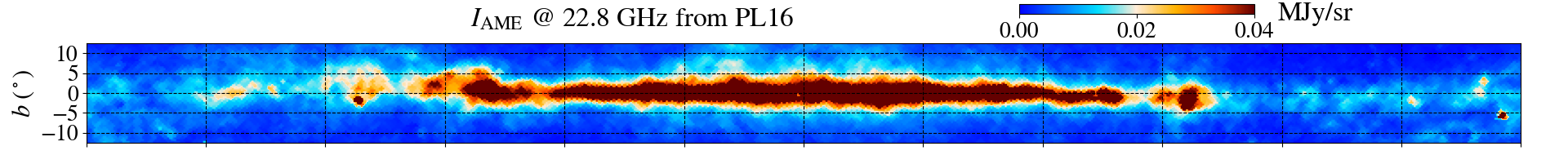}
    \includegraphics[width=1\linewidth]{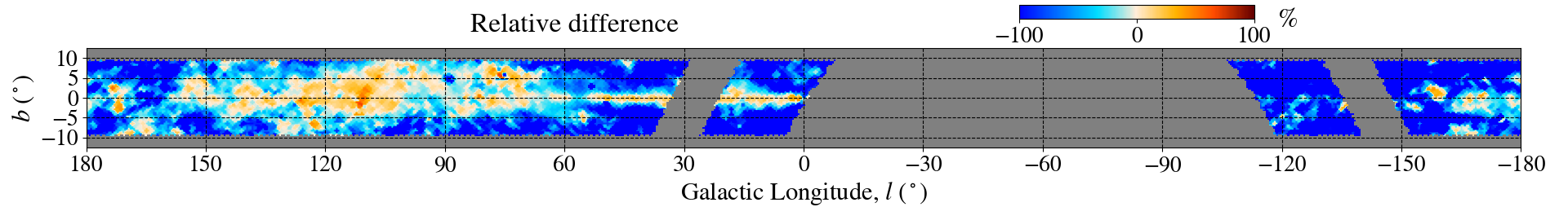}
    \caption{Comparison maps between our study and 
\citetalias{planck2016Xforegroundmaps}
for the AME. The first two are the AME 
intensity at 22.8\,GHz (from this study and 
\protect\citetalias{planck2016Xforegroundmaps}
respectively): we chose to plot the difference at a fixed, 
representative frequency owing to the difference between the 
two methodologies used to model the AME. The third one is the 
relative difference between the two, computed as the estimate 
from this study minus the one from 
\protect\citetalias{planck2016Xforegroundmaps}, divided by the 
former.}
    \label{fig:comparison_w_COMMANDER_amplitudes_AME}
\end{figure*}

\begin{figure}
    \centering
    \includegraphics[width=0.99\linewidth]{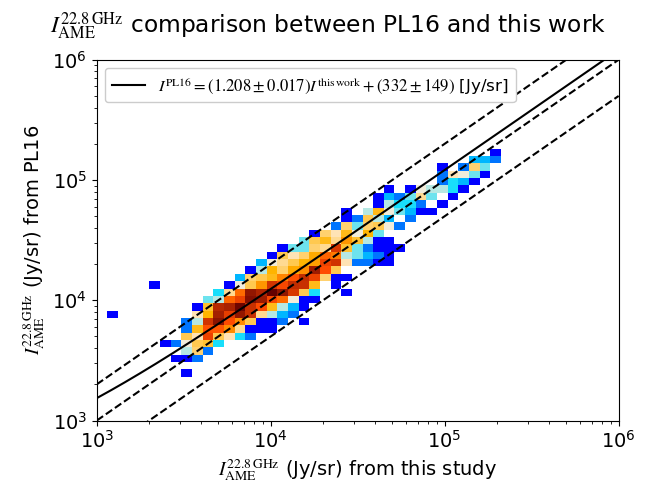}
    \caption{Correlation plot for the AME intensity estimates at 22.8\,GHz from this work and \protect\citetalias{planck2016Xforegroundmaps}. Only those pixels with $\SNRAME>2$ in this study are plotted. Dashed lines mark the 0.5, 1 and 2 comparison levels. The best linear fit is also provided and shown as a solid line. The slope is slightly higher than unity due to the excess of higher $\Iame$ values from \protect\citetalias{planck2016Xforegroundmaps} at low values ($0.5$--$1.0\times10^4$\,Jy/sr).}
    \label{fig:rtgs_asked_fig}
\end{figure}

However, when comparing the AME intensity estimates from both studies, we 
find that differences  are small for those pixels detected with 
good SNR (Fig.~\ref{fig:comparison_w_COMMANDER_amplitudes_AME}). 
Our estimates are lower as we get farther away from the 
plane. The map showing the relative difference
resembles the map tracking the AME fraction in both studies. 
This suggests that the AME amplitude reconstruction is stable in 
those pixels where the component is important. In 
Fig.~\ref{fig:rtgs_asked_fig} we show the correlation 
plot between the two $\Iame$ estimates from both studies.
A slight excess from \citetalias{planck2016Xforegroundmaps} 
is visible for those pixels with low $\Iame$ (probably those
pixels farther from the Plane).

Finally, the addition of the three frequency points from COBE-DIRBE, 
together with the previously mentioned differences with 
\citetalias{planck2016Xforegroundmaps} on the priors,
produces a shift in the dust parameters values. 
We find slightly lower (higher) values for 
$\betad$ ($\Td$), since the two parameters are known to be anticorrelated: 
both kinds of behaviour are shown in 
Fig.~\ref{fig:dust_distributions_w_COMMANDER}. 
We can also see in Fig.~\ref{fig:latitude_variations} 
how the variation of $\Td$ with latitude is similar, apart from a 
certain offset, for both this study and 
\citetalias{planck2016Xforegroundmaps}, while $\betad$ from the latter 
remains fairly constant. These differences between the two 
methodologies (in particular in the applied priors) make a quantitative 
comparison difficult.

\begin{figure}
    \centering
    \includegraphics[width=0.99\linewidth]{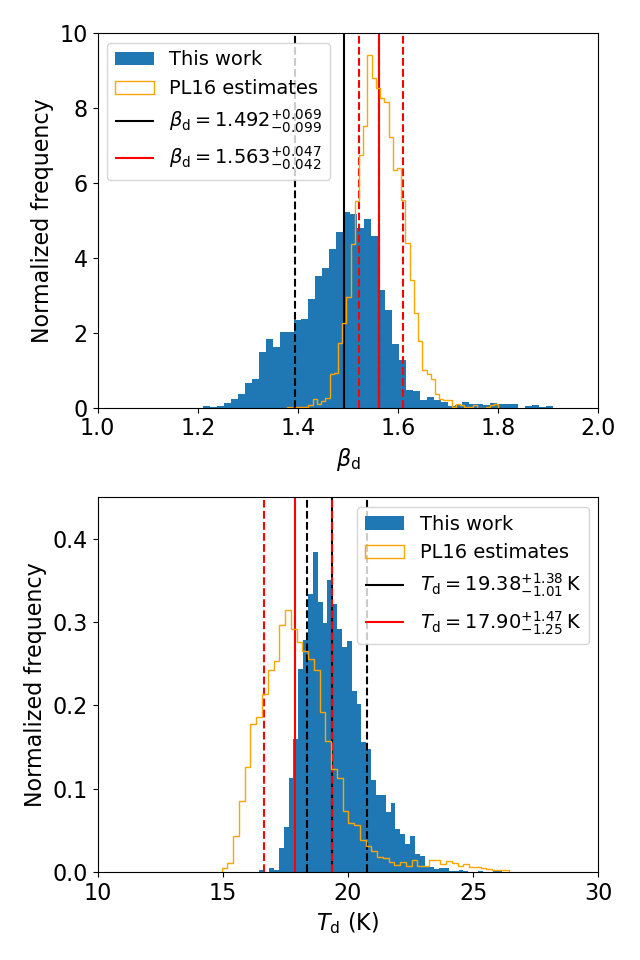}
    \caption{Comparison of the dust parameter distribution 
between \citetalias{planck2016Xforegroundmaps} and this study. We see that $\betad$ (top) 
is displaced towards lower values in comparison with  
\protect\citetalias{planck2016Xforegroundmaps}, while the behaviour of $\Td$ (bottom) is 
the opposite. In the first case, \citetalias{planck2016Xforegroundmaps} distribution 
seems to follow the Gaussian prior $N(1.55\pm0.1)$ from 
Table 4 in \protect\citetalias{planck2016Xforegroundmaps}, 
while for the second the results are far from their 
prior centre, $N(23\pm3)$ K.}
    \label{fig:dust_distributions_w_COMMANDER}
\end{figure}

\section{Conclusions}
\label{section:conclusions}
We have presented a set of ten maps for the parameters 
describing the various diffuse microwave emission 
components (synchrotron, free-free, anomalous microwave 
emission (AME), thermal dust and CMB anisotropies) 
along the Galactic plane ($|b|<10\degr$) at 1 degree 
angular scales. For this purpose, we 
introduced new data from the QUIJOTE-MFI Wide Survey
between 10 and 20\,GHz, and used a fitting methodology 
assuming minimal priors, thus removing any possible biases.
This is one of the first works to show spatial variations
in the synchrotron index along the Galactic plane
in intensity using WMAP and \textit{Planck} data. 
Spatial variations for the AME spectral parameters
are also hinted, but with reduced statistical significance.
However, the obtained median values disagree with those from
theoretical models, pointing to lower (higher) $\nuame$ ($\Wame$) 
values than first expected. We have also shown how having reliable 
data between 10 and 20\,GHz is mandatory to avoid overestimating 
synchrotron against free-free and AME.

Future studies
should focus on 
improving results through the addition of more 
frequency data. This is especially important at lower 
frequencies, where the foregrounds are heavily degenerate. 
Introducing C-BASS data at 5\,GHz
 \citep{cbass, cbassjones2018} will further improve
 the separation between synchrotron and free-free,
 and consequently AME. Repeating the analysis 
in polarisation would also be interesting, as the 
degeneracies are not expected to be as important
 (free-free and AME are negligibly polarized compared to 
synchrotron and thermal dust). However, additional work
would be required to properly correct for possible issues of
depolarisation and Faraday rotation. Low
frequency data in polarisation are much also scarcer than 
intensity data. 
Finally, repeating this analysis at higher angular 
resolutions, aiming for arcminute scales, would be 
interesting. Previous studies \citep{arcetord2020} have hinted 
that the AME relation with thermal dust (and therefore
 mid- to far-infrared surveys) could change with 
higher  angular resolution observations. However, studying the entire
 Galactic plane at high angular resolution is probably
 unrealistic, so the focus should go into smaller, resolved
 and isolated regions such as $\lambda$Orionis (\citetalias{LambdaOrionis}).

\section*{Acknowledgements}
We thank the staff of the Teide Observatory for invaluable assistance in the commissioning and operation of QUIJOTE.
The {\it QUIJOTE} experiment is being developed by the Instituto de Astrofisica de Canarias (IAC),
the Instituto de Fisica de Cantabria (IFCA), and the Universities of Cantabria, Manchester and Cambridge.
Partial financial support was provided by the Spanish Ministry of Science and Innovation 
under the projects AYA2007-68058-C03-01, AYA2007-68058-C03-02, AYA2010-21766-C03-01, AYA2010-21766-C03-02,
AYA2014-60438-P,  ESP2015-70646-C2-1-R, AYA2017-84185-P, ESP2017-83921-C2-1-R, PID2019-110610RB-C21, PID2020-120514GB-I00,
IACA13-3E-2336, IACA15-BE-3707, EQC2018-004918-P, the Severo Ochoa Programs SEV-2015-0548 and CEX2019-000920-S, the
Maria de Maeztu Program MDM-2017-0765 and by the Consolider-Ingenio project CSD2010-00064 (EPI: Exploring
the Physics of Inflation).
We acknowledge support from the ACIISI, Consejeria de Economia, Conocimiento y Empleo del Gobierno de Canarias and the European Regional Development Fund (ERDF) under grant with reference ProID2020010108.
This project has received funding from the European Union's Horizon 2020 research and innovation program under
grant agreement number 687312 (RADIOFOREGROUNDS).

We thank the anonymous referee whose comments helped 
to improve this work. We also thank Bruce Draine, Brandon
Hensley and Enrique Fernández Cancio for their useful comments.
MFT acknowledges support from from the 
Agencia Estatal de Investigación (AEI) of the Ministerio de 
Ciencia, Innovación y Universidades (MCIU) and the European 
Social Fund (ESF) under grant with reference PRE-C-2018-0067.
SEH acknowledges support from
the STFC Consolidated Grant (ST/P000649/1). FP acknowledges 
support from the Spanish State Research Agency (AEI) 
under grant number PID2019-105552RB-C43 and support from the 
Agencia Canaria de Investigación, Innovación y Sociedad de la 
Información (ACIISI) under the European FEDER (FONDO EUROPEO 
DE DESARROLLO REGIONAL) de Canarias 2014-2020 grant No. 
PROID2021010078. 
This paper made use of the IAC Supercomputing facility 
HTCondor (http://research.cs.wisc.edu/htcondor/), partly 
financed by the Ministry of Economy and Competitiveness with
FEDER funds, code IACA13-3E-2493. 
We acknowledge the use of data 
provided by the Centre d'Analyse de Données Etendues 
(CADE), a service of IRAP-UPS/CNRS (http://cade.irap.omp.eu,~\cite{CADE}).
This research has made use of the SIMBAD database,
operated at CDS, Strasbourg, France~\citep{simbad}. This work has made
use of S-band Polarisation All Sky Survey (S-PASS) data.
Some of the results in this paper have been derived using 
the healpy and {\tt HEALPix} packages~\citep{Healpix, Healpix2}.
We have also used {\tt scipy}~\citep{scipy}, 
{\tt emcee}~\citep{emcee}, {\tt numpy}~\citep{numpy},
 {\tt matplotlib}~\citep{matplotlib}, {\tt corner}~\citep{corner} 
and {\tt astropy}~\citep{astropy1, astropy2} \textsc{python} packages.

\section*{Data availability}
The set of 10 parameter maps described in Section~\ref{section:Methodology}
for the default case (with QUIJOTE-MFI data considered) can be 
downloaded from the QUIJOTE web 
page\footnote{\url{http://research.iac.es/proyecto/cmb/quijote}.}, 
as well as from the
RADIOFOREGROUNDS\footnote{\url{http://www.radioforegrounds.eu/}.} 
platform. An Explanatory Supplement describing the data format
is also available. Any other derived data products described in this paper are available upon request to the QUIJOTE collaboration.



\bibliographystyle{mnras}
\bibliography{quijote, diffuse}

\begin{thebibliography}{}
\makeatletter
\relax
\def\mn@urlcharsother{\let\do\@makeother \do\$\do\&\do\#\do\^\do\_\do\%\do\~}
\def\mn@doi{\begingroup\mn@urlcharsother \@ifnextchar [ {\mn@doi@}
  {\mn@doi@[]}}
\def\mn@doi@[#1]#2{\def\@tempa{#1}\ifx\@tempa\@empty \href
  {http://dx.doi.org/#2} {doi:#2}\else \href {http://dx.doi.org/#2} {#1}\fi
  \endgroup}
\def\mn@eprint#1#2{\mn@eprint@#1:#2::\@nil}
\def\mn@eprint@arXiv#1{\href {http://arxiv.org/abs/#1} {{\tt arXiv:#1}}}
\def\mn@eprint@dblp#1{\href {http://dblp.uni-trier.de/rec/bibtex/#1.xml}
  {dblp:#1}}
\def\mn@eprint@#1:#2:#3:#4\@nil{\def\@tempa {#1}\def\@tempb {#2}\def\@tempc
  {#3}\ifx \@tempc \@empty \let \@tempc \@tempb \let \@tempb \@tempa \fi \ifx
  \@tempb \@empty \def\@tempb {arXiv}\fi \@ifundefined
  {mn@eprint@\@tempb}{\@tempb:\@tempc}{\expandafter \expandafter \csname
  mn@eprint@\@tempb\endcsname \expandafter{\@tempc}}}

\bibitem[\protect\citeauthoryear{{Ali-Ha{\"\i}moud}}{{Ali-Ha{\"\i}moud}}{2013}]{ali-hamoud_review}
{Ali-Ha{\"\i}moud} Y.,  2013, \mn@doi [Advances in Astronomy]
  {10.1155/2013/462697}, \href
  {https://ui.adsabs.harvard.edu/abs/2013AdAst2013E...2A} {2013, 462697}

\bibitem[\protect\citeauthoryear{{Ali-Ha{\"\i}moud}, {Hirata}  \&
  {Dickinson}}{{Ali-Ha{\"\i}moud} et~al.}{2009}]{spdust1}
{Ali-Ha{\"\i}moud} Y.,  {Hirata} C.~M.,   {Dickinson} C.,  2009, \mn@doi
  [\mnras] {10.1111/j.1365-2966.2009.14599.x}, \href
  {https://ui.adsabs.harvard.edu/abs/2009MNRAS.395.1055A} {395, 1055}

\bibitem[\protect\citeauthoryear{{Alves}, {Davies}, {Dickinson}, {Davis},
  {Auld}, {Calabretta}  \& {Staveley-Smith}}{{Alves} et~al.}{2010}]{RRLs}
{Alves} M. I.~R.,  {Davies} R.~D.,  {Dickinson} C.,  {Davis} R.~J.,  {Auld}
  R.~R.,  {Calabretta} M.,   {Staveley-Smith} L.,  2010, \mn@doi [\mnras]
  {10.1111/j.1365-2966.2010.16595.x}, \href
  {https://ui.adsabs.harvard.edu/abs/2010MNRAS.405.1654A} {405, 1654}

\bibitem[\protect\citeauthoryear{{Andersen} et~al.,}{{Andersen}
  et~al.}{2023}]{beyondplanckintensity}
{Andersen} K.~J.,  et~al., 2023, \mn@doi [\aap] {10.1051/0004-6361/202243186},
  \href {https://ui.adsabs.harvard.edu/abs/2023A&A...675A..13A} {675, A13}

\bibitem[\protect\citeauthoryear{{Arce-Tord} et~al.,}{{Arce-Tord}
  et~al.}{2020}]{arcetord2020}
{Arce-Tord} C.,  et~al., 2020, \mn@doi [\mnras] {10.1093/mnras/staa1422}, \href
  {https://ui.adsabs.harvard.edu/abs/2020MNRAS.495.3482A} {495, 3482}

\bibitem[\protect\citeauthoryear{{Astropy Collaboration} et~al.,}{{Astropy
  Collaboration} et~al.}{2013}]{astropy1}
{Astropy Collaboration} et~al., 2013, \mn@doi [\aap]
  {10.1051/0004-6361/201322068}, \href
  {http://adsabs.harvard.edu/abs/2013A%26A...558A..33A} {558, A33}

\bibitem[\protect\citeauthoryear{{Astropy Collaboration} et~al.,}{{Astropy
  Collaboration} et~al.}{2018}]{astropy2}
{Astropy Collaboration} et~al., 2018, \mn@doi [\aj] {10.3847/1538-3881/aabc4f},
  \href {https://ui.adsabs.harvard.edu/abs/2018AJ....156..123A} {156, 123}

\bibitem[\protect\citeauthoryear{{Battistelli} et~al.,}{{Battistelli}
  et~al.}{2015}]{battistellispvariations}
{Battistelli} E.~S.,  et~al., 2015, \mn@doi [\apj]
  {10.1088/0004-637X/801/2/111}, \href
  {https://ui.adsabs.harvard.edu/abs/2015ApJ...801..111B} {801, 111}

\bibitem[\protect\citeauthoryear{{Battistelli} et~al.,}{{Battistelli}
  et~al.}{2019}]{m31AME}
{Battistelli} E.~S.,  et~al., 2019, \mn@doi [\apjl] {10.3847/2041-8213/ab21de},
  \href {https://ui.adsabs.harvard.edu/abs/2019ApJ...877L..31B} {877, L31}

\bibitem[\protect\citeauthoryear{Bell, Onaka, Galliano, Wu, Doi, Kaneda,
  Ishihara  \& Giard}{Bell et~al.}{2019}]{akariLOri}
Bell A.~C.,  Onaka T.,  Galliano F.,  Wu R.,  Doi Y.,  Kaneda H.,  Ishihara D.,
    Giard M.,  2019, Publications of the Astronomical Society of Japan, 71, 123

\bibitem[\protect\citeauthoryear{{Bennett} et~al.,}{{Bennett}
  et~al.}{2013}]{wmap}
{Bennett} C.~L.,  et~al., 2013, \mn@doi [\apjs] {10.1088/0067-0049/208/2/20},
  \href {https://ui.adsabs.harvard.edu/abs/2013ApJS..208...20B} {208, 20}

\bibitem[\protect\citeauthoryear{{Berkhuijsen}}{{Berkhuijsen}}{1972}]{dwingeloo}
{Berkhuijsen} E.~M.,  1972, \aaps, \href
  {https://ui.adsabs.harvard.edu/abs/1972A&AS....5..263B} {5, 263}

\bibitem[\protect\citeauthoryear{{Bianchi} et~al.,}{{Bianchi}
  et~al.}{2022}]{SRTsearches}
{Bianchi} S.,  et~al., 2022, \mn@doi [\aap] {10.1051/0004-6361/202142684},
  \href {https://ui.adsabs.harvard.edu/abs/2022A&A...658L...8B} {658, L8}

\bibitem[\protect\citeauthoryear{{Boggess} et~al.,}{{Boggess}
  et~al.}{1992}]{cobe}
{Boggess} N.~W.,  et~al., 1992, \mn@doi [\apj] {10.1086/171797}, \href
  {https://ui.adsabs.harvard.edu/abs/1992ApJ...397..420B} {397, 420}

\bibitem[\protect\citeauthoryear{{Carretti} et~al.,}{{Carretti}
  et~al.}{2019}]{carretti2019}
{Carretti} E.,  et~al., 2019, \mn@doi [\mnras] {10.1093/mnras/stz806}, \href
  {https://ui.adsabs.harvard.edu/abs/2019MNRAS.489.2330C} {489, 2330}

\bibitem[\protect\citeauthoryear{{Casassus}, {Vidal}, {Arce-Tord}, {Dickinson},
  {White}, {Burton}, {Indermuehle}  \& {Hensley}}{{Casassus}
  et~al.}{2021}]{rhoOphiuchi_arcmin}
{Casassus} S.,  {Vidal} M.,  {Arce-Tord} C.,  {Dickinson} C.,  {White} G.~J.,
  {Burton} M.,  {Indermuehle} B.,   {Hensley} B.,  2021, \mn@doi [\mnras]
  {10.1093/mnras/staa4016}, \href
  {https://ui.adsabs.harvard.edu/abs/2021MNRAS.502..589C} {502, 589}

\bibitem[\protect\citeauthoryear{{Cepeda-Arroita} et~al.,}{{Cepeda-Arroita}
  et~al.}{2021}]{LambdaOrionis}
{Cepeda-Arroita} R.,  et~al., 2021, \mn@doi [\mnras] {10.1093/mnras/stab583},
  \href {https://ui.adsabs.harvard.edu/abs/2021MNRAS.503.2927C} {503, 2927}

\bibitem[\protect\citeauthoryear{{Churchwell} et~al.,}{{Churchwell}
  et~al.}{2009}]{GLIMPSE}
{Churchwell} E.,  et~al., 2009, \mn@doi [\pasp] {10.1086/597811}, \href
  {https://ui.adsabs.harvard.edu/abs/2009PASP..121..213C} {121, 213}

\bibitem[\protect\citeauthoryear{{Chuss}, {Hensley}, {Kogut}, {Guerra}, {Nofi}
  \& {Siah}}{{Chuss} et~al.}{2022}]{chuss2022}
{Chuss} D.~T.,  {Hensley} B.~S.,  {Kogut} A.~J.,  {Guerra} J.~A.,  {Nofi}
  H.~C.,   {Siah} J.,  2022, arXiv e-prints, \href
  {https://ui.adsabs.harvard.edu/abs/2022arXiv220809049C} {p. arXiv:2208.09049}

\bibitem[\protect\citeauthoryear{{Compi{\`e}gne} et~al.,}{{Compi{\`e}gne}
  et~al.}{2011}]{compiegne2011}
{Compi{\`e}gne} M.,  et~al., 2011, \mn@doi [\aap]
  {10.1051/0004-6361/201015292}, \href
  {https://ui.adsabs.harvard.edu/abs/2011A&A...525A.103C} {525, A103}

\bibitem[\protect\citeauthoryear{{Condon} \& {Ransom}}{{Condon} \&
  {Ransom}}{2016}]{nraobook}
{Condon} J.~J.,  {Ransom} S.~M.,  2016, {Essential Radio Astronomy}

\bibitem[\protect\citeauthoryear{{Curran}}{{Curran}}{2014}]{SRCCmcmc}
{Curran} P.~A.,  2014, arXiv e-prints, \href
  {https://ui.adsabs.harvard.edu/abs/2014arXiv1411.3816C} {p. arXiv:1411.3816}

\bibitem[\protect\citeauthoryear{{Davies}, {Dickinson}, {Banday}, {Jaffe},
  {G{\'o}rski}  \& {Davis}}{{Davies} et~al.}{2006}]{davies2006}
{Davies} R.~D.,  {Dickinson} C.,  {Banday} A.~J.,  {Jaffe} T.~R.,  {G{\'o}rski}
  K.~M.,   {Davis} R.~J.,  2006, \mn@doi [\mnras]
  {10.1111/j.1365-2966.2006.10572.x}, \href
  {https://ui.adsabs.harvard.edu/abs/2006MNRAS.370.1125D} {370, 1125}

\bibitem[\protect\citeauthoryear{{\VAN{De Oliveira-Costa}{de }{de
  }{Oliveira-Costa}}, {Kogut}, {Devlin}, {Netterfield}, {Page}  \&
  {Wollack}}{{\VAN{De Oliveira-Costa}{de }{de }{Oliveira-Costa}}
  et~al.}{1997}]{firstcorrelationwdust}
{\VAN{De Oliveira-Costa}{de }{de }{Oliveira-Costa}} A.,  {Kogut} A.,  {Devlin}
  M.~J.,  {Netterfield} C.~B.,  {Page} L.~A.,   {Wollack} E.~J.,  1997, \mn@doi
  [\apjl] {10.1086/310684}, \href
  {https://ui.adsabs.harvard.edu/abs/1997ApJ...482L..17D} {482, L17}

\bibitem[\protect\citeauthoryear{{\VAN{De Oliveira-Costa}{de }{de
  }{Oliveira-Costa}}, {Tegmark}, {Guti{\'e}rrez}, {Jones}, {Davies}, {Lasenby},
  {Rebolo}  \& {Watson}}{{\VAN{De Oliveira-Costa}{de }{de }{Oliveira-Costa}}
  et~al.}{1999}]{deOliveira1999}
{\VAN{De Oliveira-Costa}{de }{de }{Oliveira-Costa}} A.,  {Tegmark} M.,
  {Guti{\'e}rrez} C.~M.,  {Jones} A.~W.,  {Davies} R.~D.,  {Lasenby} A.~N.,
  {Rebolo} R.,   {Watson} R.~A.,  1999, \mn@doi [\apjl] {10.1086/312384}, \href
  {https://ui.adsabs.harvard.edu/abs/1999ApJ...527L...9D} {527, L9}

\bibitem[\protect\citeauthoryear{{\MakeLowercase{D}e la Hoz}
  et~al.,}{{\MakeLowercase{D}e la Hoz} et~al.}{2023}]{MFIcompsep_pol}
{\MakeLowercase{D}e la Hoz} E.,  et~al., 2023, \mn@doi [\mnras]
  {10.1093/mnras/stac3020}, \href
  {https://ui.adsabs.harvard.edu/abs/2023MNRAS.519.3504D} {519, 3504}

\bibitem[\protect\citeauthoryear{{Dickinson} et~al.,}{{Dickinson}
  et~al.}{2009}]{RCW175}
{Dickinson} C.,  et~al., 2009, \mn@doi [\apj] {10.1088/0004-637X/690/2/1585},
  \href {https://ui.adsabs.harvard.edu/abs/2009ApJ...690.1585D} {690, 1585}

\bibitem[\protect\citeauthoryear{{Dickinson} et~al.,}{{Dickinson}
  et~al.}{2010}]{dickinsonorionspvariations}
{Dickinson} C.,  et~al., 2010, \mn@doi [\mnras]
  {10.1111/j.1365-2966.2010.17079.x}, \href
  {https://ui.adsabs.harvard.edu/abs/2010MNRAS.407.2223D} {407, 2223}

\bibitem[\protect\citeauthoryear{{Dickinson} et~al.,}{{Dickinson}
  et~al.}{2018}]{reviewclive2017}
{Dickinson} C.,  et~al., 2018, \mn@doi [\nar] {10.1016/j.newar.2018.02.001},
  \href {https://ui.adsabs.harvard.edu/abs/2018NewAR..80....1D} {80, 1}

\bibitem[\protect\citeauthoryear{{Dong} \& {Draine}}{{Dong} \&
  {Draine}}{2011}]{dongdraine2011fromhensley2021}
{Dong} R.,  {Draine} B.~T.,  2011, \mn@doi [\apj] {10.1088/0004-637X/727/1/35},
  \href {https://ui.adsabs.harvard.edu/abs/2011ApJ...727...35D} {727, 35}

\bibitem[\protect\citeauthoryear{{Draine}}{{Draine}}{2011}]{draine2011}
{Draine} B.~T.,  2011, {Physics of the Interstellar and Intergalactic Medium}

\bibitem[\protect\citeauthoryear{{Draine} \& {Hensley}}{{Draine} \&
  {Hensley}}{2013}]{drainehensley2013magneticpolarization}
{Draine} B.~T.,  {Hensley} B.,  2013, \mn@doi [\apj]
  {10.1088/0004-637X/765/2/159}, \href
  {https://ui.adsabs.harvard.edu/abs/2013ApJ...765..159D} {765, 159}

\bibitem[\protect\citeauthoryear{{Draine} \& {Lazarian}}{{Draine} \&
  {Lazarian}}{1998a}]{draine1998a}
{Draine} B.~T.,  {Lazarian} A.,  1998a, \mn@doi [\apjl] {10.1086/311167}, \href
  {https://ui.adsabs.harvard.edu/abs/1998ApJ...494L..19D} {494, L19}

\bibitem[\protect\citeauthoryear{{Draine} \& {Lazarian}}{{Draine} \&
  {Lazarian}}{1998b}]{draine1998b}
{Draine} B.~T.,  {Lazarian} A.,  1998b, \mn@doi [\apj] {10.1086/306387}, \href
  {https://ui.adsabs.harvard.edu/abs/1998ApJ...508..157D} {508, 157}

\bibitem[\protect\citeauthoryear{{Draine} \& {Lazarian}}{{Draine} \&
  {Lazarian}}{1999}]{draine1999magnetic}
{Draine} B.~T.,  {Lazarian} A.,  1999, \mn@doi [\apj] {10.1086/306809}, \href
  {https://ui.adsabs.harvard.edu/abs/1999ApJ...512..740D} {512, 740}

\bibitem[\protect\citeauthoryear{{Draine} \& {Li}}{{Draine} \&
  {Li}}{2007}]{pah2}
{Draine} B.~T.,  {Li} A.,  2007, \mn@doi [\apj] {10.1086/511055}, \href
  {https://ui.adsabs.harvard.edu/abs/2007ApJ...657..810D} {657, 810}

\bibitem[\protect\citeauthoryear{{Erickson}}{{Erickson}}{1957}]{historicalAME}
{Erickson} W.~C.,  1957, \mn@doi [\apj] {10.1086/146421}, \href
  {https://ui.adsabs.harvard.edu/abs/1957ApJ...126..480E} {126, 480}

\bibitem[\protect\citeauthoryear{{Eriksen}, {Jewell}, {Dickinson}, {Banday},
  {G{\'o}rski}  \& {Lawrence}}{{Eriksen} et~al.}{2008}]{eriksen2008}
{Eriksen} H.~K.,  {Jewell} J.~B.,  {Dickinson} C.,  {Banday} A.~J.,
  {G{\'o}rski} K.~M.,   {Lawrence} C.~R.,  2008, \mn@doi [\apj]
  {10.1086/525277}, \href
  {https://ui.adsabs.harvard.edu/abs/2008ApJ...676...10E} {676, 10}

\bibitem[\protect\citeauthoryear{{Fazio} et~al.,}{{Fazio}
  et~al.}{2004}]{spitzeriraccamera}
{Fazio} G.~G.,  et~al., 2004, \mn@doi [\apjs] {10.1086/422843}, \href
  {https://ui.adsabs.harvard.edu/abs/2004ApJS..154...10F} {154, 10}

\bibitem[\protect\citeauthoryear{{Finkbeiner}}{{Finkbeiner}}{2003}]{finkbeinerHalpha}
{Finkbeiner} D.~P.,  2003, \mn@doi [\apjs] {10.1086/374411}, \href
  {https://ui.adsabs.harvard.edu/abs/2003ApJS..146..407F} {146, 407}

\bibitem[\protect\citeauthoryear{{Fixsen}}{{Fixsen}}{2009}]{fixsen2009}
{Fixsen} D.~J.,  2009, \mn@doi [\apj] {10.1088/0004-637X/707/2/916}, \href
  {https://ui.adsabs.harvard.edu/abs/2009ApJ...707..916F} {707, 916}

\bibitem[\protect\citeauthoryear{Foreman-Mackey}{Foreman-Mackey}{2016}]{corner}
Foreman-Mackey D.,  2016, \mn@doi [Journal of Open Source Software]
  {10.21105/joss.00024}, 1, 24

\bibitem[\protect\citeauthoryear{{Foreman-Mackey}, {Hogg}, {Lang}  \&
  {Goodman}}{{Foreman-Mackey} et~al.}{2013}]{emcee}
{Foreman-Mackey} D.,  {Hogg} D.~W.,  {Lang} D.,   {Goodman} J.,  2013, \mn@doi
  [\pasp] {10.1086/670067}, \href
  {https://ui.adsabs.harvard.edu/abs/2013PASP..125..306F} {125, 306}

\bibitem[\protect\citeauthoryear{{G{\'e}nova-Santos}, {Rebolo},
  {Rubi{\~n}o-Mart{\'\i}n}, {L{\'o}pez-Caraballo}  \&
  {Hildebrandt}}{{G{\'e}nova-Santos} et~al.}{2011}]{rgspleiades}
{G{\'e}nova-Santos} R.,  {Rebolo} R.,  {Rubi{\~n}o-Mart{\'\i}n} J.~A.,
  {L{\'o}pez-Caraballo} C.~H.,   {Hildebrandt} S.~R.,  2011, \mn@doi [\apj]
  {10.1088/0004-637X/743/1/67}, \href
  {https://ui.adsabs.harvard.edu/abs/2011ApJ...743...67G} {743, 67}

\bibitem[\protect\citeauthoryear{{G{\'e}nova-Santos}
  et~al.,}{{G{\'e}nova-Santos} et~al.}{2015}]{Perseus}
{G{\'e}nova-Santos} R.,  et~al., 2015, \mn@doi [\mnras]
  {10.1093/mnras/stv1405}, \href
  {http://esoads.eso.org/abs/2015MNRAS.452.4169G} {452, 4169}

\bibitem[\protect\citeauthoryear{{G{\'e}nova-Santos}
  et~al.,}{{G{\'e}nova-Santos} et~al.}{2017}]{W44}
{G{\'e}nova-Santos} R.,  et~al., 2017, \mn@doi [\mnras]
  {10.1093/mnras/stw2503}, \href
  {http://esoads.eso.org/abs/2017MNRAS.464.4107G} {464, 4107}

\bibitem[\protect\citeauthoryear{{G{\'e}nova-Santos},
  {Rubi{\~n}o-Mart{\'{\i}}n}  et~al.}{{G{\'e}nova-Santos}
  et~al.}{2023}]{mfipipeline}
{G{\'e}nova-Santos} R.~T.,  {Rubi{\~n}o-Mart{\'{\i}}n} J.~A.,   et~al., 2023,
  \mnras, in prep.

\bibitem[\protect\citeauthoryear{{G{\'o}rski}, {Hivon}, {Banday}, {Wand elt},
  {Hansen}, {Reinecke}  \& {Bartelmann}}{{G{\'o}rski} et~al.}{2005}]{Healpix}
{G{\'o}rski} K.~M.,  {Hivon} E.,  {Banday} A.~J.,  {Wand elt} B.~D.,  {Hansen}
  F.~K.,  {Reinecke} M.,   {Bartelmann} M.,  2005, \mn@doi [\apj]
  {10.1086/427976}, \href
  {https://ui.adsabs.harvard.edu/abs/2005ApJ...622..759G} {622, 759}

\bibitem[\protect\citeauthoryear{{Haffner}, {Reynolds}, {Tufte}, {Madsen},
  {Jaehnig}  \& {Percival}}{{Haffner} et~al.}{2003}]{WHAM}
{Haffner} L.~M.,  {Reynolds} R.~J.,  {Tufte} S.~L.,  {Madsen} G.~J.,  {Jaehnig}
  K.~P.,   {Percival} J.~W.,  2003, \mn@doi [\apjs] {10.1086/378850}, \href
  {https://ui.adsabs.harvard.edu/abs/2003ApJS..149..405H} {149, 405}

\bibitem[\protect\citeauthoryear{{Harper} et~al.,}{{Harper}
  et~al.}{2022}]{cbassSH2022}
{Harper} S.~E.,  et~al., 2022, \mn@doi [\mnras] {10.1093/mnras/stac1210}, \href
  {https://ui.adsabs.harvard.edu/abs/2022MNRAS.513.5900H} {513, 5900}

\bibitem[\protect\citeauthoryear{Harris et~al.,}{Harris et~al.}{2020}]{numpy}
Harris C.~R.,  et~al., 2020, \mn@doi [Nature] {10.1038/s41586-020-2649-2}, 585,
  357

\bibitem[\protect\citeauthoryear{{Haslam}, {Salter}, {Stoffel}  \&
  {Wilson}}{{Haslam} et~al.}{1982}]{haslam1982}
{Haslam} C.~G.~T.,  {Salter} C.~J.,  {Stoffel} H.,   {Wilson} W.~E.,  1982,
  \aaps, \href {https://ui.adsabs.harvard.edu/abs/1982A&AS...47....1H} {47, 1}

\bibitem[\protect\citeauthoryear{{Hauser} et~al.,}{{Hauser}
  et~al.}{1998}]{cobe-dirbe}
{Hauser} M.~G.,  et~al., 1998, \mn@doi [\apj] {10.1086/306379}, \href
  {https://ui.adsabs.harvard.edu/abs/1998ApJ...508...25H} {508, 25}

\bibitem[\protect\citeauthoryear{{Hazumi} et~al.,}{{Hazumi}
  et~al.}{2020}]{litebird2020}
{Hazumi} M.,  et~al., 2020, in Society of Photo-Optical Instrumentation
  Engineers (SPIE) Conference Series. p. 114432F (\mn@eprint {arXiv}
  {2101.12449}), \mn@doi{10.1117/12.2563050}

\bibitem[\protect\citeauthoryear{{Hensley} \& {Draine}}{{Hensley} \&
  {Draine}}{2017}]{hensley2017}
{Hensley} B.~S.,  {Draine} B.~T.,  2017, \mn@doi [\apj]
  {10.3847/1538-4357/aa5c37}, \href
  {https://ui.adsabs.harvard.edu/abs/2017ApJ...836..179H} {836, 179}

\bibitem[\protect\citeauthoryear{{Hensley}, {Murphy}  \& {Staguhn}}{{Hensley}
  et~al.}{2015}]{NGC6946AME3}
{Hensley} B.,  {Murphy} E.,   {Staguhn} J.,  2015, \mn@doi [\mnras]
  {10.1093/mnras/stv287}, \href
  {https://ui.adsabs.harvard.edu/abs/2015MNRAS.449..809H} {449, 809}

\bibitem[\protect\citeauthoryear{{Hensley}, {Draine}  \& {Meisner}}{{Hensley}
  et~al.}{2016}]{hensley2016}
{Hensley} B.~S.,  {Draine} B.~T.,   {Meisner} A.~M.,  2016, \mn@doi [\apj]
  {10.3847/0004-637X/827/1/45}, \href
  {https://ui.adsabs.harvard.edu/abs/2016ApJ...827...45H} {827, 45}

\bibitem[\protect\citeauthoryear{{Hensley}, {Murray}  \& {Dodici}}{{Hensley}
  et~al.}{2022}]{hensley2022}
{Hensley} B.~S.,  {Murray} C.~E.,   {Dodici} M.,  2022, \mn@doi [\apj]
  {10.3847/1538-4357/ac5cbd}, \href
  {https://ui.adsabs.harvard.edu/abs/2022ApJ...929...23H} {929, 23}

\bibitem[\protect\citeauthoryear{{Hildebrandt}, {Rebolo},
  {Rubi{\~n}o-Mart{\'\i}n}, {Watson}, {Guti{\'e}rrez}, {Hoyland}  \&
  {Battistelli}}{{Hildebrandt} et~al.}{2007}]{cosmosomas2007}
{Hildebrandt} S.~R.,  {Rebolo} R.,  {Rubi{\~n}o-Mart{\'\i}n} J.~A.,  {Watson}
  R.~A.,  {Guti{\'e}rrez} C.~M.,  {Hoyland} R.~J.,   {Battistelli} E.~S.,
  2007, \mn@doi [\mnras] {10.1111/j.1365-2966.2007.12380.x}, \href
  {https://ui.adsabs.harvard.edu/abs/2007MNRAS.382..594H} {382, 594}

\bibitem[\protect\citeauthoryear{{Hoang} \& {Lazarian}}{{Hoang} \&
  {Lazarian}}{2016}]{hoanglazarian2016magneticpolarization}
{Hoang} T.,  {Lazarian} A.,  2016, \mn@doi [\apj] {10.3847/0004-637X/821/2/91},
  \href {https://ui.adsabs.harvard.edu/abs/2016ApJ...821...91H} {821, 91}

\bibitem[\protect\citeauthoryear{Hunter}{Hunter}{2007}]{matplotlib}
Hunter J.~D.,  2007, \mn@doi [Computing in Science \& Engineering]
  {10.1109/MCSE.2007.55}, 9, 90

\bibitem[\protect\citeauthoryear{Irfan}{Irfan}{2014}]{irfanthesis}
Irfan M.,  2014, PhD thesis, The University of Manchester

\bibitem[\protect\citeauthoryear{{Irfan} et~al.,}{{Irfan} et~al.}{2015}]{cbass}
{Irfan} M.~O.,  et~al., 2015, \mn@doi [\mnras] {10.1093/mnras/stv212}, \href
  {https://ui.adsabs.harvard.edu/abs/2015MNRAS.448.3572I} {448, 3572}

\bibitem[\protect\citeauthoryear{{Jonas}, {Baart}  \& {Nicolson}}{{Jonas}
  et~al.}{1998}]{hartrao}
{Jonas} J.~L.,  {Baart} E.~E.,   {Nicolson} G.~D.,  1998, \mn@doi [\mnras]
  {10.1046/j.1365-8711.1998.01367.x}, \href
  {https://ui.adsabs.harvard.edu/abs/1998MNRAS.297..977J} {297, 977}

\bibitem[\protect\citeauthoryear{{Jones} et~al.,}{{Jones}
  et~al.}{2018}]{cbassjones2018}
{Jones} M.~E.,  et~al., 2018, \mn@doi [\mnras] {10.1093/mnras/sty1956}, \href
  {https://ui.adsabs.harvard.edu/abs/2018MNRAS.480.3224J} {480, 3224}

\bibitem[\protect\citeauthoryear{{Kogut}, {Banday}, {Bennett}, {Gorski},
  {Hinshaw}, {Smoot}  \& {Wright}}{{Kogut} et~al.}{1996}]{kogut1996}
{Kogut} A.,  {Banday} A.~J.,  {Bennett} C.~L.,  {Gorski} K.~M.,  {Hinshaw} G.,
  {Smoot} G.~F.,   {Wright} E.~I.,  1996, \mn@doi [\apjl] {10.1086/310072},
  \href {https://ui.adsabs.harvard.edu/abs/1996ApJ...464L...5K} {464, L5}

\bibitem[\protect\citeauthoryear{{Leitch}, {Readhead}, {Pearson}  \&
  {Myers}}{{Leitch} et~al.}{1997}]{leitch1997}
{Leitch} E.~M.,  {Readhead} A.~C.~S.,  {Pearson} T.~J.,   {Myers} S.~T.,  1997,
  \mn@doi [\apjl] {10.1086/310823}, \href
  {https://ui.adsabs.harvard.edu/abs/1997ApJ...486L..23L} {486, L23}

\bibitem[\protect\citeauthoryear{{Li} \& {Draine}}{{Li} \&
  {Draine}}{2001}]{pah1}
{Li} A.,  {Draine} B.~T.,  2001, \mn@doi [\apj] {10.1086/323147}, \href
  {https://ui.adsabs.harvard.edu/abs/2001ApJ...554..778L} {554, 778}

\bibitem[\protect\citeauthoryear{{Linden}, {Murphy}, {Dong}, {Momjian},
  {Kennicutt}, {Meier}, {Schinnerer}  \& {Turner}}{{Linden}
  et~al.}{2020}]{radiostarformationAMEsurvey}
{Linden} S.~T.,  {Murphy} E.~J.,  {Dong} D.,  {Momjian} E.,  {Kennicutt} R.~C.
  J.,  {Meier} D.~S.,  {Schinnerer} E.,   {Turner} J.~L.,  2020, \mn@doi
  [\apjs] {10.3847/1538-4365/ab8a4d}, \href
  {https://ui.adsabs.harvard.edu/abs/2020ApJS..248...25L} {248, 25}

\bibitem[\protect\citeauthoryear{{L{\'o}pez-Caraballo},
  {Rubi{\~n}o-Mart{\'\i}n}, {Rebolo}  \&
  {G{\'e}nova-Santos}}{{L{\'o}pez-Caraballo} et~al.}{2011}]{caraballo2011}
{L{\'o}pez-Caraballo} C.~H.,  {Rubi{\~n}o-Mart{\'\i}n} J.~A.,  {Rebolo} R.,
  {G{\'e}nova-Santos} R.,  2011, \mn@doi [\apj] {10.1088/0004-637X/729/1/25},
  \href {https://ui.adsabs.harvard.edu/abs/2011ApJ...729...25L} {729, 25}

\bibitem[\protect\citeauthoryear{{Mathis}, {Mezger}  \& {Panagia}}{{Mathis}
  et~al.}{1983}]{G0mathis1983}
{Mathis} J.~S.,  {Mezger} P.~G.,   {Panagia} N.,  1983, \aap, \href
  {https://ui.adsabs.harvard.edu/abs/1983A&A...128..212M} {128, 212}

\bibitem[\protect\citeauthoryear{{Miville-Desch{\^e}nes} \&
  {Lagache}}{{Miville-Desch{\^e}nes} \& {Lagache}}{2005}]{iris}
{Miville-Desch{\^e}nes} M.-A.,  {Lagache} G.,  2005, \mn@doi [\apjs]
  {10.1086/427938}, \href
  {https://ui.adsabs.harvard.edu/abs/2005ApJS..157..302M} {157, 302}

\bibitem[\protect\citeauthoryear{{Murphy} et~al.,}{{Murphy}
  et~al.}{2010}]{NGC6946AME}
{Murphy} E.~J.,  et~al., 2010, \mn@doi [\apjl] {10.1088/2041-8205/709/2/L108},
  \href {https://ui.adsabs.harvard.edu/abs/2010ApJ...709L.108M} {709, L108}

\bibitem[\protect\citeauthoryear{{Murphy}, {Linden}, {Dong}, {Hensley},
  {Momjian}, {Helou}  \& {Evans}}{{Murphy} et~al.}{2018}]{NGC4125AME}
{Murphy} E.~J.,  {Linden} S.~T.,  {Dong} D.,  {Hensley} B.~S.,  {Momjian} E.,
  {Helou} G.,   {Evans} A.~S.,  2018, \mn@doi [\apj]
  {10.3847/1538-4357/aac5f5}, \href
  {https://ui.adsabs.harvard.edu/abs/2018ApJ...862...20M} {862, 20}

\bibitem[\protect\citeauthoryear{{Neugebauer} et~al.,}{{Neugebauer}
  et~al.}{1984}]{iras1}
{Neugebauer} G.,  et~al., 1984, \mn@doi [\apjl] {10.1086/184209}, \href
  {https://ui.adsabs.harvard.edu/abs/1984ApJ...278L...1N} {278, L1}

\bibitem[\protect\citeauthoryear{{Orlando} \& {Strong}}{{Orlando} \&
  {Strong}}{2013}]{orlando2013}
{Orlando} E.,  {Strong} A.,  2013, \mn@doi [\mnras] {10.1093/mnras/stt1718},
  \href {https://ui.adsabs.harvard.edu/abs/2013MNRAS.436.2127O} {436, 2127}

\bibitem[\protect\citeauthoryear{{Paradis}, {Dobashi}, {Shimoikura},
  {Kawamura}, {Onishi}, {Fukui}  \& {Bernard}}{{Paradis} et~al.}{2012}]{CADE}
{Paradis} D.,  {Dobashi} K.,  {Shimoikura} T.,  {Kawamura} A.,  {Onishi} T.,
  {Fukui} Y.,   {Bernard} J.~P.,  2012, \mn@doi [\aap]
  {10.1051/0004-6361/201118740}, \href
  {https://ui.adsabs.harvard.edu/abs/2012A&A...543A.103P} {543, A103}

\bibitem[\protect\citeauthoryear{{Peel}, {Dickinson}, {Davies}, {Clements}  \&
  {Beswick}}{{Peel} et~al.}{2011}]{peelAMEgalaxies}
{Peel} M.~W.,  {Dickinson} C.,  {Davies} R.~D.,  {Clements} D.~L.,   {Beswick}
  R.~J.,  2011, \mn@doi [\mnras] {10.1111/j.1745-3933.2011.01108.x}, \href
  {https://ui.adsabs.harvard.edu/abs/2011MNRAS.416L..99P} {416, L99}

\bibitem[\protect\citeauthoryear{{Peel}, {Genova-Santos}, {Dickinson}, {Leahy},
  {L{\'o}pez-Caraballo}, {Fern{\'a}ndez-Torreiro}, {Rubi{\~n}o-Mart{\'\i}n}  \&
  {Spencer}}{{Peel} et~al.}{2022}]{fastcc}
{Peel} M.~W.,  {Genova-Santos} R.,  {Dickinson} C.,  {Leahy} J.~P.,
  {L{\'o}pez-Caraballo} C.,  {Fern{\'a}ndez-Torreiro} M.,
  {Rubi{\~n}o-Mart{\'\i}n} J.~A.,   {Spencer} L.~D.,  2022, \mn@doi [Research
  Notes of the American Astronomical Society] {10.3847/2515-5172/aca6eb}, \href
  {https://ui.adsabs.harvard.edu/abs/2022RNAAS...6..252P} {6, 252}

\bibitem[\protect\citeauthoryear{{Planck Collaboration} et~al.,}{{Planck
  Collaboration} et~al.}{2011}]{planck2011XXnewlight}
{Planck Collaboration} et~al., 2011, \mn@doi [\aap]
  {10.1051/0004-6361/201116470}, \href
  {https://ui.adsabs.harvard.edu/abs/2011A&A...536A..20P} {536, A20}

\bibitem[\protect\citeauthoryear{{Planck Collaboration} et~al.,}{{Planck
  Collaboration} et~al.}{2014a}]{planckbetamm}
{Planck Collaboration} et~al., 2014a, \mn@doi [\aap]
  {10.1051/0004-6361/201322367}, \href
  {https://ui.adsabs.harvard.edu/abs/2014A&A...564A..45P} {564, A45}

\bibitem[\protect\citeauthoryear{{Planck Collaboration} et~al.,}{{Planck
  Collaboration} et~al.}{2014b}]{planck2015galacticcloudsAME}
{Planck Collaboration} et~al., 2014b, \mn@doi [\aap]
  {10.1051/0004-6361/201322612}, \href
  {https://ui.adsabs.harvard.edu/abs/2014A&A...565A.103P} {565, A103}

\bibitem[\protect\citeauthoryear{{Planck Collaboration} et~al.,}{{Planck
  Collaboration} et~al.}{2014c}]{planck2014LFIcalibration}
{Planck Collaboration} et~al., 2014c, \mn@doi [\aap]
  {10.1051/0004-6361/201321527}, \href
  {https://ui.adsabs.harvard.edu/abs/2014A&A...571A...5P} {571, A5}

\bibitem[\protect\citeauthoryear{{Planck Collaboration} et~al.,}{{Planck
  Collaboration} et~al.}{2014d}]{planck2013XI}
{Planck Collaboration} et~al., 2014d, \mn@doi [\aap]
  {10.1051/0004-6361/201323195}, \href
  {https://ui.adsabs.harvard.edu/abs/2014A&A...571A..11P} {571, A11}

\bibitem[\protect\citeauthoryear{{Planck Collaboration} et~al.,}{{Planck
  Collaboration} et~al.}{2015a}]{planckancillarydata}
{Planck Collaboration} et~al., 2015a, \mn@doi [\aap]
  {10.1051/0004-6361/201424434}, \href
  {https://ui.adsabs.harvard.edu/abs/2015A&A...580A..13P} {580, A13}

\bibitem[\protect\citeauthoryear{{Planck Collaboration} et~al.,}{{Planck
  Collaboration} et~al.}{2015b}]{planckM31}
{Planck Collaboration} et~al., 2015b, \mn@doi [\aap]
  {10.1051/0004-6361/201424643}, \href
  {https://ui.adsabs.harvard.edu/abs/2015A&A...582A..28P} {582, A28}

\bibitem[\protect\citeauthoryear{{Planck Collaboration} et~al.,}{{Planck
  Collaboration} et~al.}{2016a}]{planck2016Xforegroundmaps}
{Planck Collaboration} et~al., 2016a, \mn@doi [\aap]
  {10.1051/0004-6361/201525967}, \href
  {https://ui.adsabs.harvard.edu/abs/2016A&A...594A..10P} {594, A10}

\bibitem[\protect\citeauthoryear{{Planck Collaboration} et~al.,}{{Planck
  Collaboration} et~al.}{2016b}]{planckemissivityestimates}
{Planck Collaboration} et~al., 2016b, \mn@doi [\aap]
  {10.1051/0004-6361/201526803}, \href
  {https://ui.adsabs.harvard.edu/abs/2016A&A...594A..25P} {594, A25}

\bibitem[\protect\citeauthoryear{{Planck Collaboration} et~al.,}{{Planck
  Collaboration} et~al.}{2018}]{planck}
{Planck Collaboration} et~al., 2018, arXiv e-prints, \href
  {https://ui.adsabs.harvard.edu/abs/2018arXiv180706205P} {p. arXiv:1807.06205}

\bibitem[\protect\citeauthoryear{{Planck Collaboration} et~al.,}{{Planck
  Collaboration} et~al.}{2020}]{npipe}
{Planck Collaboration} et~al., 2020, \mn@doi [\aap]
  {10.1051/0004-6361/202038073}, \href
  {https://ui.adsabs.harvard.edu/abs/2020A&A...643A..42P} {643, A42}

\bibitem[\protect\citeauthoryear{{Platania}, {Burigana}, {Maino}, {Caserini},
  {Bersanelli}, {Cappellini}  \& {Mennella}}{{Platania}
  et~al.}{2003}]{plataniamaps}
{Platania} P.,  {Burigana} C.,  {Maino} D.,  {Caserini} E.,  {Bersanelli} M.,
  {Cappellini} B.,   {Mennella} A.,  2003, \mn@doi [\aap]
  {10.1051/0004-6361:20031125}, \href
  {https://ui.adsabs.harvard.edu/abs/2003A&A...410..847P} {410, 847}

\bibitem[\protect\citeauthoryear{{Poidevin} et~al.,}{{Poidevin}
  et~al.}{2019}]{Taurus}
{Poidevin} F.,  et~al., 2019, \mn@doi [\mnras] {10.1093/mnras/sty3462}, \href
  {https://ui.adsabs.harvard.edu/abs/2019MNRAS.486..462P} {486, 462}

\bibitem[\protect\citeauthoryear{{Poidevin} et~al.,}{{Poidevin}
  et~al.}{2023}]{AMEwidesurvey}
{Poidevin} F.,  et~al., 2023, \mn@doi [\mnras] {10.1093/mnras/stac3151}, \href
  {https://ui.adsabs.harvard.edu/abs/2023MNRAS.519.3481P} {519, 3481}

\bibitem[\protect\citeauthoryear{{Reich}}{{Reich}}{1982}]{reich1982}
{Reich} W.,  1982, \aaps, \href
  {https://ui.adsabs.harvard.edu/abs/1982A&AS...48..219R} {48, 219}

\bibitem[\protect\citeauthoryear{{Reich} \& {Reich}}{{Reich} \&
  {Reich}}{1986}]{reich1986}
{Reich} P.,  {Reich} W.,  1986, \aaps, \href
  {https://ui.adsabs.harvard.edu/abs/1986A&AS...63..205R} {63, 205}

\bibitem[\protect\citeauthoryear{{Reich} \& {Reich}}{{Reich} \&
  {Reich}}{1988}]{reich1988}
{Reich} P.,  {Reich} W.,  1988, \aaps, \href
  {https://ui.adsabs.harvard.edu/abs/1988A&AS...74....7R} {74, 7}

\bibitem[\protect\citeauthoryear{{Reich}, {Reich}  \& {Fuerst}}{{Reich}
  et~al.}{1990}]{effelsberg1}
{Reich} W.,  {Reich} P.,   {Fuerst} E.,  1990, \aaps, \href
  {https://ui.adsabs.harvard.edu/abs/1990A&AS...83..539R} {83, 539}

\bibitem[\protect\citeauthoryear{{Reich}, {Reich}  \& {Furst}}{{Reich}
  et~al.}{1997}]{effelsberg2}
{Reich} P.,  {Reich} W.,   {Furst} E.,  1997, \mn@doi [\aaps]
  {10.1051/aas:1997274}, \href
  {https://ui.adsabs.harvard.edu/abs/1997A&AS..126..413R} {126, 413}

\bibitem[\protect\citeauthoryear{{Reich}, {Testori}  \& {Reich}}{{Reich}
  et~al.}{2001}]{villaelisa}
{Reich} P.,  {Testori} J.~C.,   {Reich} W.,  2001, \mn@doi [\aap]
  {10.1051/0004-6361:20011000}, \href
  {https://ui.adsabs.harvard.edu/abs/2001A&A...376..861R} {376, 861}

\bibitem[\protect\citeauthoryear{{Remazeilles}, {Dickinson}, {Banday},
  {Bigot-Sazy}  \& {Ghosh}}{{Remazeilles} et~al.}{2015}]{remazeilleshaslam}
{Remazeilles} M.,  {Dickinson} C.,  {Banday} A.~J.,  {Bigot-Sazy} M.~A.,
  {Ghosh} T.,  2015, \mn@doi [\mnras] {10.1093/mnras/stv1274}, \href
  {https://ui.adsabs.harvard.edu/abs/2015MNRAS.451.4311R} {451, 4311}

\bibitem[\protect\citeauthoryear{{Rennie} et~al.,}{{Rennie}
  et~al.}{2022}]{comapVI}
{Rennie} T.~J.,  et~al., 2022, \mn@doi [\apj] {10.3847/1538-4357/ac63c8}, \href
  {https://ui.adsabs.harvard.edu/abs/2022ApJ...933..187R} {933, 187}

\bibitem[\protect\citeauthoryear{{Rubi{\~n}o-Mart{\'\i}n}
  et~al.,}{{Rubi{\~n}o-Mart{\'\i}n} et~al.}{2010}]{quijote2010}
{Rubi{\~n}o-Mart{\'\i}n} J.~A.,  et~al., 2010, \mn@doi [Astrophysics and Space
  Science Proceedings] {10.1007/978-3-642-11250-8_12}, \href
  {https://ui.adsabs.harvard.edu/abs/2010ASSP...14..127R} {14, 127}

\bibitem[\protect\citeauthoryear{{Rubi{\~n}o-Mart{\'\i}n},
  {L{\'o}pez-Caraballo}, {G{\'e}nova-Santos}  \&
  {Rebolo}}{{Rubi{\~n}o-Mart{\'\i}n} et~al.}{2012}]{JARM2012}
{Rubi{\~n}o-Mart{\'\i}n} J.~A.,  {L{\'o}pez-Caraballo} C.~H.,
  {G{\'e}nova-Santos} R.,   {Rebolo} R.,  2012, \mn@doi [Advances in Astronomy]
  {10.1155/2012/351836}, \href
  {https://ui.adsabs.harvard.edu/abs/2012AdAst2012E..40R} {2012, 351836}

\bibitem[\protect\citeauthoryear{{Rubi{\~n}o-Mart{\'\i}n}
  et~al.,}{{Rubi{\~n}o-Mart{\'\i}n} et~al.}{2023}]{mfiwidesurvey}
{Rubi{\~n}o-Mart{\'\i}n} J.~A.,  et~al., 2023, \mn@doi [\mnras]
  {10.1093/mnras/stac3439}, \href
  {https://ui.adsabs.harvard.edu/abs/2023MNRAS.519.3383R} {519, 3383}

\bibitem[\protect\citeauthoryear{{Rybicki} \& {Lightman}}{{Rybicki} \&
  {Lightman}}{1979}]{rybickilightman}
{Rybicki} G.~B.,  {Lightman} A.~P.,  1979, {Radiative processes in
  astrophysics}

\bibitem[\protect\citeauthoryear{{Scaife} et~al.,}{{Scaife}
  et~al.}{2010}]{NGC6946AME2}
{Scaife} A. M.~M.,  et~al., 2010, \mn@doi [\mnras]
  {10.1111/j.1745-3933.2010.00878.x}, \href
  {https://ui.adsabs.harvard.edu/abs/2010MNRAS.406L..45S} {406, L45}

\bibitem[\protect\citeauthoryear{{Schlegel}, {Finkbeiner}  \&
  {Davis}}{{Schlegel} et~al.}{1998}]{100micronmap}
{Schlegel} D.~J.,  {Finkbeiner} D.~P.,   {Davis} M.,  1998, \mn@doi [\apj]
  {10.1086/305772}, \href
  {https://ui.adsabs.harvard.edu/abs/1998ApJ...500..525S} {500, 525}

\bibitem[\protect\citeauthoryear{{Silsbee}, {Ali-Ha{\"\i}moud}  \&
  {Hirata}}{{Silsbee} et~al.}{2011}]{spdust2}
{Silsbee} K.,  {Ali-Ha{\"\i}moud} Y.,   {Hirata} C.~M.,  2011, \mn@doi [\mnras]
  {10.1111/j.1365-2966.2010.17882.x}, \href
  {https://ui.adsabs.harvard.edu/abs/2011MNRAS.411.2750S} {411, 2750}

\bibitem[\protect\citeauthoryear{{Smoot} et~al.,}{{Smoot} et~al.}{1990}]{dmr}
{Smoot} G.,  et~al., 1990, \mn@doi [\apj] {10.1086/169154}, \href
  {https://ui.adsabs.harvard.edu/abs/1990ApJ...360..685S} {360, 685}

\bibitem[\protect\citeauthoryear{{Stevenson}}{{Stevenson}}{2014}]{stevenson2014}
{Stevenson} M.~A.,  2014, \mn@doi [\apj] {10.1088/0004-637X/781/2/113}, \href
  {https://ui.adsabs.harvard.edu/abs/2014ApJ...781..113S} {781, 113}

\bibitem[\protect\citeauthoryear{{Tibbs}, {Scaife}, {Dickinson}, {Paladini},
  {Davies}, {Davis}, {Grainge}  \& {Watson}}{{Tibbs}
  et~al.}{2013}]{tibbsspvariations}
{Tibbs} C.~T.,  {Scaife} A.~M.~M.,  {Dickinson} C.,  {Paladini} R.,  {Davies}
  R.~D.,  {Davis} R.~J.,  {Grainge} K.~J.~B.,   {Watson} R.~A.,  2013, \mn@doi
  [\apj] {10.1088/0004-637X/768/2/98}, \href
  {https://ui.adsabs.harvard.edu/abs/2013ApJ...768...98T} {768, 98}

\bibitem[\protect\citeauthoryear{{Tibbs}, {Israel}, {Laureijs}, {Tauber},
  {Partridge}, {Peel}  \& {Fauvet}}{{Tibbs} et~al.}{2018}]{planckM33}
{Tibbs} C.~T.,  {Israel} F.~P.,  {Laureijs} R.~J.,  {Tauber} J.~A.,
  {Partridge} B.,  {Peel} M.~W.,   {Fauvet} L.,  2018, \mn@doi [\mnras]
  {10.1093/mnras/sty824}, \href
  {https://ui.adsabs.harvard.edu/abs/2018MNRAS.477.4968T} {477, 4968}

\bibitem[\protect\citeauthoryear{{Todorovi{\'c}} et~al.,}{{Todorovi{\'c}}
  et~al.}{2010}]{VSAGalacticplanesurvey}
{Todorovi{\'c}} M.,  et~al., 2010, \mn@doi [\mnras]
  {10.1111/j.1365-2966.2010.16809.x}, \href
  {https://ui.adsabs.harvard.edu/abs/2010MNRAS.406.1629T} {406, 1629}

\bibitem[\protect\citeauthoryear{{Tramonte} et~al.,}{{Tramonte}
  et~al.}{2023}]{W51}
{Tramonte} D.,  et~al., 2023, \mn@doi [\mnras] {10.1093/mnras/stac3502}, \href
  {https://ui.adsabs.harvard.edu/abs/2023MNRAS.519.3432T} {519, 3432}

\bibitem[\protect\citeauthoryear{{Vidal}, {Dickinson}, {Harper}, {Casassus}  \&
  {Witt}}{{Vidal} et~al.}{2020}]{vidalLDN180}
{Vidal} M.,  {Dickinson} C.,  {Harper} S.~E.,  {Casassus} S.,   {Witt} A.~N.,
  2020, \mn@doi [\mnras] {10.1093/mnras/staa1186}, \href
  {https://ui.adsabs.harvard.edu/abs/2020MNRAS.495.1122V} {495, 1122}

\bibitem[\protect\citeauthoryear{Virtanen et~al.,}{Virtanen
  et~al.}{2020}]{scipy}
Virtanen P.,  et~al., 2020, \mn@doi [Nature Methods]
  {10.1038/s41592-019-0686-2}, \href {https://rdcu.be/b08Wh} {17, 261}

\bibitem[\protect\citeauthoryear{{Watson}, {Rebolo}, {Rubi{\~n}o-Mart{\'\i}n},
  {Hildebrandt}, {Guti{\'e}rrez}, {Fern{\'a}ndez-Cerezo}, {Hoyland}  \&
  {Battistelli}}{{Watson} et~al.}{2005}]{cosmosomas2005}
{Watson} R.~A.,  {Rebolo} R.,  {Rubi{\~n}o-Mart{\'\i}n} J.~A.,  {Hildebrandt}
  S.,  {Guti{\'e}rrez} C.~M.,  {Fern{\'a}ndez-Cerezo} S.,  {Hoyland} R.~J.,
  {Battistelli} E.~S.,  2005, \mn@doi [\apjl] {10.1086/430519}, \href
  {https://ui.adsabs.harvard.edu/abs/2005ApJ...624L..89W} {624, L89}

\bibitem[\protect\citeauthoryear{{Wehus} et~al.,}{{Wehus}
  et~al.}{2017}]{wehus+2017}
{Wehus} I.~K.,  et~al., 2017, \mn@doi [\aap] {10.1051/0004-6361/201525659},
  \href {https://ui.adsabs.harvard.edu/abs/2017A&A...597A.131W} {597, A131}

\bibitem[\protect\citeauthoryear{{Wenger} et~al.,}{{Wenger}
  et~al.}{2000}]{simbad}
{Wenger} M.,  et~al., 2000, \mn@doi [\aaps] {10.1051/aas:2000332}, \href
  {https://ui.adsabs.harvard.edu/abs/2000A&AS..143....9W} {143, 9}

\bibitem[\protect\citeauthoryear{{Werner} et~al.,}{{Werner}
  et~al.}{2004}]{spitzertelescope}
{Werner} M.~W.,  et~al., 2004, \mn@doi [\apjs] {10.1086/422992}, \href
  {https://ui.adsabs.harvard.edu/abs/2004ApJS..154....1W} {154, 1}

\bibitem[\protect\citeauthoryear{{Wheelock} et~al.,}{{Wheelock}
  et~al.}{1994}]{iras2_issa}
{Wheelock} S.~L.,  et~al., 1994, {IRAS sky survey atlas: Explanatory
  supplement}, NASA STI/Recon Technical Report N

\bibitem[\protect\citeauthoryear{{Xie} \& {Ho}}{{Xie} \&
  {Ho}}{2021}]{oct21PAHinQSOs}
{Xie} Y.,  {Ho} L.~C.,  2021, arXiv e-prints, \href
  {https://ui.adsabs.harvard.edu/abs/2021arXiv211009705X} {p. arXiv:2110.09705}

\bibitem[\protect\citeauthoryear{{Ysard}, {Juvela}  \& {Verstraete}}{{Ysard}
  et~al.}{2011}]{ysard2011}
{Ysard} N.,  {Juvela} M.,   {Verstraete} L.,  2011, \mn@doi [\aap]
  {10.1051/0004-6361/201117394}, \href
  {https://ui.adsabs.harvard.edu/abs/2011A&A...535A..89Y} {535, A89}

\bibitem[\protect\citeauthoryear{{Ysard}, {Miville-Desch{\^e}nes}, {Verstraete}
   \& {Jones}}{{Ysard} et~al.}{2022}]{ysard2022}
{Ysard} N.,  {Miville-Desch{\^e}nes} M.-A.,  {Verstraete} L.,   {Jones} A.~P.,
  2022, arXiv e-prints, \href
  {https://ui.adsabs.harvard.edu/abs/2022arXiv220501400Y} {p. arXiv:2205.01400}

\bibitem[\protect\citeauthoryear{Zonca, Singer, Lenz, Reinecke, Rosset, Hivon
  \& Gorski}{Zonca et~al.}{2019}]{Healpix2}
Zonca A.,  Singer L.,  Lenz D.,  Reinecke M.,  Rosset C.,  Hivon E.,   Gorski
  K.,  2019, \mn@doi [Journal of Open Source Software] {10.21105/joss.01298},
  4, 1298

\makeatother
\end{thebibliography}

\clearpage \pagebreak
\section*{ONLINE MATERIAL}
\appendix
\section{Recalibration factors for low-frequency surveys}
\label{section:appendix_recalibration}
\begin{figure}
    \centering
    \includegraphics[width=1\linewidth]{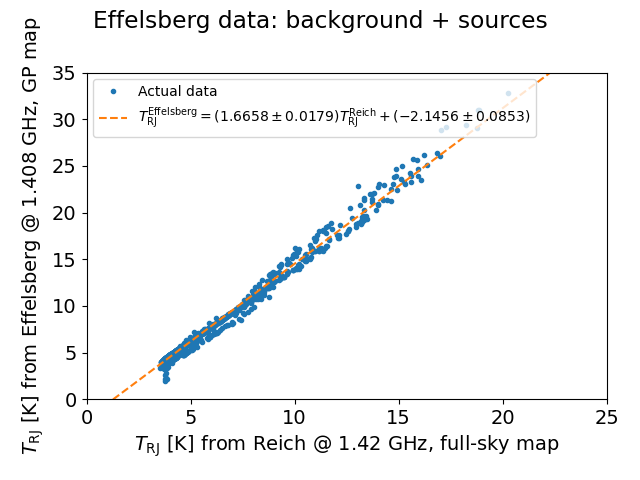}
    \includegraphics[width=1\linewidth]{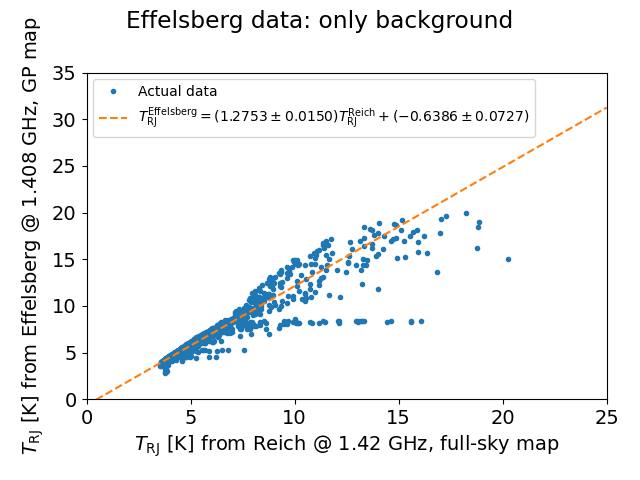}
    \includegraphics[width=1\linewidth]{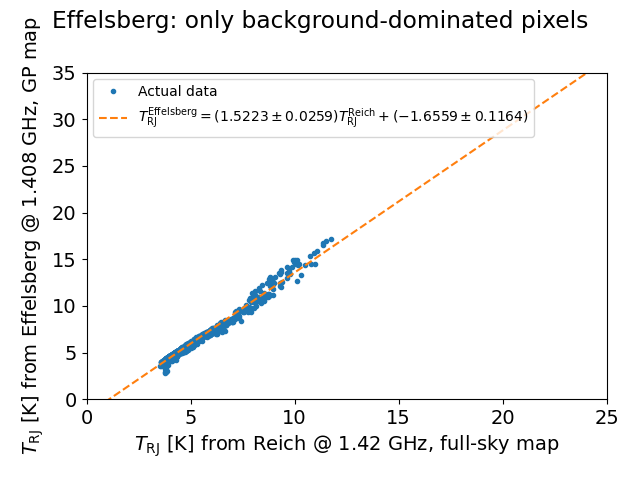}
    \caption{Comparison between the Effelsberg 1408 MHz \citep{effelsberg1, effelsberg2}
and \protect\cite{reich1982, reich1986, villaelisa} surveys 
when taking into account (top) the sum of both the Effelsberg
1408 MHz survey diffuse and puoint components; (middle) 
the diffuse component only; and (bottom) the same but only
with those pixels with residual punctual importance 
($<10\,\%$ the diffuse one) considered. This one would
be the one most similar to studying pure diffuse
emission, so closest to our case. We see that the slope 
(which is equivalent to the recalibration factor required
for the \protect\cite{reich1982, reich1986, villaelisa} 
survey) is lower when we increase the weight of diffuse 
emission.}
    \label{fig:reich_recalibration}
\end{figure}

\begin{figure}
    \centering
    \includegraphics[width=1\linewidth]{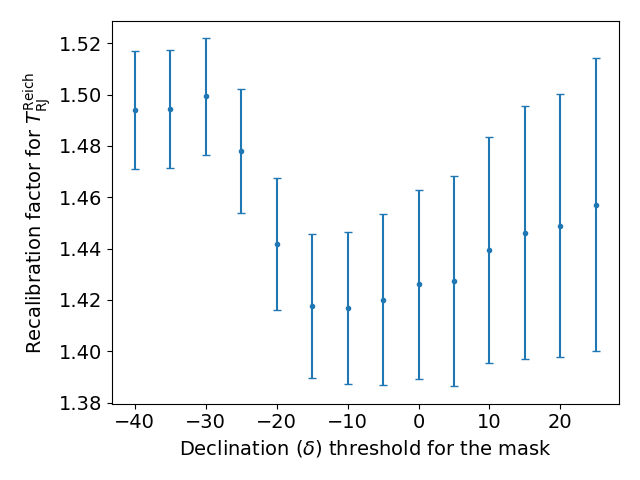}
    \caption{Variation of the recalibration factor 
for the \protect\cite{reich1982, reich1986, villaelisa} 
survey when changing the 
threshold for the declination mask. We see that, when 
taking into account low-declination pixels (which cover 
the Galactic centre), the value is offseted to larger 
values. However, even when a modest mask is applied, the 
value decreases to a constant value close to 1.4--1.45.}
    \label{fig:recalibration_factor_variation}
\end{figure}

To evaluate the \cite{reich1982, reich1986, villaelisa}
recalibration factor 
when studying large-scale diffuse emission, we used 
the desourced Effelsberg 1408\,MHz survey 
\citep{effelsberg1,effelsberg2}. This 
 survey is in fact calibrated using the previous
 1420\,MHz survey, so both should be equal. We 
study those pixels with $|b|<3\degr$ (to avoid 
border effects due to the smoothing to $1\degr$ 
scales) and $\delta>-20\degr$ (as lower declinations
 are not visible in the \citealp{effelsberg1} survey). When
 we add the source map to the former and perform a 
regular TTplot with the 1420 MHz one, we find that 
the latter should be multiplied by a factor of 1.67 to reach the 
emission of the former (top of Fig.~\ref{fig:reich_recalibration}),
 without taking into account the different nominal 
frequencies between the two (which could produce a 
difference as large as a 3\,\%). This result is a 
little higher than the value of 1.55 reported in the literature 
mentioned in Section~\ref{section:low_frequency_surveys}.
Nevertheless, if we use the 
desourced maps taking into account only those pixels 
where the diffuse emission is dominant (we selected
those where the source map from \citealp{effelsberg1, effelsberg2} 
had an emission 
$\leq$10\,\% of that of the desourced), that 
factor decreases to 1.52 (bottom of 
Fig.~\ref{fig:reich_recalibration}). Finally, we 
must bear in mind that really bright pixels 
close to the Galactic centre might be contaminating
this factor. When applying masks discarding those 
low-declination pixels close to the Galactic centre 
(Fig.~\ref{fig:recalibration_factor_variation}), we end 
up with even lower values (around $\sim$1.45). Owing to the
 addition of the nominal frequency difference, we
 decided to apply a 1.4 recalibration factor (this 
should be taken as an upper limit, as we shall still 
have some emission coming from point sources). 

\begin{figure}
\includegraphics[width=1\linewidth]{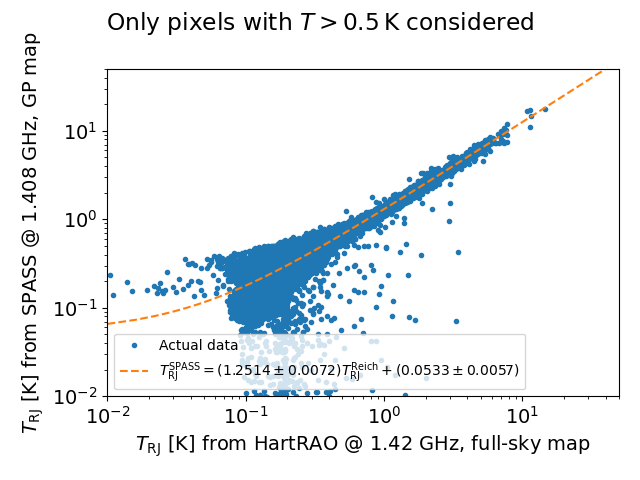}
\caption{Calculation for the recalibration factor 
due to the main-beam full-beam transformation for 
HartRAO survey, using its TTplot with the SPASS survey.}
\label{fig:TTplot_HartRAOvsSPASS}
\end{figure}

For the HartRAO~\citep{hartrao} survey, we built 
the TTplot between this survey and that of  
SPASS~\citep{carretti2019} since the nominal frequencies 
are similar for both (2.303\,GHz and 
2.326\,GHz respectively). This means that the difference 
between the two would be within 2\,\,\% (for a 
pure free-free spectrum with a temperature index of $=-2.1$) 
and 3\,\,\% (for a synchrotron index  of$=-3.1$). The 
result is shown in Figure~\ref{fig:TTplot_HartRAOvsSPASS}. 
Apart from this issue, the \cite{hartrao} survey has
 polarisation leakage into the intensity signal. The 
overall calibration uncertainty is increased accordingly 
from the nominal 5\,\,\% value to 20\,\,\%, 
embedding also the difference due to the nominal frequency
displacement mentioned before. The \cite{reich1982, 
reich1986, villaelisa} survey
calibration uncertainty is also increased from 5\,\,\% 
to 20\,\,\%, owing to the addition of its recalibration 
factor. All the previously mentioned data  are available 
in the LAMBDA, with the exception of the {\tt HEALPix} version 
of the Effelsberg survey, which is available through the 
Centre d'Analyse de Données Etendue (CADE) 
website.\footnote{http://cade.irap.omp.eu}

\section{MCMC convergence criterion}
\label{section:appendix_convergence}
As we are fitting a SED for each of the pixels 
in the selection defined in Section~\ref{section:region_definition}, 
reducing the computation time for each one is 
critical. That is why, when computing the MCMC 
chains, we do not use the same number of steps for 
every pixel. Instead, we run a short MCMC with 
32 chains and just $N=2500$ steps. We then 
compute the autocorrelation times, $\tau$, for 
every parameter in the returned set. The chains 
will have converged when none of these times 
exceeds $N/50$, $N$ being the length  of the chains at any 
time. As this first guess of $\tau$ is usually
 underestimated due to the chains' short lengths, we 
resume the MCMC for the maximum $\tau$ value 
multiplied by 50$\cdot$3 steps more. 

\begin{figure*}
\includegraphics[width=1\linewidth]{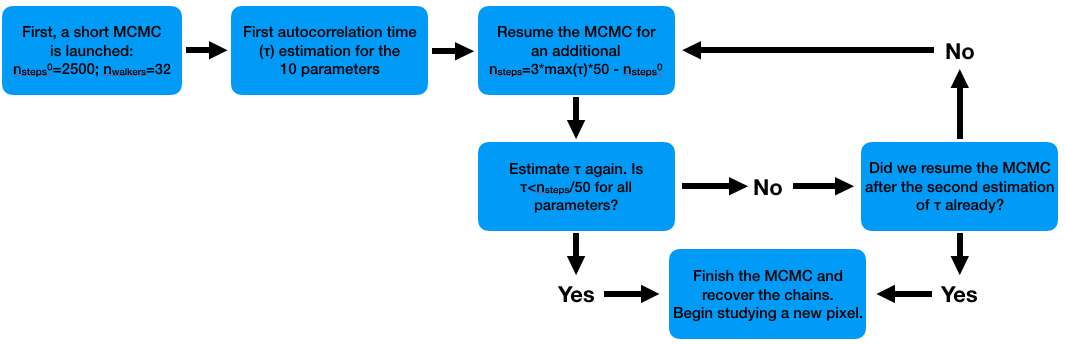}
\caption{Diagram showing how convergence is considered 
to be achieved when performing the MCMC in this paper.}
\label{fig:autocorrelation_diagram}
\end{figure*}

Pixels with higher signal-to-noise ratios will 
have lower $\tau$ values, so they take 
less time to compute, thus allowing the computation node 
to process a new pixel. In the case that $\tau>N/50$ 
is still true after this second run, the process 
is repeated once more. If $\tau>N/50$ continues 
to be true after the third MCMC run, the pixel 
is left as it is: this is the case for the pixels 
with the lowest signal-to-noise ratios, which are 
dominated by degeneracies (mainly between synchrotron 
and free-free or AME, and free-free and the CMB): that 
induces their $\tau$ values to increase with the 
chain lengths, thus preventing them from converging in a 
reasonable amount or time (or even at all). These 
pixels will be mostly discarded when running the 
analysis, so this issue is not critical. A diagram 
showing how this is computed can be seen in 
Figure~\ref{fig:autocorrelation_diagram}.

\section{Maps after applying the $\SNRAME>2$ threshold}
\label{section:appendix_snr}
\begin{figure*}
    \centering
    \includegraphics[width=1\linewidth]{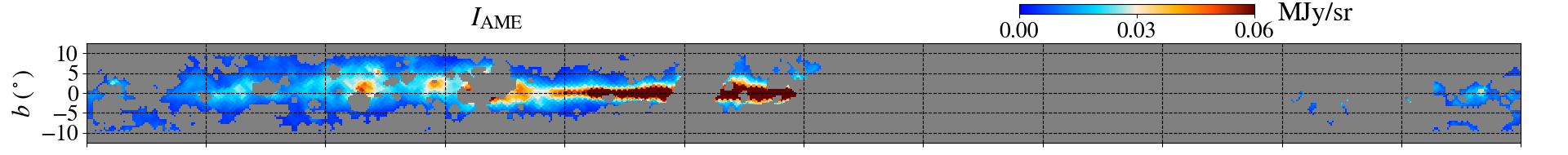}
    \includegraphics[width=1\linewidth]{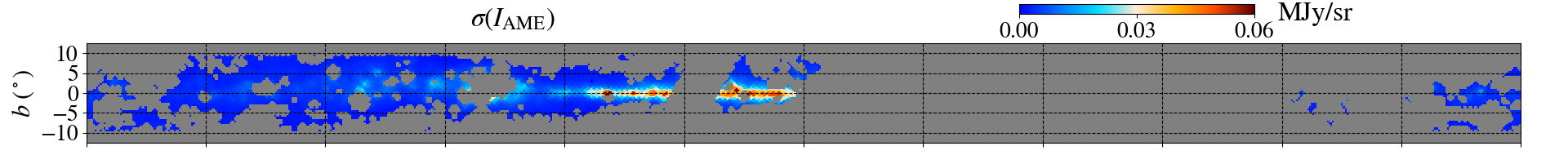}
    \includegraphics[width=1\linewidth]{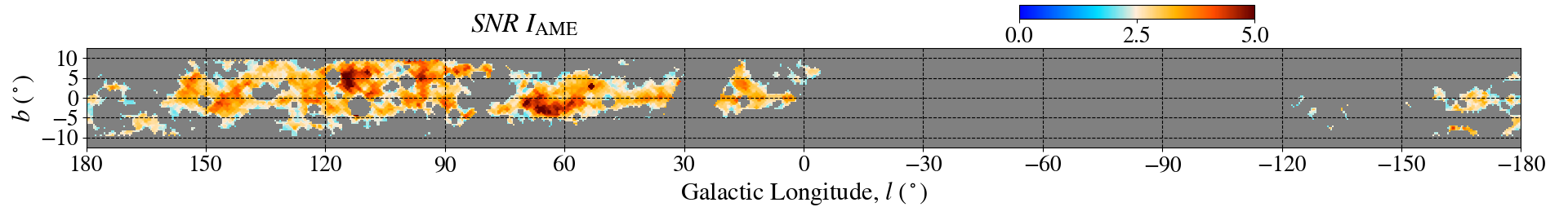}
    \caption{AME intensity map ($I_{\rm AME}$),
 with pixels with $\SNRAME<2$ masked.}
    \label{fig:snr_map_Aame_gt_3}
\end{figure*}

\begin{figure*}
    \centering
    \includegraphics[width=1\linewidth]{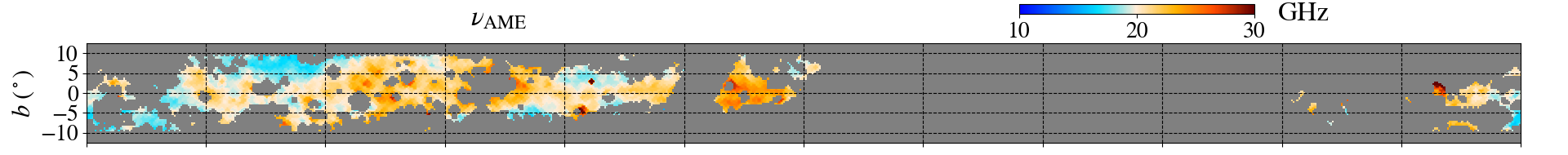}
    \includegraphics[width=1\linewidth]{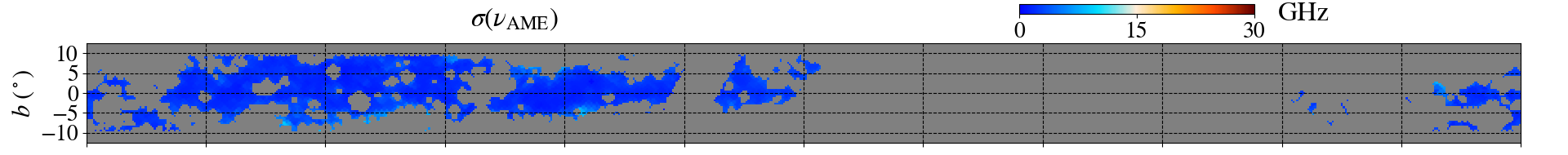}
    \includegraphics[width=1\linewidth]{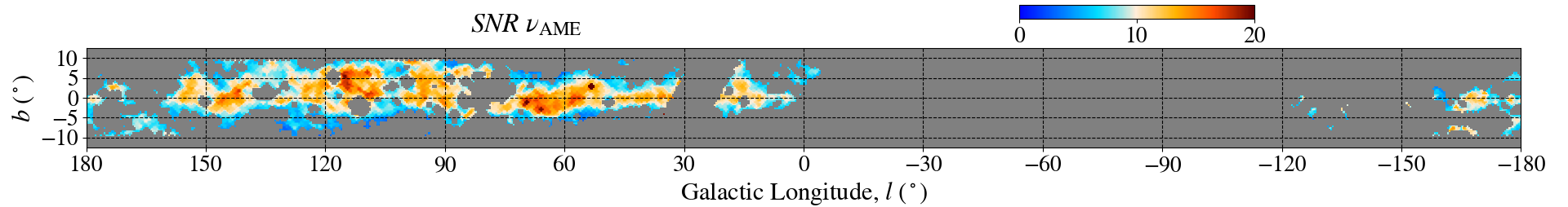}
    \caption{AME peak frequency ($\nuame$) map, 
with pixels with $\SNRAME<2$ masked.}
    \label{fig:snr_map_nuame_gt_3}
\end{figure*}

\begin{figure*}
    \centering
    \includegraphics[width=1\linewidth]{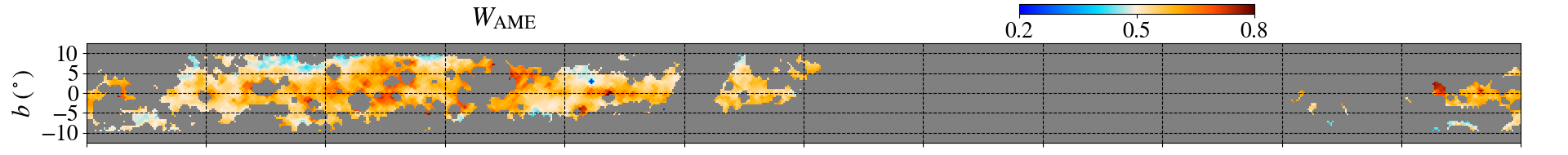}
    \includegraphics[width=1\linewidth]{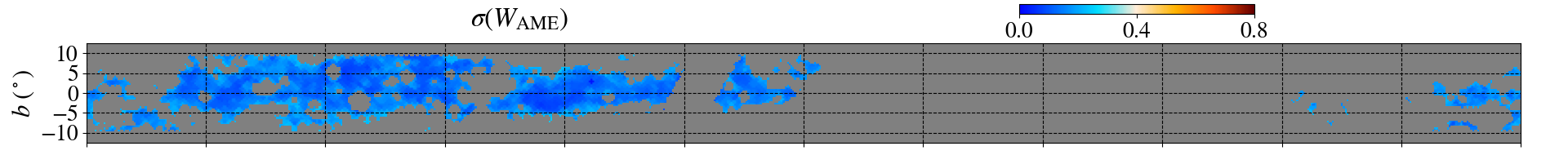}
    \includegraphics[width=1\linewidth]{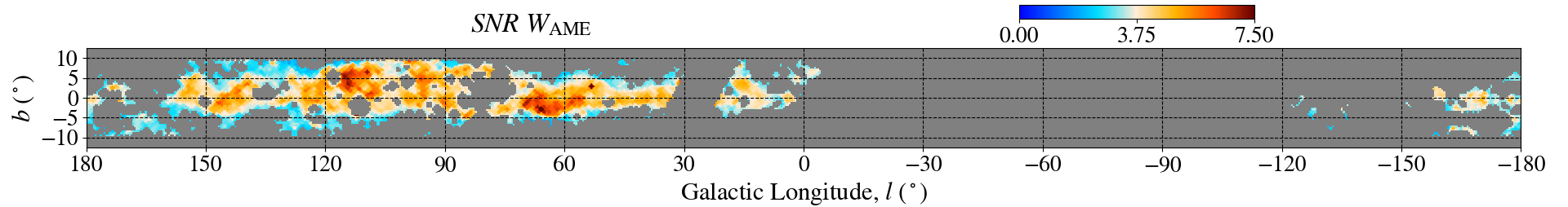}
    \caption{AME distribution width ($\Wame$) map, 
with pixels with $\SNRAME<2$ masked.}
    \label{fig:snr_map_Wame_gt_3}
\end{figure*}

\begin{figure*}
    \includegraphics[width=1\linewidth]{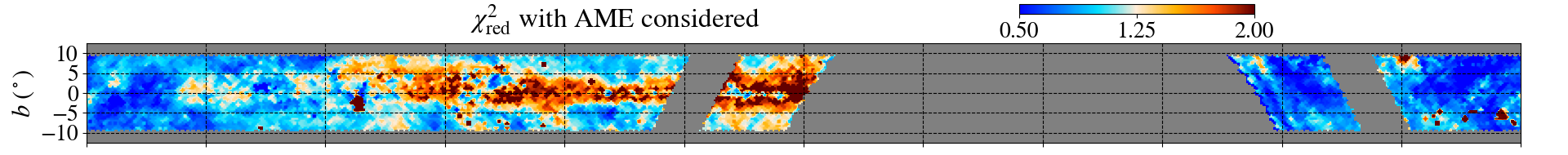}
    \includegraphics[width=1\linewidth]{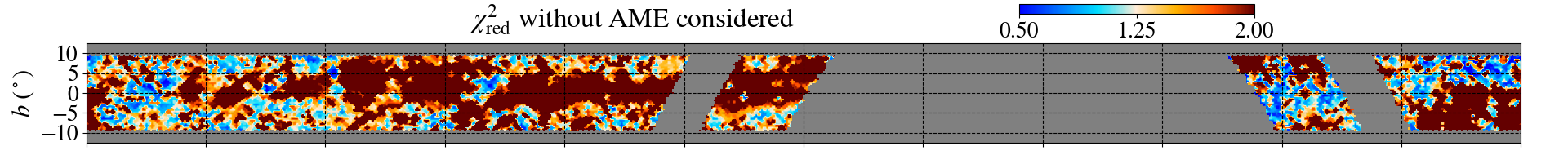}
    \includegraphics[width=1\linewidth]{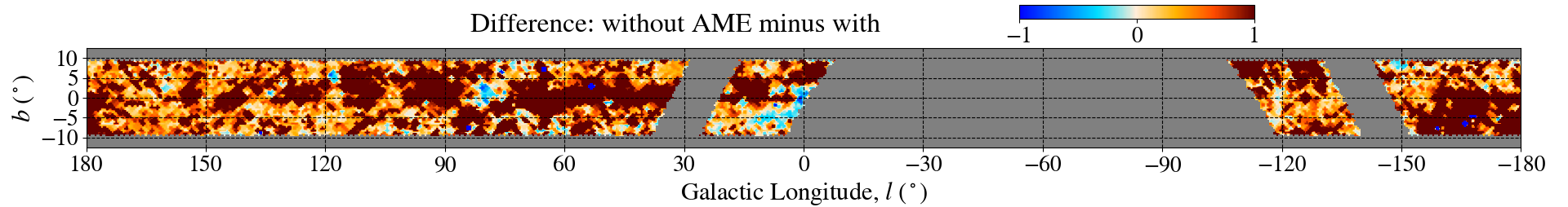}
    \caption{$\chi^2_{\rm red}$ results along the maps, both when
    an AME component is considered in the fit (top) and also with
    that same component missing (middle). In the bottom panel, the 
    difference (calculated as the result for the case without an
    AME component minus the case with it) between the two is shown.
    Values larger than 0 trace pixels with worst defined fits for
    the case without AME. We can see that that is the case for most
    of the Galactic plane, except for e.g. the Cygnus region (
    $l\approx80\degr$, strongly dominated by free-free emission).}
    \label{fig:chisq_w_wo_AMEcomparison}
\end{figure*}

In Figures~\ref{fig:snr_map_Aame_gt_3},~\ref{fig:snr_map_nuame_gt_3} 
and~\ref{fig:snr_map_Wame_gt_3} we show the maps
 for $\Iame$, $\nuame$ and $\Wame$ respectively, 
with pixels with $\SNRAME<2$ masked. This is the 
selection used for most of the analysis described 
in the text. We chose this threshold instead
 of the more common AME residual at 28.4\,GHz 
because of the high flux density variability we find between 
Galactic regions: we are interested in AME-dominated 
regions, not in those with the highest photometry
 residual at 28.4\,GHz (the Galactic centre and 
the Cygnus region would dominate the analysis in 
that case, while having low AME fractions).
Although more than half of the pixels show $\SNRAME$
values lower than 2, this does not mean that the
fit is better without an AME component. Instead, we 
have to use a goodness-of-fit statistic to discriminate
between the different models. In fact, when 
comparing the $\chi^2_{\rm red}$ map from the fit
considering AME and the one without it, the latter
is better only for those pixels dominated by 
free-free (mainly within the Cygnus region). This is
shown in Fig.~\ref{fig:chisq_w_wo_AMEcomparison}.
When focusing on the $\SNRAME>2$ sample, the preferred 
model (for 98 per cent of the pixels) is the one with 
AME, as one would expect.


\section{Linear regression fits and correlations between SED parameters}
\label{section:correlation_fits}
\begin{figure}
    \centering
    \includegraphics[width=1\linewidth]{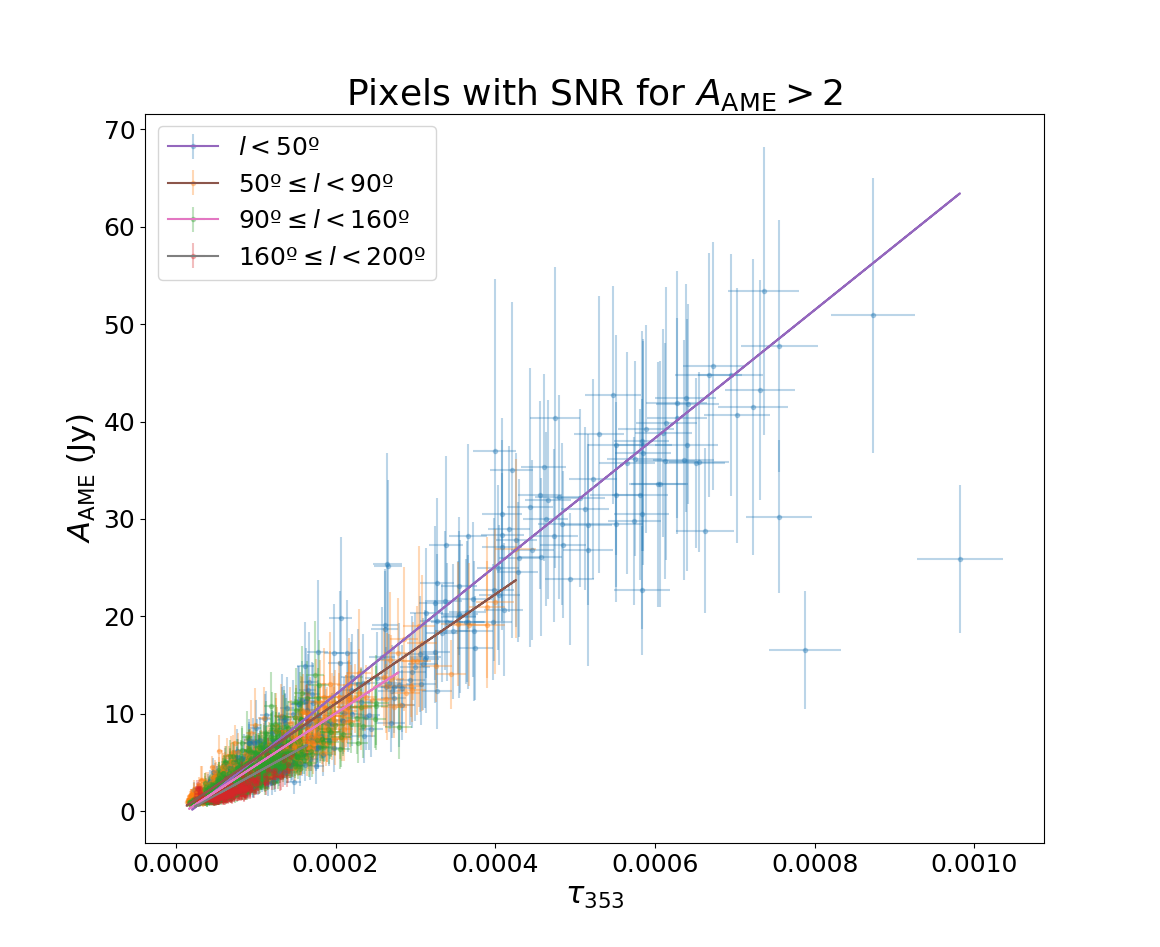}
    \begin{tabular}{cccc}
    \hline 
     \multicolumn{4}{c}{$\Aame$ [Jy] vs. $10^4\tau_{353}$}  \\
    \hline
     Region                        & SRCC                       & Slope                      & Intercept                  \\
     All pixels                    & $0.90\pm0.03$            & $5.76\pm0.02$        & $-0.69\pm0.01$             \\
     $l<50\degr$                  & $0.96\pm0.02$            & $6.58\pm0.03$        & $-1.16\pm0.03$             \\
     $50\degr\leq l<90\degr$     & $0.94\pm0.02$            & $5.61\pm0.03$        & $-0.17\pm0.01$             \\
     $90\degr\leq l < 160\degr$  & $0.90\pm0.03$            & $5.34\pm0.02$        & $-0.61\pm0.01$             \\
     $160\degr\leq l < 200\degr$ & $0.84\pm0.10$            & $4.51\pm0.06$        & $-0.58\pm0.03$             \\
    \hline
    \end{tabular}
    \caption{AME flux density amplitude ($\Aame$) 
versus dust opacity ($\taud$). First order fits are 
provided, along with the correspondent value for the Spearman 
Rank Correlation Coefficient (SRCC). 
Top: fits obtained using every pixel with $\SNRAME>2$,
 but separated into their longitude sectors. Bottom: results
 for the fits shown in the previous figure, plus the one using
 all $\SNRAME>2$ pixels together. We see that 
the fit obtained with all the pixels is closer to those 
in the regions $50\degr<l<90\degr$ and $90\degr<l<160\degr$, which 
was expected, as both regions together account for almost 
$\simeq68\,\%$ of the pixels in the sample. Slopes from the
$|l|<50\degr$ and anticentre regions are not consistent 
with these two (differences $\geq10\sigma$).}
    \label{fig:correlation_bw_Aame_and_dustopacity}
\end{figure}

\begin{figure}
    \centering
    \includegraphics[width=1\linewidth]{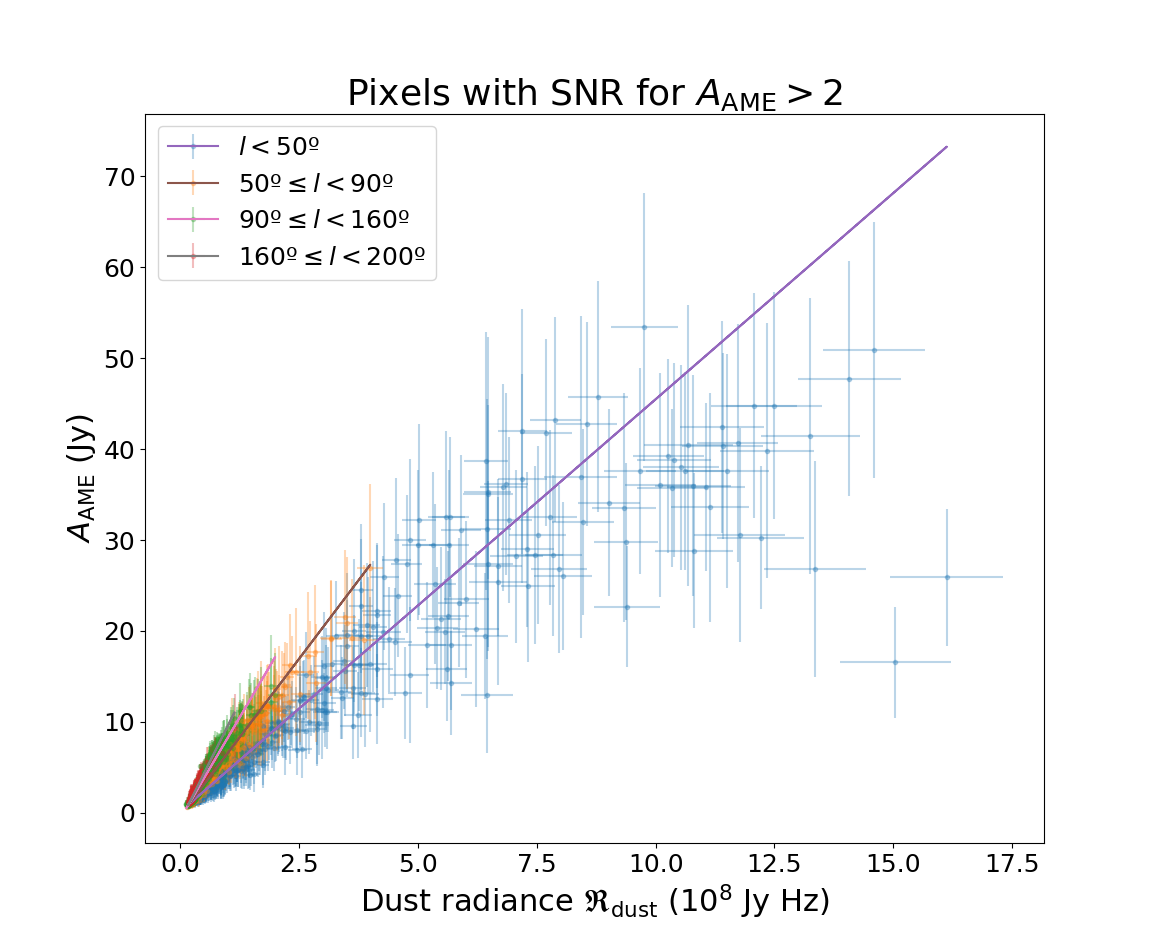}
    \begin{tabular}{cccc}
    \hline
    \multicolumn{4}{c}{$\Aame$ [Jy] vs. Dust radiance ($\mathfrak{R}_{\rm dust}$) [$10^8$ Jy Hz]} \\
    \hline
     Region                                                                 & SRCC                                                                   & Slope                                                                  & Intercept                                                              \\
     All pixels                                                             & $0.95\pm0.03$                                                        & $7.55\pm0.02$                                                          & $-0.29\pm0.01$                                                         \\
     $l<50\degr$                                                           & $0.98\pm0.02$                                                        & $4.53\pm0.02$                                                          & $0.13\pm0.01$                                                          \\
     $50\degr\leq l<90\degr$                                              & $0.98\pm0.02$                                                        & $6.90\pm0.04$                                                          & $-0.31\pm0.01$                                                         \\
     $90\degr\leq l < 160\degr$                                           & $0.98\pm0.03$                                                        & $8.89\pm0.04$                                                          & $-0.62\pm0.01$                                                         \\
     $160\degr\leq l < 200\degr$                                          & $0.90\pm0.11$                                                        & $10.47\pm0.16$                                                         & $-0.90\pm0.04$                                                         \\
    \hline
    \end{tabular}
    \caption{AME flux density amplitude ($\Aame$) versus 
dust radiance ($\mathfrak{R}$). 
Top: fits obtained using every pixel with $\SNRAME>2$,
 but separated into their longitude sectors. Bottom: results
 for the fits shown in the previous figure, plus the one using
 all $\SNRAME>2$ pixels together. Here, none of the slopes seem consistent 
against each other. The slope flattens as we get closer
 to the Galactic centre, hinting at a defect of $\Aame$ as
 dust signal increases. This is especially clear around 
$10^9$\,Jy Hz, where the relation seems to bend.}
    \label{fig:correlation_bw_Aame_and_dustradiance}
\end{figure}

\begin{figure}
    \centering
    \includegraphics[width=1\linewidth]{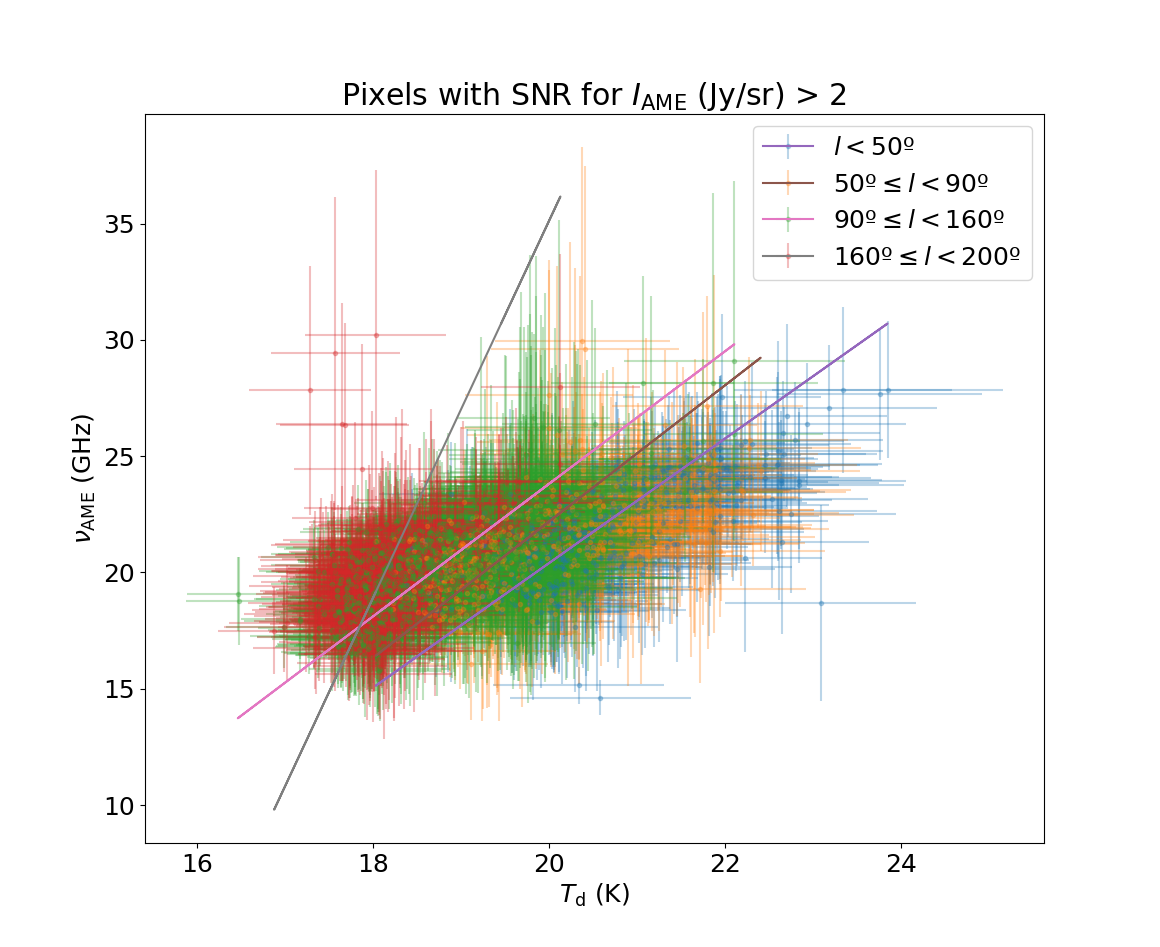}
    \begin{tabular}{cccc}
    \hline 
    \multicolumn{4}{c}{$\nuame$ [GHz] vs. $\Td$ [K]} \\
    \hline
     Region                        & SRCC                 & Slope                &    Intercept            \\
     All pixels                    & $0.63\pm0.11$      & $3.10\pm0.08$        & $-38.47\pm1.57$      \\
     $l<50\degr$                  & $0.69\pm0.13$      & $2.68\pm0.18$        & $-33.16\pm3.74$      \\
     $50\degr\leq l<90\degr$     & $0.66\pm0.16$      & $2.93\pm0.21$        & $-36.40\pm4.09$    \\
     $90\degr\leq l < 160\degr$  & $0.66\pm0.15$      & $2.85\pm0.12$        & $-33.20\pm2.31$      \\
     $160\degr\leq l < 200\degr$ & $0.52\pm0.13$      & $8.10\pm1.51$        & $-126.99\pm27.36$    \\
    \hline
    \end{tabular}
    \caption{AME peak frequency ($\nuame$) versus dust 
temperature ($\Td$). 
Top: fits obtained using every pixel with $\SNRAME>2$,
 but separated into their longitude sectors. Bottom: results
 for the fits shown in the previous figure, plus the one using
 all $\SNRAME>2$ pixels together. The fit for the last
 region, $160\degr<l<200\degr$, covering the anticentre, 
recovers a very different value, probably owing to the low 
number of pixels and their scatter: this result is 
unreliable, as the anticentre region has fewer 
detected pixels, with the lowest detected SNR.Opposite to previous $\Iame$ vs.\ dust correlations, here
the differences between the slopes are consistent to within 
1$\sigma$.}
    \label{fig:correlation_bw_Td_and_nuame}
\end{figure}

In this appendix we show in more detail the most 
interesting correlations between parameters. First, 
the well-known correlation between AME emission (or 
amplitude) and thermal dust is studied, both
 when using the dust opacity 
(Fig.~\ref{fig:correlation_bw_Aame_and_dustopacity}) 
and the dust radiance 
(Fig.~\ref{fig:correlation_bw_Aame_and_dustradiance}) 
as tracers for the latter. The correlation is 
good in all cases, but when fitting the data to a 
first order polynomial, the slope turns steeper/flatter
when approaching the Galactic Centre
(higher $\Iame$ values with increasing 
$\taud$/$\mathfrak{R}$). The AME peak frequency, 
$\nuame$ and dust temperature, $\Td$, are also 
correlated (Fig.~\ref{fig:correlation_bw_Td_and_nuame}), 
the SRCC index being similar for every region 
but the anticentre. The slope in the polynomial 
fits match between $l<50\degr$ and $90\degr<l<160\degr$, 
and then between $50\degr<l<90\degr$ and 
$160\degr<l<200\degr$. This could be due to different 
types of environments being the most important in the 
first two regions, and then another ones for the other
 two.

\begin{figure}
    \centering
    \includegraphics[width=1\linewidth]{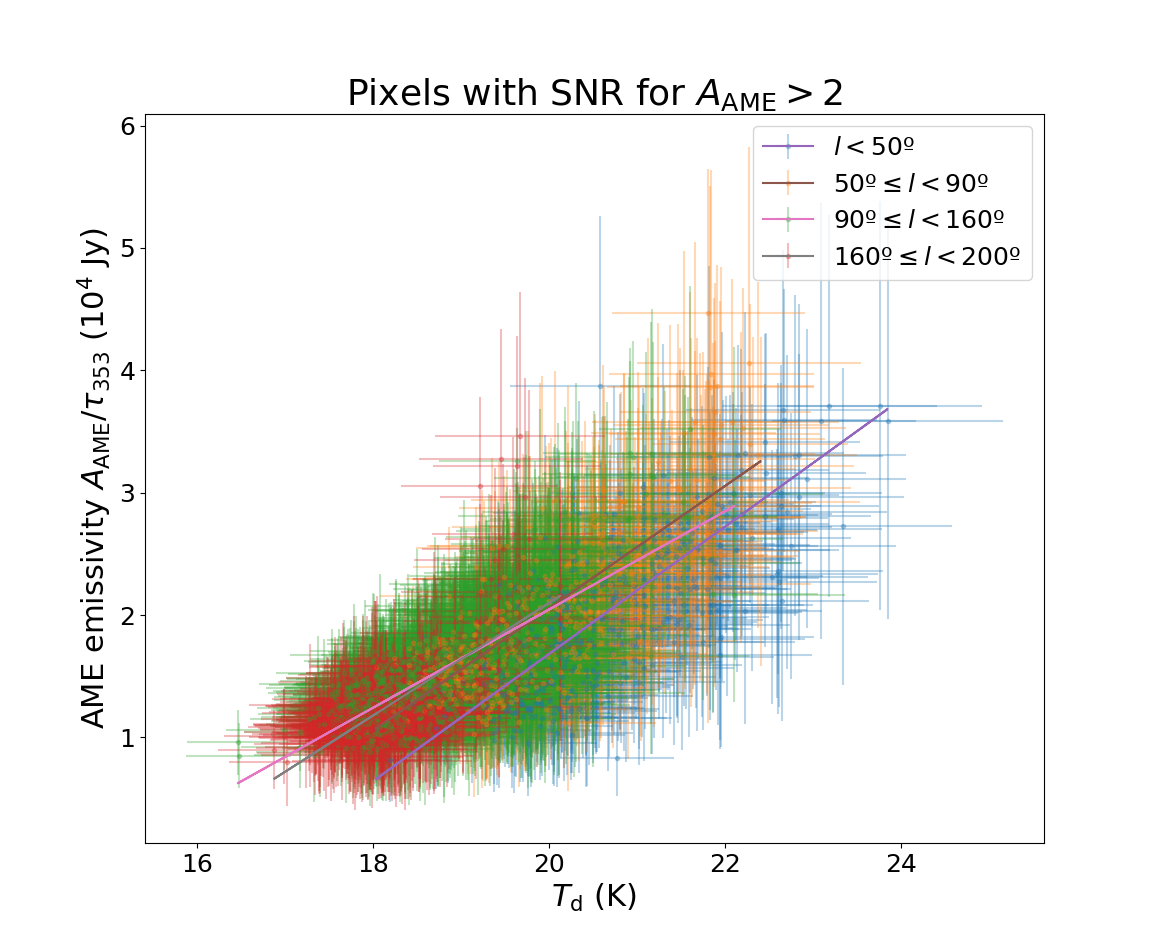}
    \begin{tabular}{cccc}
    \hline 
    \multicolumn{4}{c}{$\frac{A_{\rm AME}}{10^4\tau_{353}}$ [Jy] vs. $\Td$ [K]} \\
    \hline
     Region                                     & SRCC                                       & Slope                                      & Intercept                                  \\
     All pixels                                 & $0.76\pm0.12$                            & $1.22\pm0.03$                              & $-19.19\pm0.54$                            \\
     $l<50\degr$                               & $0.74\pm0.14$                            & $1.84\pm0.12$                              & $-32.78\pm2.47$                            \\
     $50\degr\leq l<90\degr$                  & $0.73\pm0.14$                            & $1.65\pm0.10$                              & $-27.54\pm2.02$                            \\
     $90\degr\leq l < 160\degr$               & $0.72\pm0.16$                            & $1.36\pm0.06$                              & $-21.58\pm1.12$                            \\
     $160\degr\leq l < 200\degr$              & $0.52\pm0.15$                            & $1.83\pm0.22$                              & $-29.99\pm3.98$                            \\
    \hline
    \end{tabular}
    \caption{Relation between AME emissivity 
($A_{\rm AME}/\tau_{353}$) and dust temperature ($\Td$). 
Top: fits obtained using every pixel with $\SNRAME>2$,
 but separated into their longitude sectors. Bottom: results
 for the fits shown in the previous figure, plus the one using
 all $\SNRAME>2$ pixels together. 
 Differences are a a little greater here than in the $\Td$ vs.\ 
 $\nuame$ relation, but the slopes from the two main areas 
$50\degr<l<90\degr$ and $90\degr<l<160\degr$ are consistent to 
 within $2\sigma$.}
    \label{fig:correlation_bw_AMEemissivity_and_Td}
\end{figure}

\begin{figure}
    \centering
    \includegraphics[width=1\linewidth]{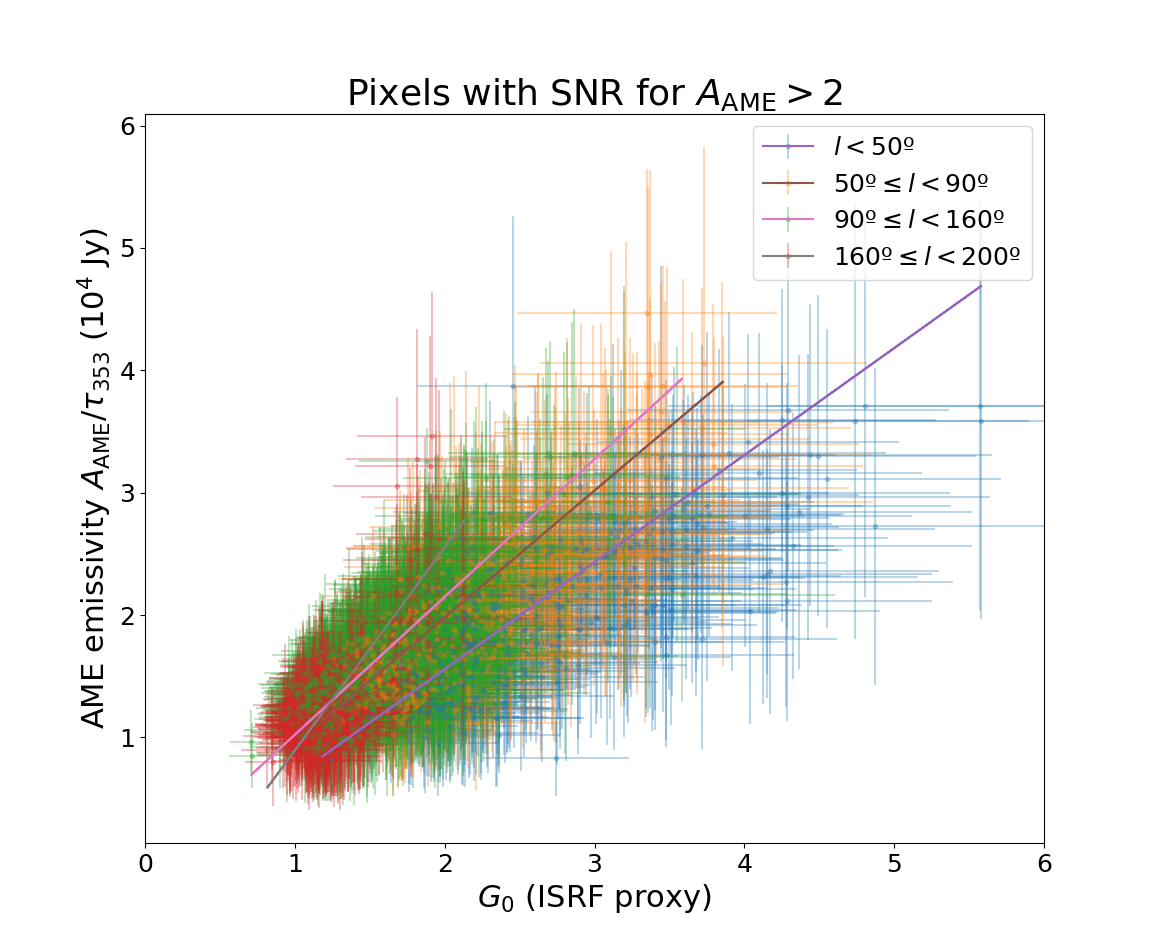}
    \begin{tabular}{cccc}
    \hline 
    \multicolumn{4}{c}{$\frac{A_{\rm AME}}{10^4\tau_{353}}$ [Jy] vs. $G_0$} \\
    \hline
     Region                                      & SRCC                                        & Slope                                       & Intercept                                   \\
     All pixels                                  & $0.76\pm0.12$                             & $2.96\pm0.07$                               & $-0.60\pm0.10$                              \\
     $l<50\degr$                                & $0.75\pm0.13$                             & $3.06\pm0.20$                               & $-2.32\pm0.45$                              \\
     $50\degr\leq l<90\degr$                   & $0.74\pm0.14$                             & $3.39\pm0.22$                               & $-1.54\pm0.40$                              \\
     $90\degr\leq l < 160\degr$                & $0.73\pm0.16$                             & $3.61\pm0.15$                               & $-1.27\pm0.21$                              \\
     $160\degr\leq l < 200\degr$               & $0.53\pm0.15$                             & $5.45\pm0.71$                               & $-3.37\pm0.83$                              \\
    \hline
    \end{tabular}
    \caption{Relation between AME emissivity 
($A_{\rm AME}/\tau_{353}$) and ISRF strength proxy 
$G_0$. 
Top: fits obtained using every pixel with $\SNRAME>2$,
 but separated into their longitude sectors. Bottom: results
 for the fits shown in the previous figure, plus the one using
 all $\SNRAME>2$ pixels together. The result is really 
similar to that using $\Td$ instead of $G_0$, as
 expected ($\Td$ guides $G_0$ behaviour), but the slopes 
are more consistent between regions ($\leq0.5\sigma$,
 apart from the anticentre region).}
    \label{fig:correlation_bw_AMEemissivity_and_G0}
\end{figure}

The correlation between the AME emissivity (the ratio
 between $\Iame$ and $\taud$) and the dust temperature, 
$\Td$, is shown in Fig.~\ref{fig:correlation_bw_AMEemissivity_and_Td}. 
In the case that AME and thermal dust emission 
come from the same carriers, both of these should be
 proportional to the column density in the region observed. 
Thus, when dividing one by the other, that dependence is 
erased, and regions with different densities along the
 line-of-sight can be compared with each other with more
 confidence. In fact, this is important in our work, as 
Galactic plane pixels present a lot of very different
 environments along their lines of sight. Now, the slopes 
for the different latitude bins are much more similar  
than for the previous correlations. The same result 
is achieved when using $G_0$, a proxy for the strength of 
the interstellar radiation field (or ISRF; 
Fig.~\ref{fig:correlation_bw_AMEemissivity_and_G0}), 
as it is mainly driven by $\Td$.

In Table~\ref{table:appendix_full_SRCC_table}  we present
all the correlations between the various
parameters considered in this study. We report the 
absence of correlations for some of the parameters, 
e.g.\ $\betad$. $\Wame$ has weak (0.5$<$SRCC$<$0.6)
 correlations with a few parameters ($EM$, $\nuame$ 
and $\mathfrak{R}_{\rm AME}$), but all show large
 uncertainties (> 0.2), so it is not clear if the
 correlation exists. Correlation with $\nuame$, however,  
has been proposed in the past from theoretical 
studies.


\begin{table*}
  \centering
  \caption{Spearman Rank Correlation Coefficients (SRCCs) 
values between all the parameters considered in this study,
after applying the $\SNRAME>2$ pixel threshold.
 We see how $\Wame$ and $\betad$, for example, lack 
correlations ($>0.6$) with any other parameter ($\Wame$ 
seems to be slightly correlated with $\nuame$, but the 
uncertainty is large: $0.58\pm0.27$).}
    \begin{tabular}{ccccccccc}
    \hline
     & $A_{\rm 1\,GHz}$ (Jy)  & $\alphasyn$      & EM (pc cm$^{-6}$)  & $\Aame$ (Jy)  & $\nuame$ (GHz)   & $\Wame$          & $\tau_{353}$    & $\betad$         \\
    \hline
     $A_{\rm 1\,GHz}$ (Jy)                    & -       & 0.59 $\pm$ 0.11 & 0.61 $\pm$ 0.08    & 0.71 $\pm$ 0.03      & 0.33 $\pm$ 0.04  & 0.20 $\pm$ 0.09  & 0.59 $\pm$ 0.02 & 0.38 $\pm$ 0.04  \\
     $\alphasyn$                               & 0.59 $\pm$ 0.11       & - & 0.71 $\pm$ 0.18    & 0.68 $\pm$ 0.13      & 0.44 $\pm$ 0.12  & 0.28 $\pm$ 0.14  & 0.62 $\pm$ 0.10 & 0.23 $\pm$ 0.05  \\
     EM (pc cm$^{-6}$)                       & 0.61 $\pm$ 0.08       & 0.71 $\pm$ 0.18 & -    & 0.90 $\pm$ 0.14      & 0.42 $\pm$ 0.08  & 0.50 $\pm$ 0.23  & 0.87 $\pm$ 0.11 & 0.05 $\pm$ 0.03  \\
     $\Aame$ (Jy)                     & 0.71 $\pm$ 0.03       & 0.68 $\pm$ 0.13 & 0.90 $\pm$ 0.14    & -      & 0.42 $\pm$ 0.05  & 0.37 $\pm$ 0.15  & 0.90 $\pm$ 0.02 & 0.21 $\pm$ 0.02  \\
     $\nuame$ (GHz)                         & 0.33 $\pm$ 0.04       & 0.44 $\pm$ 0.12 & 0.42 $\pm$ 0.08    & 0.42 $\pm$ 0.05      & -  & 0.58 $\pm$ 0.27  & 0.23 $\pm$ 0.02 & -0.11 $\pm$ 0.06 \\
     $\Wame$                                  & 0.20 $\pm$ 0.09       & 0.28 $\pm$ 0.14 & 0.50 $\pm$ 0.23    & 0.37 $\pm$ 0.15      & 0.58 $\pm$ 0.27  & -  & 0.22 $\pm$ 0.09 & -0.44 $\pm$ 0.21 \\
     $\tau_{353}$                             & 0.59 $\pm$ 0.02       & 0.62 $\pm$ 0.10 & 0.87 $\pm$ 0.11    & 0.90 $\pm$ 0.03      & 0.23 $\pm$ 0.02  & 0.22 $\pm$ 0.09  & - & 0.17 $\pm$ 0.02  \\
     $\betad$                                 & 0.38 $\pm$ 0.04       & 0.23 $\pm$ 0.05 & 0.05 $\pm$ 0.03    & 0.21 $\pm$ 0.02      & -0.11 $\pm$ 0.06 & -0.44 $\pm$ 0.21 & 0.17 $\pm$ 0.02 & -  \\
     $\Td$ (K)                              & 0.56 $\pm$ 0.03       & 0.30 $\pm$ 0.06 & 0.25 $\pm$ 0.03    & 0.29 $\pm$ 0.02      & 0.63 $\pm$ 0.11  & 0.41 $\pm$ 0.17  & 0.00 $\pm$ 0.02 & 0.07 $\pm$ 0.02  \\
     $\Delta T_{\rm CMB}$ (K)            & 0.10 $\pm$ 0.02       & 0.12 $\pm$ 0.03 & 0.01 $\pm$ 0.02    & 0.07 $\pm$ 0.02      & 0.13 $\pm$ 0.03  & -0.07 $\pm$ 0.03 & 0.04 $\pm$ 0.02 & 0.19 $\pm$ 0.03  \\
     $\Aame / \tau_{353}$         & 0.48 $\pm$ 0.07       & 0.36 $\pm$ 0.10 & 0.37 $\pm$ 0.07    & 0.51 $\pm$ 0.07      & 0.64 $\pm$ 0.16  & 0.44 $\pm$ 0.20  & 0.14 $\pm$ 0.02 & 0.19 $\pm$ 0.03  \\
     $S_{\rm TD,\ peak}$ (Jy)               & 0.80 $\pm$ 0.17       & 0.69 $\pm$ 0.22 & 0.88 $\pm$ 0.27    & 0.96 $\pm$ 0.22      & 0.37 $\pm$ 0.09  & 0.29 $\pm$ 0.16  & 0.91 $\pm$ 0.19 & 0.27 $\pm$ 0.04  \\
     $\mathfrak{R}_{\rm dust}$ (Jy Hz) & 0.83 $\pm$ 0.01       & 0.69 $\pm$ 0.12 & 0.87 $\pm$ 0.11    & 0.95 $\pm$ 0.03      & 0.40 $\pm$ 0.04  & 0.32 $\pm$ 0.12  & 0.88 $\pm$ 0.01 & 0.27 $\pm$ 0.03  \\
     $\mathfrak{R}_{\rm AME}$ (Jy Hz)         & 0.68 $\pm$ 0.11       & 0.68 $\pm$ 0.20 & 0.91 $\pm$ 0.23    & 0.97 $\pm$ 0.17      & 0.58 $\pm$ 0.14  & 0.55 $\pm$ 0.26  & 0.84 $\pm$ 0.13 & 0.10 $\pm$ 0.02  \\
     $G_0$                                    & 0.57 $\pm$ 0.03       & 0.31 $\pm$ 0.06 & 0.25 $\pm$ 0.03    & 0.30 $\pm$ 0.02      & 0.63 $\pm$ 0.11  & 0.41 $\pm$ 0.17  & 0.01 $\pm$ 0.02 & 0.09 $\pm$ 0.02  \\
     $S_{\rm AME}^{\rm 28.4\ GHz}$  (Jy)           & 0.77 $\pm$ 0.02       & 0.79 $\pm$ 0.15 & 0.91 $\pm$ 0.13    & 0.88 $\pm$ 0.04      & 0.45 $\pm$ 0.05  & 0.40 $\pm$ 0.16  & 0.82 $\pm$ 0.02 & 0.19 $\pm$ 0.02  \\
    \hline
    \end{tabular}
  \label{table:appendix_full_SRCC_table}
 \end{table*}

\begin{table*}
  \centering
  \contcaption{}
    \begin{tabular}{ccccccccc}
    \hline
     & $\Td$ (K)      & $\Delta T_{\rm CMB}$ (K)  & $A_{\rm AME} / \tau_{353}$   & $S_{\rm TD,\ peak}$ (Jy)  & $\mathfrak{R}_{\rm dust}$ (Jy Hz)   & $\mathfrak{R}_{\rm AME}$ (Jy Hz)   & $G_0$           & $S_{\rm AME}^{\rm 28.4\ GHz}$  (Jy)  \\
    \hline
     $A_{\rm 1\,GHz}$ (Jy)                     & 0.56 $\pm$ 0.03 & 0.10 $\pm$ 0.02            & 0.48 $\pm$ 0.07                   & 0.80 $\pm$ 0.17            & 0.83 $\pm$ 0.01                            & 0.68 $\pm$ 0.11                    & 0.57 $\pm$ 0.03 & 0.77 $\pm$ 0.02                 \\
     $\alphasyn$                               & 0.30 $\pm$ 0.06 & 0.12 $\pm$ 0.03            & 0.36 $\pm$ 0.10                   & 0.69 $\pm$ 0.22            & 0.69 $\pm$ 0.12                            & 0.68 $\pm$ 0.20                    & 0.31 $\pm$ 0.06 & 0.79 $\pm$ 0.15                 \\
     EM (pc cm$^{-6}$)                       & 0.25 $\pm$ 0.03 & 0.01 $\pm$ 0.02            & 0.37 $\pm$ 0.07                   & 0.88 $\pm$ 0.27            & 0.87 $\pm$ 0.11                            & 0.91 $\pm$ 0.23                    & 0.25 $\pm$ 0.03 & 0.91 $\pm$ 0.13                 \\
     $\Aame$ (Jy)                    & 0.29 $\pm$ 0.02 & 0.07 $\pm$ 0.02            & 0.51 $\pm$ 0.07                   & 0.96 $\pm$ 0.22            & 0.95 $\pm$ 0.03                            & 0.97 $\pm$ 0.17                    & 0.30 $\pm$ 0.02 & 0.88 $\pm$ 0.04                 \\
     $\nuame$ (GHz)                        & 0.63 $\pm$ 0.11 & 0.13 $\pm$ 0.03            & 0.64 $\pm$ 0.16                   & 0.37 $\pm$ 0.09            & 0.40 $\pm$ 0.04                            & 0.58 $\pm$ 0.14                    & 0.63 $\pm$ 0.11 & 0.45 $\pm$ 0.05                 \\
     $\Wame$                                  & 0.41 $\pm$ 0.17 & -0.07 $\pm$ 0.04           & 0.44 $\pm$ 0.21                   & 0.29 $\pm$ 0.16            & 0.32 $\pm$ 0.13                            & 0.55 $\pm$ 0.26                    & 0.41 $\pm$ 0.17 & 0.40 $\pm$ 0.16                 \\
     $\tau_{353}$                             & 0.00 $\pm$ 0.02 & 0.04 $\pm$ 0.02            & 0.14 $\pm$ 0.03                   & 0.91 $\pm$ 0.19            & 0.88 $\pm$ 0.01                            & 0.84 $\pm$ 0.13                    & 0.01 $\pm$ 0.02 & 0.82 $\pm$ 0.02                 \\
     $\betad$                                 & 0.07 $\pm$ 0.02 & 0.19 $\pm$ 0.03            & 0.19 $\pm$ 0.03                   & 0.27 $\pm$ 0.04            & 0.27 $\pm$ 0.03                            & 0.10 $\pm$ 0.02                    & 0.09 $\pm$ 0.02 & 0.19 $\pm$ 0.02                 \\
     $\Td$ (K)                              & - & 0.09 $\pm$ 0.02            & 0.76 $\pm$ 0.12                   & 0.33 $\pm$ 0.06            & 0.40 $\pm$ 0.03                            & 0.39 $\pm$ 0.07                    & 1.00 $\pm$ 0.09 & 0.37 $\pm$ 0.02                 \\
     $\Delta T_{\rm CMB}$ (K)                & 0.09 $\pm$ 0.02 & -            & 0.16 $\pm$ 0.03                   & 0.06 $\pm$ 0.02            & 0.06 $\pm$ 0.02                            & 0.07 $\pm$ 0.02                    & 0.09 $\pm$ 0.02 & 0.25 $\pm$ 0.03                 \\
     $\Aame / \tau_{353}$          & 0.76 $\pm$ 0.12 & 0.16 $\pm$ 0.03            & -                   & 0.40 $\pm$ 0.11            & 0.45 $\pm$ 0.05                            & 0.59 $\pm$ 0.14                    & 0.76 $\pm$ 0.12 & 0.43 $\pm$ 0.06                 \\
     $S_{\rm TD,\ peak}$ (Jy)               & 0.33 $\pm$ 0.06 & 0.06 $\pm$ 0.02            & 0.40 $\pm$ 0.11                   & -            & 1.00 $\pm$ 0.22                            & 0.92 $\pm$ 0.30                    & 0.34 $\pm$ 0.06 & 0.91 $\pm$ 0.20                 \\
     $\mathfrak{R}_{\rm dust}$ (Jy Hz) & 0.40 $\pm$ 0.02 & 0.06 $\pm$ 0.02            & 0.45 $\pm$ 0.05                   & 1.00 $\pm$ 0.22            & -                            & 0.92 $\pm$ 0.14                    & 0.40 $\pm$ 0.02 & 0.91 $\pm$ 0.02                 \\
     $\mathfrak{R}_{\rm AME}$ (Jy Hz)        & 0.39 $\pm$ 0.07 & 0.07 $\pm$ 0.02            & 0.59 $\pm$ 0.14                   & 0.92 $\pm$ 0.30            & 0.92 $\pm$ 0.14                            & -                    & 0.40 $\pm$ 0.06 & 0.88 $\pm$ 0.15                 \\
     $G_0$                                    & 1.00 $\pm$ 0.09 & 0.09 $\pm$ 0.02            & 0.76 $\pm$ 0.12                   & 0.34 $\pm$ 0.05            & 0.40 $\pm$ 0.02                            & 0.40 $\pm$ 0.06                    & - & 0.38 $\pm$ 0.02                 \\
     $S_{\rm AME}^{\rm 28.4\ GHz}$  (Jy)           & 0.37 $\pm$ 0.02 & 0.25 $\pm$ 0.03            & 0.43 $\pm$ 0.06                   & 0.91 $\pm$ 0.20            & 0.91 $\pm$ 0.02                            & 0.88 $\pm$ 0.15                    & 0.38 $\pm$ 0.02 & -                 \\
    \hline
    \end{tabular}
  \label{table:appendix_full_SRCC_table}
 \end{table*}

\section{Computation of SRCC uncertainties}
\label{section:appendix_SRCC}
We followed the procedure described in \cite{SRCCmcmc}. For that, 
we took the pairs $x_i-y_i$ and reordered them randomly 
(being $x$ and $y$ the two variables we want to assess their 
correlation from), but maintining the pair together. To 
take into account the individual uncertainties from the points, 
we introduce a random offset to each sample of both parameters, 
so $x_i'=x_i+\Delta x_i$ and $y_i'=y_i+\Delta y_i$ form our new
pair. These offsets are taken from a Gaussian distribution
scaled to $\sigma_{x_i}$ and $\sigma_{y_i}$ values, respectively.
Finally, we estimate the variance of the SRCC as the sum
of the squared difference between the SRCC value from the
real data and the SRCC values obtained after the reordering 
of the data plus the addition of the offsets:
\begin{equation}
    \sigma^2(SRCC) = \frac{1}{N}\sum_i^N\left[SRCC(x,y) - SRCC(x',y')\right]^2
\end{equation}
where $x=x_1,...,x_i,...,x_M$, $y=y_1,...,y_i,...,y_M$ 
(being $M$ the number of samples, or pairs)
and $N$ is the number of different $(x',y')$ sets produced.

\bsp
\clearpage
\end{document}